\documentclass{aa}
\usepackage{txfonts}
\usepackage{graphicx}
\usepackage{hyperref}
\usepackage{xcolor}
\usepackage{overpic}
\usepackage{CJK}

\usepackage[verbose]{placeins}
\newcommand{\XMM}{ XMM-{\em Newton}}

\newcommand{\egs}{erg cm$^{-2}$ s$^{-1}$}
\newcommand{\N}[1]{$N_\textrm{H}$}
\newcommand{\arcdeg}{\mbox{$^\circ$}}
\newcommand{\ledit}[1]{#1}
\newcommand{\edit}[1]{#1}
\newcommand{\TL}[1]{#1}

\begin{document}
\begin{CJK*}{UTF8}{gkai}
\title{Establishing the X-ray source detection strategy for eROSITA with simulations}

   \author{Teng~Liu (刘腾) \thanks{email: liu@mpe.mpg.de}\inst{\ref{inst:mpe}}\and
     Andrea~Merloni\inst{\ref{inst:mpe}}\and
     Johan~Comparat\inst{\ref{inst:mpe}}\and
Kirpal~Nandra\inst{\ref{inst:mpe}}\and
Jeremy~Sanders\inst{\ref{inst:mpe}}\and
Georg~Lamer\inst{\ref{inst:aip}}\and
Johannes~Buchner\inst{\ref{inst:mpe}}\and
Tom~Dwelly\inst{\ref{inst:mpe}}\and
Michael~Freyberg\inst{\ref{inst:mpe}}\and
Adam~Malyali\inst{\ref{inst:mpe}}\and
Antonis~Georgakakis\inst{\ref{inst:athens}}\and
Mara~Salvato\inst{\ref{inst:mpe}}\and
Hermann~Brunner\inst{\ref{inst:mpe}}\and
Marcella~Brusa\inst{\ref{inst:bologna1},\ref{inst:bologna2}}\and
Matthias~Klein\inst{\ref{inst:mpe}}\and
Vittorio~Ghirardini\inst{\ref{inst:mpe}}\and
Nicolas~Clerc\inst{\ref{inst:irap}}\and
Florian~Pacaud\inst{\ref{inst:bonn}}\and
Esra~Bulbul\inst{\ref{inst:mpe}}\and
Ang~Liu (刘昂)\inst{\ref{inst:mpe}}\and
Axel~Schwope\inst{\ref{inst:aip}}\and
Jan~Robrade\inst{\ref{inst:hamburg}}\and
J\"orn~Wilms\inst{\ref{inst:remeis}}\and
Thomas~Dauser\inst{\ref{inst:remeis}}\and
Miriam~E.~Ramos-Ceja\inst{\ref{inst:mpe}}\and
Thomas~H.~Reiprich\inst{\ref{inst:bonn}}\and
Thomas~Boller\inst{\ref{inst:mpe}}\and
Julien~Wolf\inst{\ref{inst:mpe}}
}

\institute{Max-Planck-Institut f\"ur extraterrestrische Physik, Giessenbachstra{\ss}e 1, D-85748 Garching bei M\"unchen, Germany\label{inst:mpe}
\and Leibniz-Institut f\"ur Astrophysik, An der Sternwarte 16, 14482 Potsdam, Germany\label{inst:aip}
\and Institute for Astronomy and Astrophysics, National Observatory of Athens, V. Paulou and I. Metaxa 11532, Greece\label{inst:athens}
\and Dipartimento di Fisica e Astronomia, Universit\`a di Bologna, Via Piero Gobetti 93/2, 40129 Bologna, Italy\label{inst:bologna1}
\and INAF - Osservatorio di Astrofisica e Scienza dello Spazio di Bologna, Via Piero Gobetti 93/3, 40129 Bologna, Italy\label{inst:bologna2}
\and IRAP, Universite de Toulouse, CNRS, UPS, CNES, 31028 Toulouse, France\label{inst:irap}
\and Argelander-Institut für Astronomie, Rheinische Friedrich-Wilhelms-Universität Bonn, Auf dem Hügel 71, 53121 Bonn, Germany\label{inst:bonn}
\and Hamburger Sternwarte, University of Hamburg, Gojenbergsweg 112, 21029 Hamburg, Germany\label{inst:hamburg}
\and Dr.~Karl Remeis-Sternwarte \& Erlangen Centre for Astroparticle Physics, Sternwartstr.~7, 96049 Bamberg, Germany \label{inst:remeis}}

   \date{} 

  \abstract
  {The eROSITA X-ray telescope on board the Spectrum-Roentgen-Gamma (SRG) satellite has started to \TL{detect} new X-ray sources over the full sky at an unprecedented rate. Understanding the \TL{performance and} selection function of the source detection is important for the subsequent scientific analysis of the eROSITA catalogs.}
   {Through simulations, we test and optimize the eROSITA source detection procedures, and we characterize the detected catalog quantitatively.}
   {Taking the eROSITA Final Equatorial-Depth Survey (eFEDS) as an example, we ran extensive photon-event simulations \edit{based on} our best knowledge of the instrument characteristics, the background spectrum, and the population of astronomical X-ray sources. We \edit{introduce a method of analyzing source detection completeness, purity, and efficiency} based on the origin of each photon.}
   {\edit{According to the source detection efficiency measured in the simulation, we chose a two-pronged strategy to build eROSITA X-ray catalogs, creating a main catalog using only the most sensitive band (0.2-2.3 keV) and an independent hard-band-selected catalog using multiband detection in a range up to 5 keV.
     Because our mock data are highly representative of the real eFEDS data, we used the mock catalogs to measure the completeness and purity of the eFEDS catalogs as a function of multiple parameters, such as detection likelihood, flux, and luminosity.
     These measurements provide a basis for choosing the eFEDS catalog selection thresholds.
     The mock catalogs (available with this paper) can be used to construct the selection function of active galactic nuclei and galaxy clusters.
     A direct comparison of the output and input mock catalogs also gives rise to a correction curve that converts the raw point-source flux distribution into the intrinsic number counts distribution.
   }
       }
   {}

\titlerunning{eROSITA simulation}
\authorrunning{Liu et al.}
   \maketitle
\end{CJK*}
\section{Introduction}
Significant developments have been made in X-ray surveys \TL{in the past decades}. More and more X-ray sources are detected, resolving an increasingly larger fraction of the cosmic X-ray background. With the largest grasp \TL{in the soft X-ray band} of current X-ray imaging telescopes, the eROSITA telescope is detecting new X-ray sources at an unprecedented rate. It is expected to detect millions of active galactic nuclei (AGN) and \TL{ more than 10000 galaxy clusters} in the ongoing eROSITA all-sky survey \citep[eRASS;][]{Merloni2012, Predehl2021}.
To maximize the impact of the survey, it is important to choose a source detection strategy that is appropriate and ideally optimized for the main scientific applications of the survey. 

\TL{
  The eROSITA is extremely efficient in X-ray imaging surveys not only because of its large grasp, but also because of its scanning observation mode.
  During the eRASS surveys, it continuously scans the sky and covers the full sky every six months.
  Ahead of the four-year eRASS survey, we performed the eROSITA Final Equatorial-Depth Survey \citep[eFEDS;][]{Brunner2021}, which was designed as a prototype survey of eRASS.
  Using raster-scanning mode, the eFEDS survey observes a 140~degree$^2$ field centered at RA 136\arcdeg, DEC 1.5\arcdeg{} in $\sim100$ hours. With a scanning speed of $\sim13\arcsec/s$, the field has a relatively uniform exposure depth of $\sim$2.2~ks (0.2--2.3~keV vignetted depth $\sim$1.2~ks), which is about 50\% higher than the final depth of the four-year eRASS survey at the same position.
  The average point spread function (PSF) in the scanning mode has a half-energy width of $\sim$26\arcsec{} at 1.5 keV \citep{Predehl2021}.
  The eFEDS survey is currently the largest continuous X-ray survey and highly representative of the final eRASS data in extragalactic regions.
  Therefore, we can investigate the eROSITA source detection through this survey.
  }

  \TL{
The detection of astrophysical sources in imaging X-ray surveys is challenging because the ultimate sensitivity of these surveys generally probes the low count-rate regime, leading to high Poisson fluctuations. In addition, the spatial resolution of X-ray telescopes is often relatively high compared to the sky density of potential X-ray emitting sources. 
The eROSITA Science Analysis Software System \citep[eSASS; version eSASS\_users201009;][]{Brunner2021} is employed in the source detection. It provides at least two methods of defining the source detection likelihood: PSF-fitting likelihood, based on maximum likelihood image fitting with the PSF model, and a Poissonian likelihood, based on aperture count extraction.
\ledit{The detection} likelihood is always defined as corresponding to a probability of being false (probability $=$ exp(-likelihood)). However, in practice, the false rate cannot be accurately predicted theoretically.
The performance of source detection algorithms depends on instrumental characteristics, observing strategy, and background, combined with the nature of the target sources (e.g., brightness, spectral shape, extent, and number density). In dealing with these effects, every algorithm has adjustable parameters that need to be optimized. Realistically, the source detection problem is sufficiently complex that, to fully characterize an X-ray detection scheme or catalog in terms of completeness and purity, realistic simulations are needed. 
In this work, we simulate the eFEDS survey with two goals: 1) to investigate and optimize the eROSITA source detection strategy, and 2) to quantify the completeness and purity of the eFEDS catalogs.
}

Simulation tests have often been used in previous X-ray surveys \citep[e.g.,][]{LaMassa2013, Liu2013} and were employed for eROSITA pre-launch in order to forecast the instrument capabilities for detecting galaxy clusters and AGN \citep{Clerc2018}.
Typically, simulation results are analyzed at the catalog level, that is, by comparing the input and output catalogs on the basis of source positions and fluxes.
In the case of deep X-ray surveys with relatively modest spatial resolution, the source detection process is not just a question of distinguishing real sources from background fluctuations. The blending of point sources and the unknown profile of extended sources \ledit{introduce} additional complexity and uncertainty, which are hard to quantify at the catalog level.
To address these issues, we here analyze simulation results at the event level.
We use the SIXTE software to simulate the X-ray event files. SIXTE is the official eROSITA end-to-end simulator \citep[][;provided by ECAP/Remeis observatory\footnote{https://www.sternwarte.uni-erlangen.de/research/sixte/}]{Dauser2019}.
It takes an input catalog to create photons, which are then propagated through the mirrors and incident on the detector. It uses the measured energy-dependent PSF and vignetting to simulate the mirrors as close as possible to the flying detector. The incident photons create a charge cloud and are then read out as single events and are then reconstructed. This approach allows correctly predicting the split patterns and also including detector effects such as pile-up \citep{Dauser2019}.
Because of the nature of these simulations, the origin of each X-ray photon in the event list is known, encoded in the ID of the input source or ID of the background component. Based on the origin of each photon, we introduce a strategy of attributing detected sources to input sources, which could reveal any potential source blending or misclassification \ledit{(extended or} unresolved). This allows us to analyze the properties of the detected sources at a more detailed level.

The content of the paper can be summarized as follows: in \S~\ref{sec:method} we use the measurements of the eROSITA background based on the real eFEDS data to create a mock eFEDS dataset.
\TL{
In \S~\ref{sec:DET_dis} we describe the method of characterizing the nature of detected sources and measure the purity of the eFEDS catalog in various manners, for instance, the fraction of spurious sources, the fraction of misclassified clusters, and the fraction of misclassified AGN.
In \S~\ref{sec:efficiency} we investigate the selection function of AGN and clusters in terms of detected fraction as a function of source properties and the source detection efficiency in terms of completeness--contamination parametric curve. We optimize the eFEDS source detection according to the source detection efficiency.
  In \S~\ref{sec:ape_logNlogS} we test the construction of point-source number counts.
  The results are summarized in \S~\ref{sec:conclusion}.
}

\section{Simulating eFEDS data} \label{sec:method}
\subsection{Input}
To simulate the eFEDS source detection, we began with the inputs we describe below.
\begin{enumerate}
\item Hardware characteristics\\
The currently available updated eROSITA calibration files were used, including the 2D PSF model version 190219v05, the vignetting model version 4.0, and the normalized single-pattern\TL{ redistribution matrix file (RMF)} version 20170725 \citep[][]{Dennerl2020,Brunner2021}.
\item Observing strategy\\
We used the telescope attitude (at every instant) of the eFEDS cleaned event file (version c001) that was used to build the eFEDS catalogs \citep{Brunner2021}.
\TL{As a result of the raster-scanning mode, the field has a sharp drop in exposure depth at the border, where it has not only a much lower exposure depth, but also a different vignetting and spatial resolution from the main field.}
We simulated the full eFEDS field, but the tests for catalog completeness and purity were made within the inner 90\% area (126.6 degree$^2$) region where the 0.2-2.3 keV vignetted exposure value is above 500s. In this region, the exposure depth of eFEDS is relatively flat. The number count analysis of the real eFEDS catalog is also limited to this region \citep{Brunner2021}.
\item Input sources and their background\\
As input sources, we have one cluster catalog and two point-source catalogs, one for AGN and one for stars.
Based on the Universe N-body simulations for the Investigation of Theoretical models from galaxy surveys
\citep[UNIT;][]{Chuang2019}, \citet{Comparat2019} and \citet{Comparat2020} created a full-sky light cone \ledit{for} AGN and clusters, extending to redshift 6.1.
We extracted $18$ nonoverlapping regions \TL{of the eFEDS-field shape} from the full-sky mock catalogs. 
Particularly for AGN, the adopted spectral model is composed of an absorbed power law with $\Gamma=1.9$ plus three additional weak components (\texttt{TBabs(plcabs + constant*powerlaw + pexrav + zgauss)} in \texttt{Xspec}), including a soft scattered power law with the same $\Gamma,$ but $1\%$ normalization, a cold reflection component with reflection scaling factor 1, and a narrow 6.4 keV Gaussian emission line with an equivalent width of 0.3 keV against an unabsorbed power law.
The AGN population follows a realistic luminosity function and obscuring column density distributions as measured by previous X-ray surveys \citep{Comparat2019}.
The soft fluxes of the input AGN catalog extend to much lower values $\sim 10^{-17}$ \egs \citep{Comparat2019} than the source detection limit of eFEDS ($\sim 7\times10^{-15}$ \egs).

The cluster fluxes and surface brightness profiles were constructed based on the dark matter halo properties\TL{ as described in \citet{Comparat2020}.
We summarize the method here. 
Given a set of dark matter halo properties (mass, redshift, ellipticity, and offset parameter), we assigned an X-ray emissivity profile and image \TL{ to each halo using a profile generator trained} on a set of observed clusters. 
For the complete cluster population, the emissivity profiles reproduce the distribution and scatter as a function of scale as measured in \citet{Ghirardini2019}. 
The nature of the core (cool vs. noncool) devised by the central part of the profile was mapped on the offset parameter following a tentative physical correlation \citep{Seppi2021}. 
Simulated clusters are isothermal. 
Compared to \citet{Comparat2020}, we lower the mass limit to $M_{500c}=10^{13}M_\odot$. $M_{500c}$ is the halo mass inside the radius $R_{500c}$ , where the mass density is 500 times the universe critical density.
This low limit leads to a strong Malmquist bias on the X-ray luminosities of clusters, as discussed in \S~4.1.1 of \citet{Comparat2020}.
We performed an empirical correction to the luminosities in order to align the scaling relation between luminosity and stellar mass with that measured by \citet{Anderson2015MNRAS.449.3806A}.
The obtained population reproduces the cluster X-ray luminosity function and \TL{number counts (see \S~\ref{sec:mockvsreal})}.}
A typical Galactic \N{H} of $3.26\times10^{20}$ cm$^{-2}$ was adopted for the X-ray spectra of the AGN and clusters.
The stellar catalog follows the logN-logS observed in the eFEDS field.

\TL{As described below in \S~\ref{sec:bkg}}, we measured the background spectra from the real eFEDS data and decomposed the vignetted and unvignetted components. They were simulated separately using the proper vignetting models.
The mock and real data are compared in detail in \S~\ref{sec:mockvsreal}.
\item Software settings\\
We used the same source detection pipeline as was used to build the real eFEDS catalogs \citep{Brunner2021}.
\end{enumerate}

Based on these four inputs, we used sixte-2.6.2 to create mock event files.
Each input source and each background component has a unique ID, and each simulated photon has a flag of the input ID.

\subsection{eFEDS background spectra}
\label{sec:bkg}

Background plays a crucial role in the detection of faint sources.
We considered the background as two components, a vignetted and an unvignetted component.
The vignetted component corresponds to the X-ray photons from the sky, that is, diffuse Galactic X-ray emission and cosmic X-ray background, which are transmitted through the mirror before hitting the CCD.
The unvignetted component is mainly due to high-energy particles, which hit the camera and generate secondary X-ray emissions inside the camera.
Electronic noise also contributes to the unvignetted component.

\begin{figure}[hptb]
\centering
\includegraphics[width=\columnwidth]{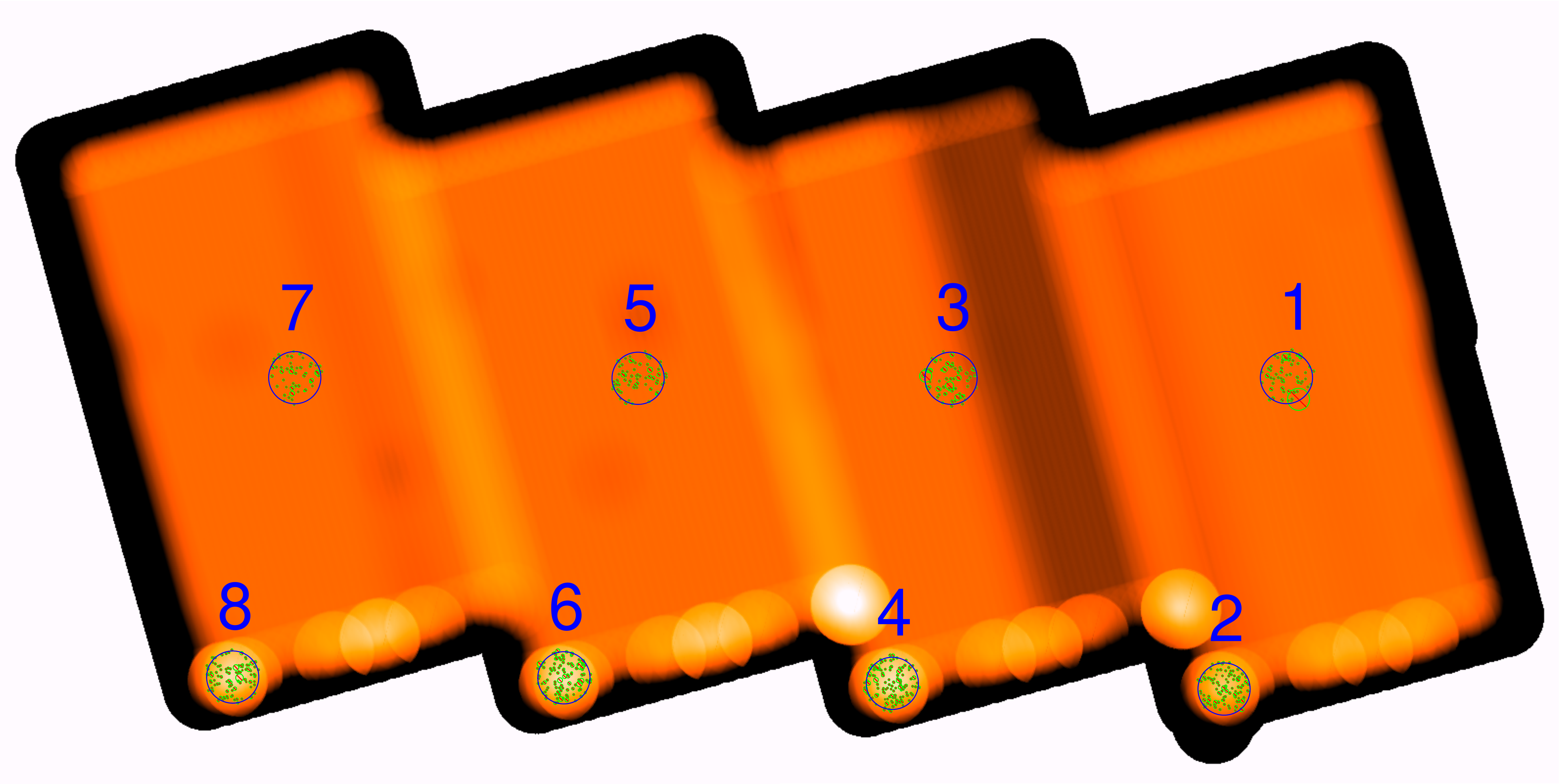}
\includegraphics[width=0.8\columnwidth]{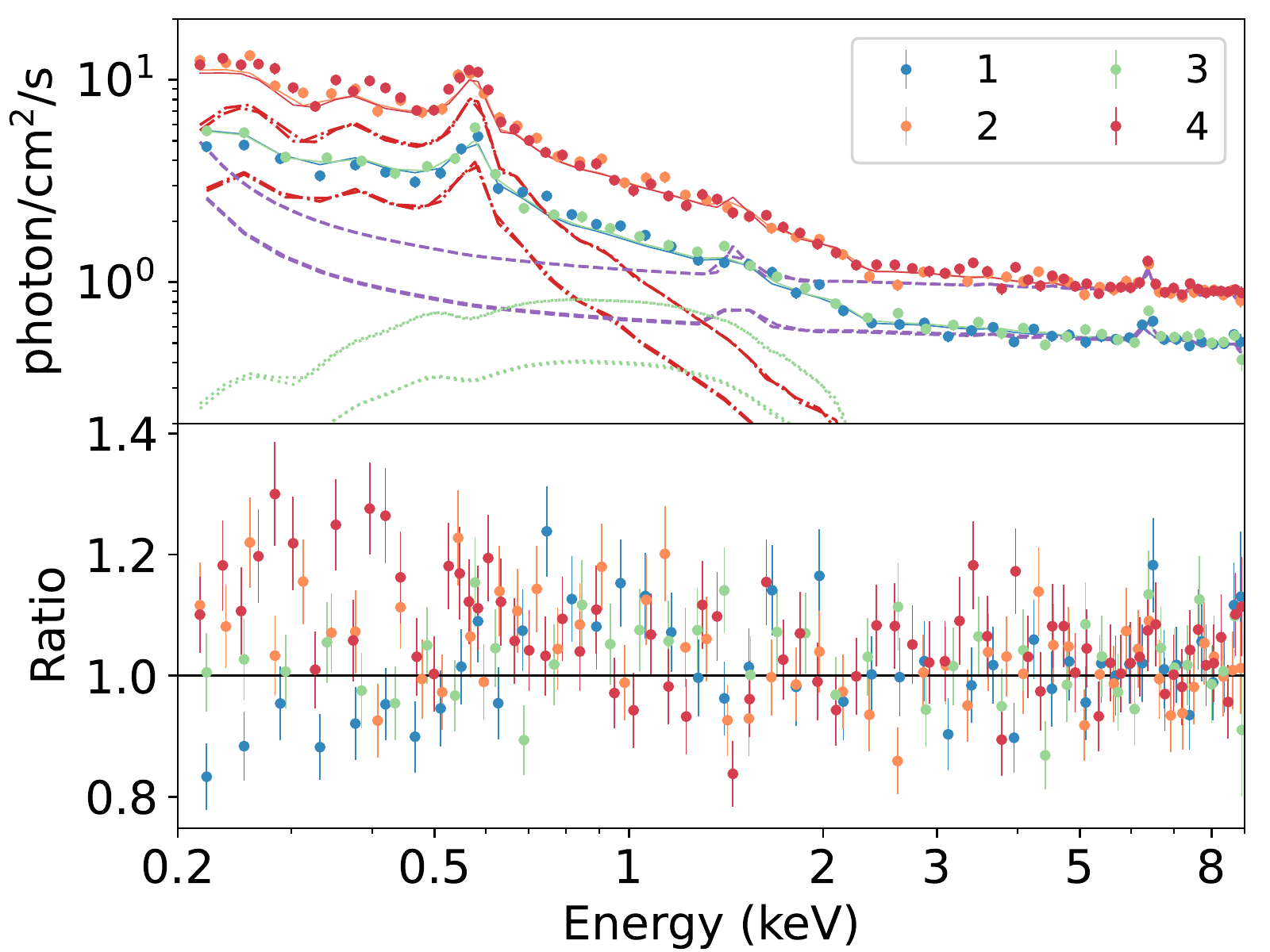}
\includegraphics[width=0.8\columnwidth]{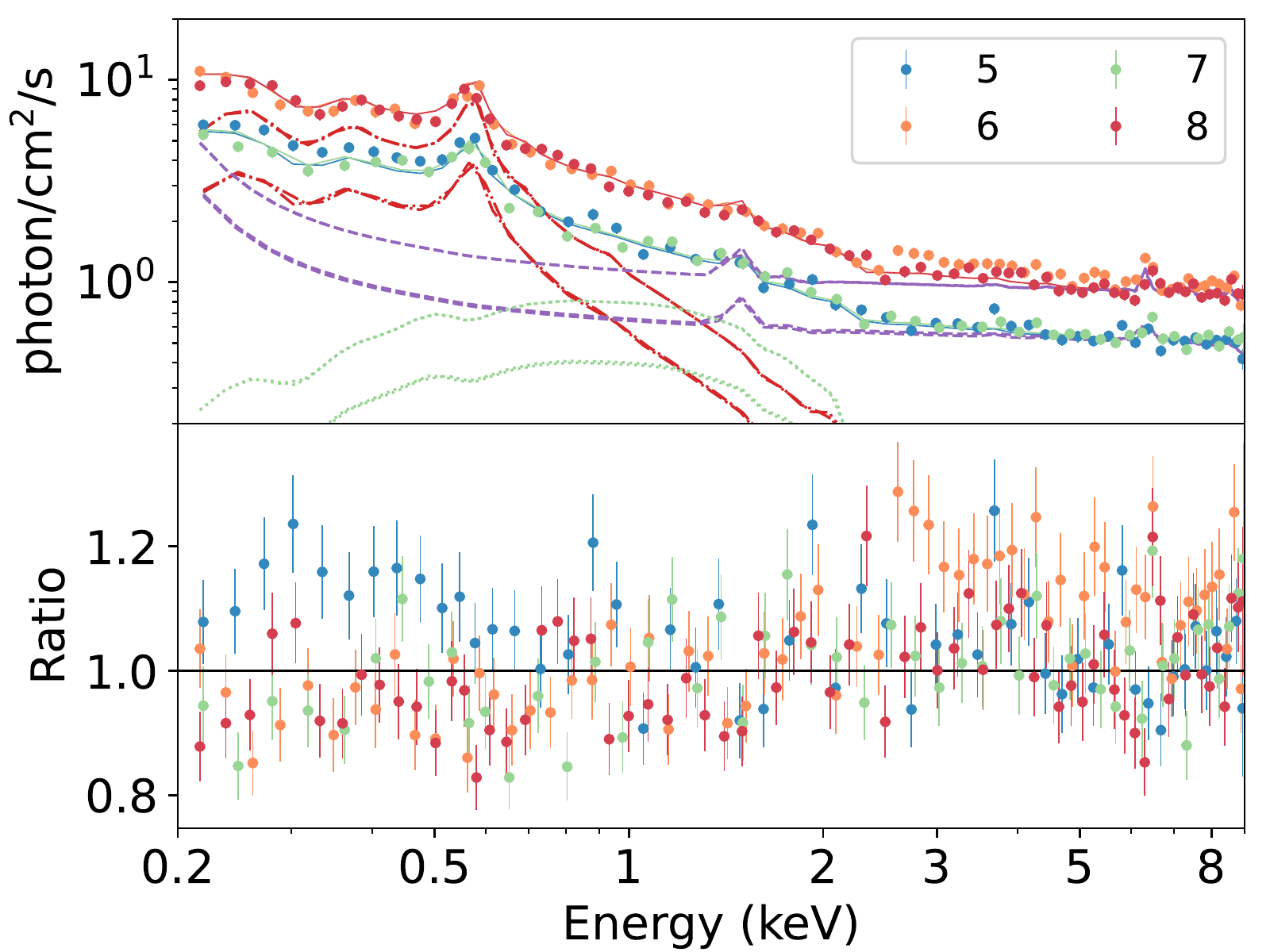}
\caption{
  Eight background extraction regions with radius 20$\arcmin$  overlaid on the 0.2-2.3 keV exposure map of eFEDS (top).
  To avoid overlapping, the eight spectra are displayed separately in the two lower panels in terms of the data (points), the model (solid lines), and the data-to-model ratio.
  The purple, green, and red lines indicate the averaged particle background, the mock cosmic X-ray background, and the diffuse Galactic X-ray background, respectively.
  \edit{All the lines correspond to the same model with the same flux for the eight spectra\protect\footnotemark.}
\label{fig:bkg}
}
\end{figure}

In order to make the mock data as representative as possible of the real data, we measured the background from the eFEDS data (version c001) that were used to build the eFEDS catalogs \citep{Brunner2021}, adopting all the valid events \TL{(single, double, triple, and quadruple patterns)} and all the seven telescopes.
\citet{Liu2021_AGN} extracted the spectra of all the eFEDS sources. We used their source-excluding regions and extracted background spectra from eight circular source-free regions with a radius of 20\arcmin \ as displayed in Fig.~\ref{fig:bkg}.
The eight regions were chosen to be evenly distributed in the field and are representative of both the region in pure scanning-mode (regions 1, 3, 5, and 7) and of the region in which the pointing stays still for a long time (regions 2, 4, 6, and 8).
These two types of regions have different scanning strategies and thus different vignetting.

\footnotetext{\edit{The count rate in the spectra was calculated based on the total exposure time, which is longer than the effective exposure depth in survey mode because the extraction region is not always covered by the moving FOV. The difference (by a factor of $\sim 2$) between regions 1, 3, 5, and 7 and regions 2, 4, 6, and 8 reflects the different effective exposure depths. We did correct for the count rate because the uncorrected data look better for the representation.}}

Because our input AGN catalog has a very low flux limit ($\sim 10^{-17}$\egs), \edit{some faint sources do not contribute any signal at all, while a large number of them contribute a few photons that are too few to make the source detectable.
  These photons from undetectable AGN compose a mock cosmic X-ray background.
  This is not necessarily the same as the real cosmic X-ray background \citep{Brandt2005,Brandt2015}, as it depends on the assumption of the AGN luminosity function adopted when creating the mock AGN catalog \citep{Comparat2020}.
  However, in the background spectral model described above, the real cosmic X-ray background is already included, thus this mock cosmic X-ray background is a duplication.
  Therefore, we measured this background component in order to exclude it from the background spectral model.}
Because we simulated the source and background events separately, we created a special version of mock data with source events alone.
We performed a similar background extraction from the pure-source-event data as was done above for the real data, that is, we ran the source detection on it and then extracted a background spectrum from a source-free region. This background component from the undetectable input AGN can be fit with a partially covering absorbed power law with $\Gamma=1.41$, \N{H}$=3.5\times 10^{21}$ cm$^{-2}$, and a covering factor of $52\%$.
\edit{This spectral component was only measured in order to be excluded from our background model}, so that this component is not duplicated in the simulation.

Then we fit each of the real background spectra with three components, as shown in Fig.~\ref{fig:bkg}.
The first component was the cosmic X-ray component measured above from the undetectable input AGN.
The second component was the particle background.
We adopted the phenomenological spectral shape of the eROSITA Filter Wheel Closed (FWC) data and normalized it to the eFEDS background spectrum in the 4.5-9 keV band. In this band, the spectrum is dominated by particle background and can be well fit with the FWC spectral model.
Then we fixed the parameters of the above two components and fit the residual signal, which is the foreground diffuse X-ray emission, using a phenomenological model composed of a power law, an APEC \citep{Smith2001} plasma model, and a Gaussian component.

The eROSITA background has been found to be relatively constant in time and thus flat in the scanned field \citep[][]{Brunner2021, Predehl2021}.
In Fig.~\ref{fig:bkg} we also test the background variability by comparing the eight background spectra with the averaged model. The background is highly constant in the hard band. It is only relatively higher in region 6 by a factor of $<$20\%.
In the very soft band below 0.5 keV, the background is relatively more variable, which might be caused by the spatial variability of the Galactic emission and/or the time variability of light leak \citep{Predehl2021}.
\TL{Overall, the variability has a moderate amplitude \ledit{and its spatial scales are much larger} than the PSF size}, thus the impact on source detection is small. We therefore assumed a flat background in the simulation.
The particle background components measured from the eight regions were averaged and converted into a SIXTE particle-background file (available with the SIXTE package).
The foreground diffuse X-ray background components from the eight regions were also averaged, creating an X-ray background spectral model.
We ran SIXTE to simulate the particle background events and X-ray background events separately, and we finally merged them into the source signal event file.

\subsection{Mock versus real data}
\label{sec:mockvsreal}
\begin{figure*}[hptb]
\begin{center}
  \begin{overpic}[width=0.3\textwidth]{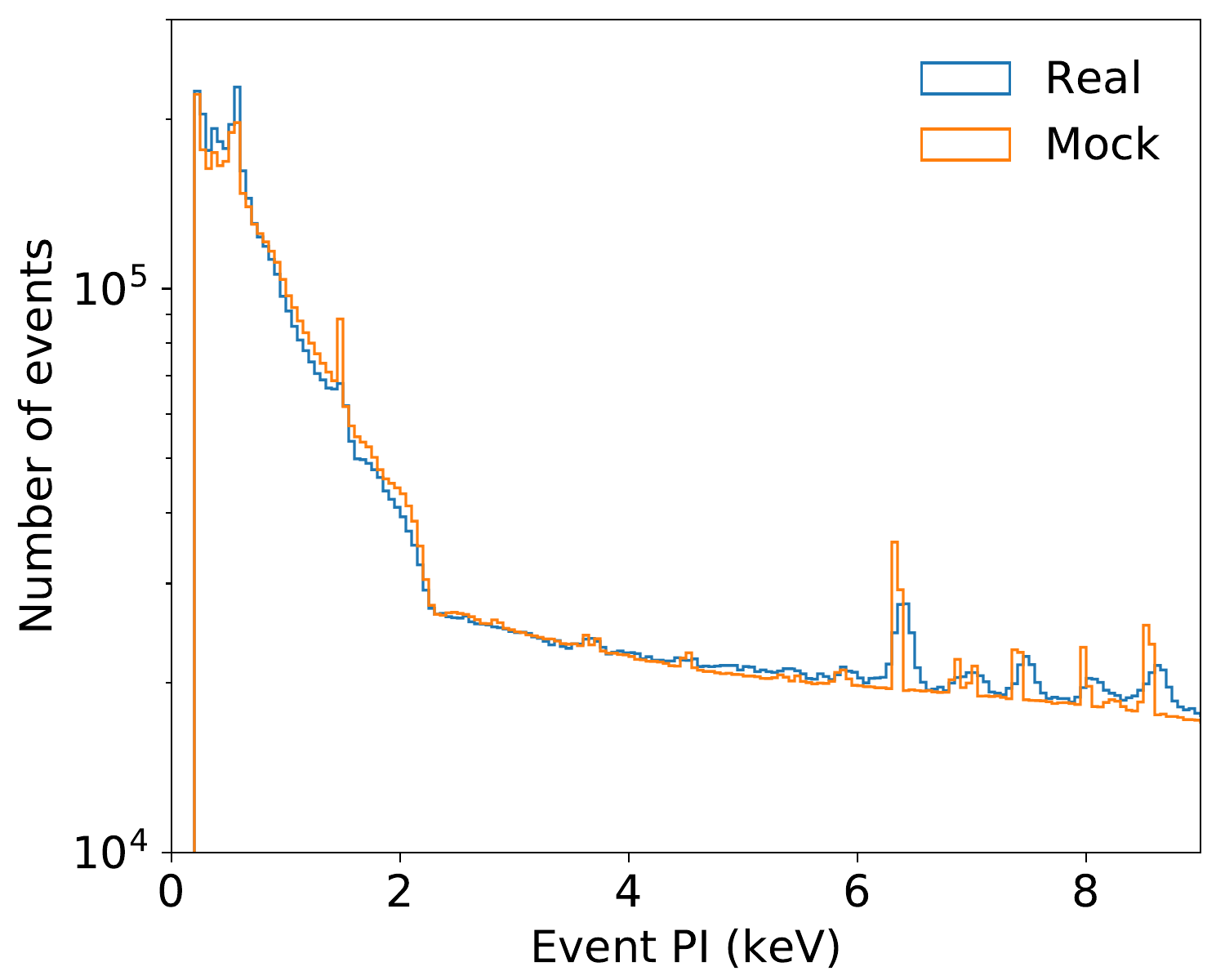}
    \put(20,20){(1)} \end{overpic}
  \begin{overpic}[width=0.3\textwidth]{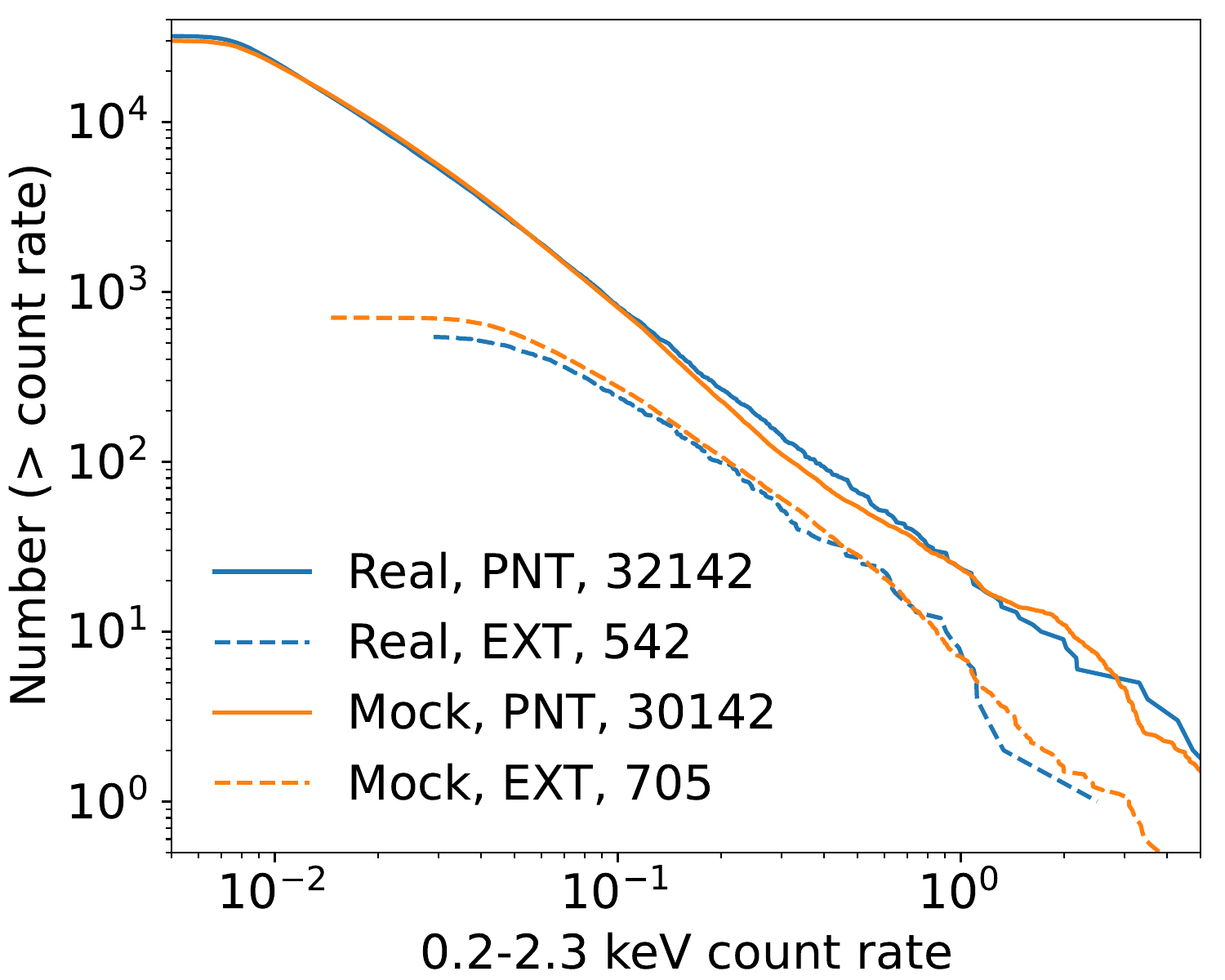}
    \put(80,70){(3)} \end{overpic}
  \begin{overpic}[width=0.3\textwidth]{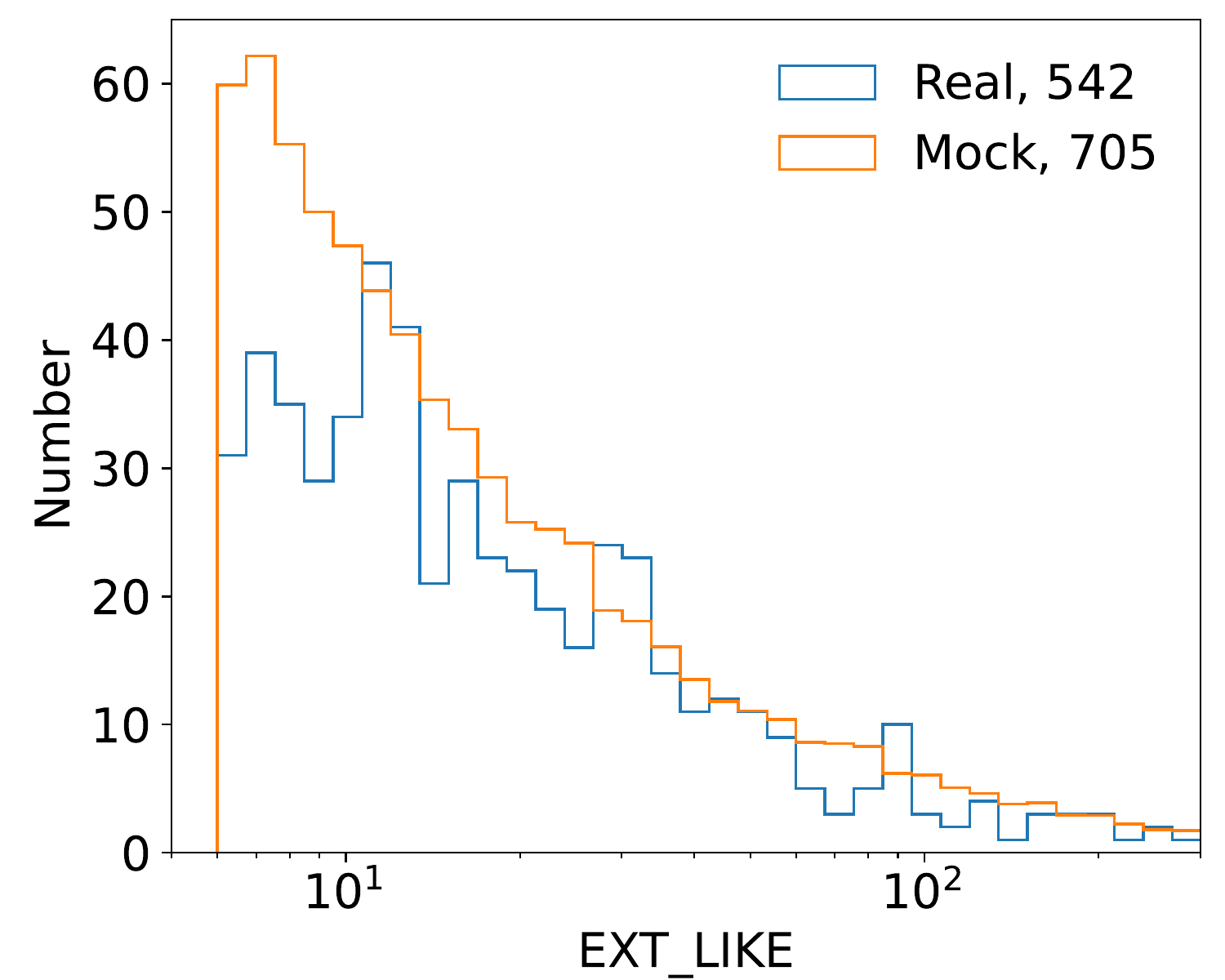}
    \put(80,50){(5)} \end{overpic}
  \begin{overpic}[width=0.3\textwidth]{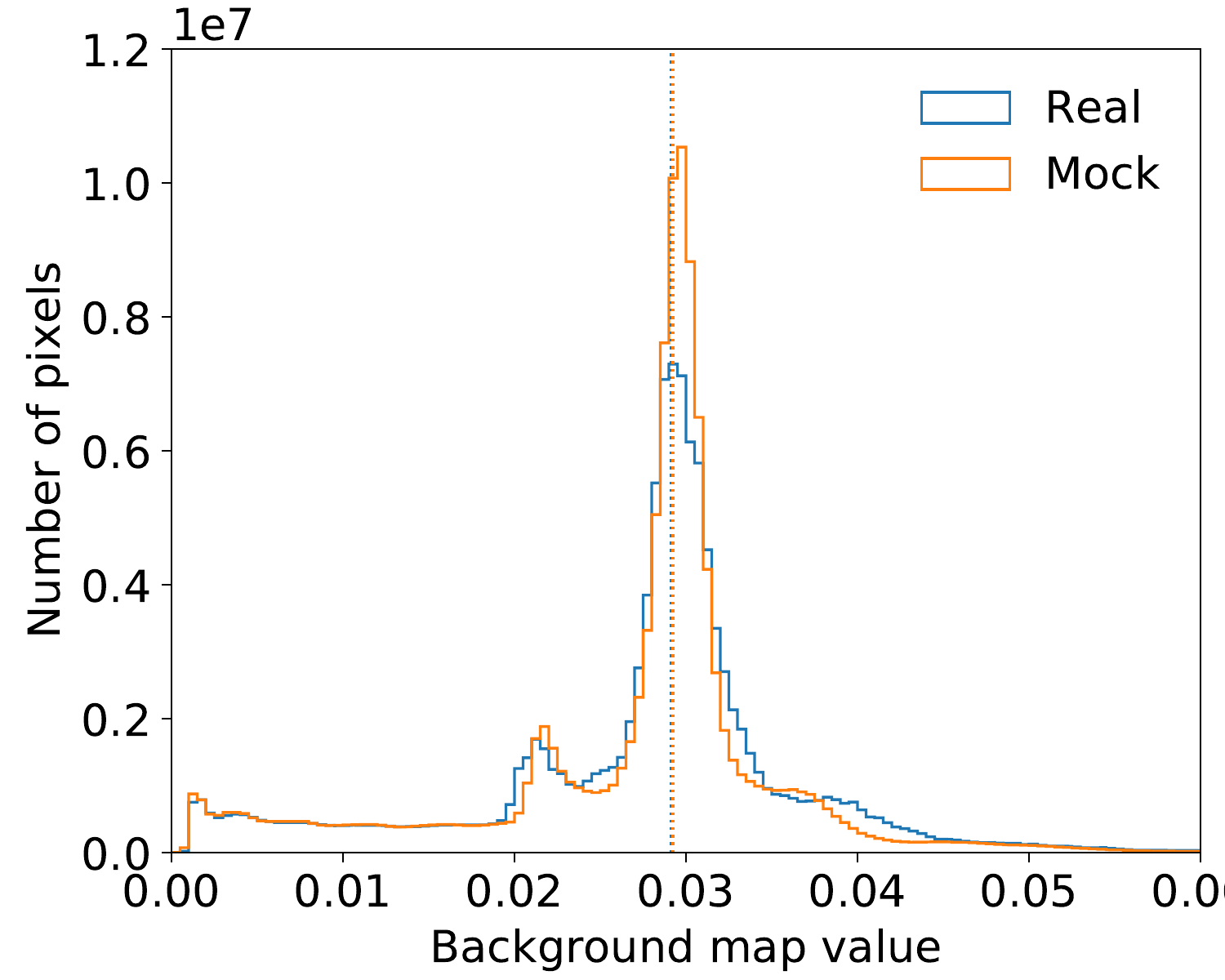}
    \put(20,20){(2)} \end{overpic}
  \begin{overpic}[width=0.3\textwidth]{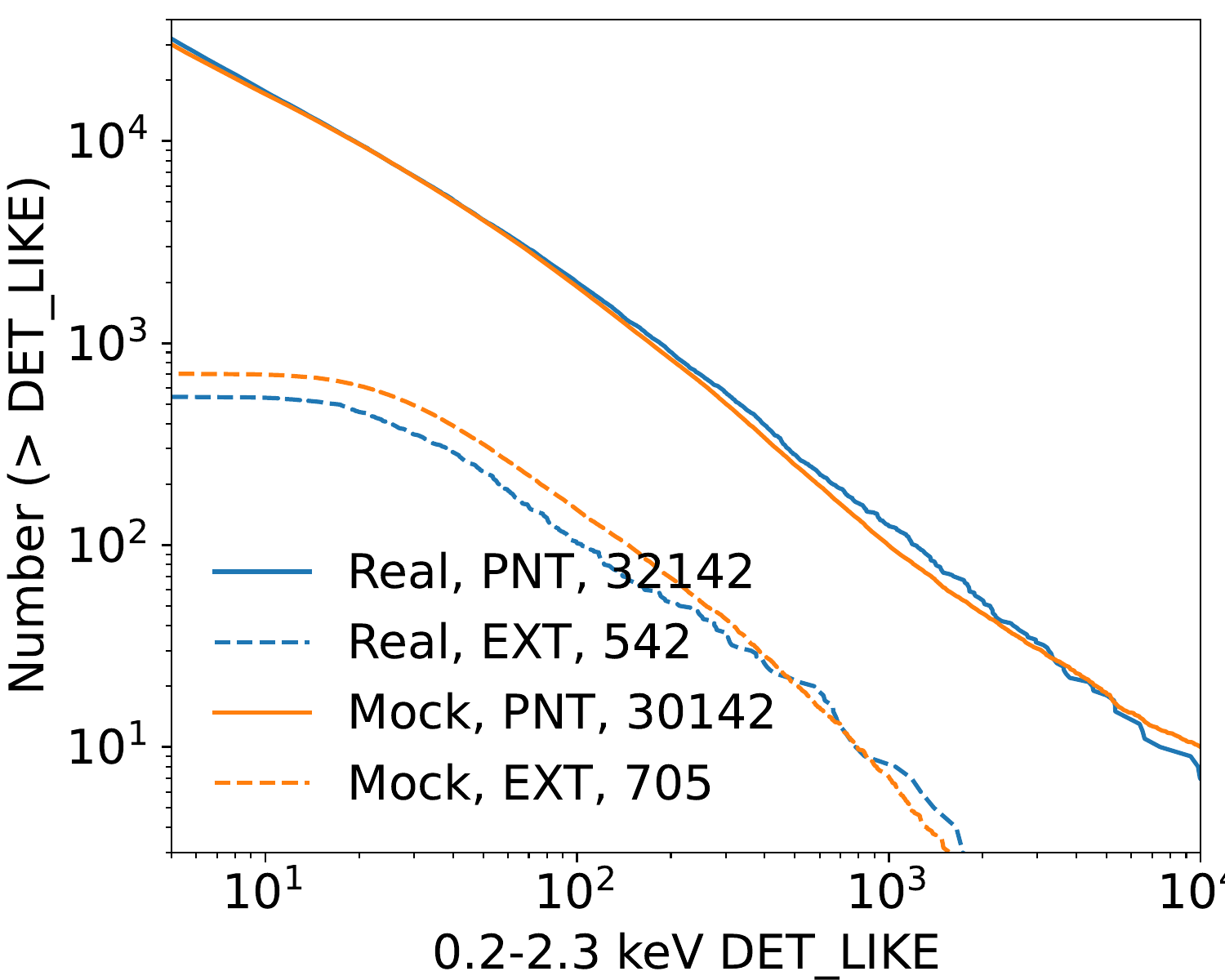}
    \put(80,70){(4)} \end{overpic}
  \begin{overpic}[width=0.3\textwidth]{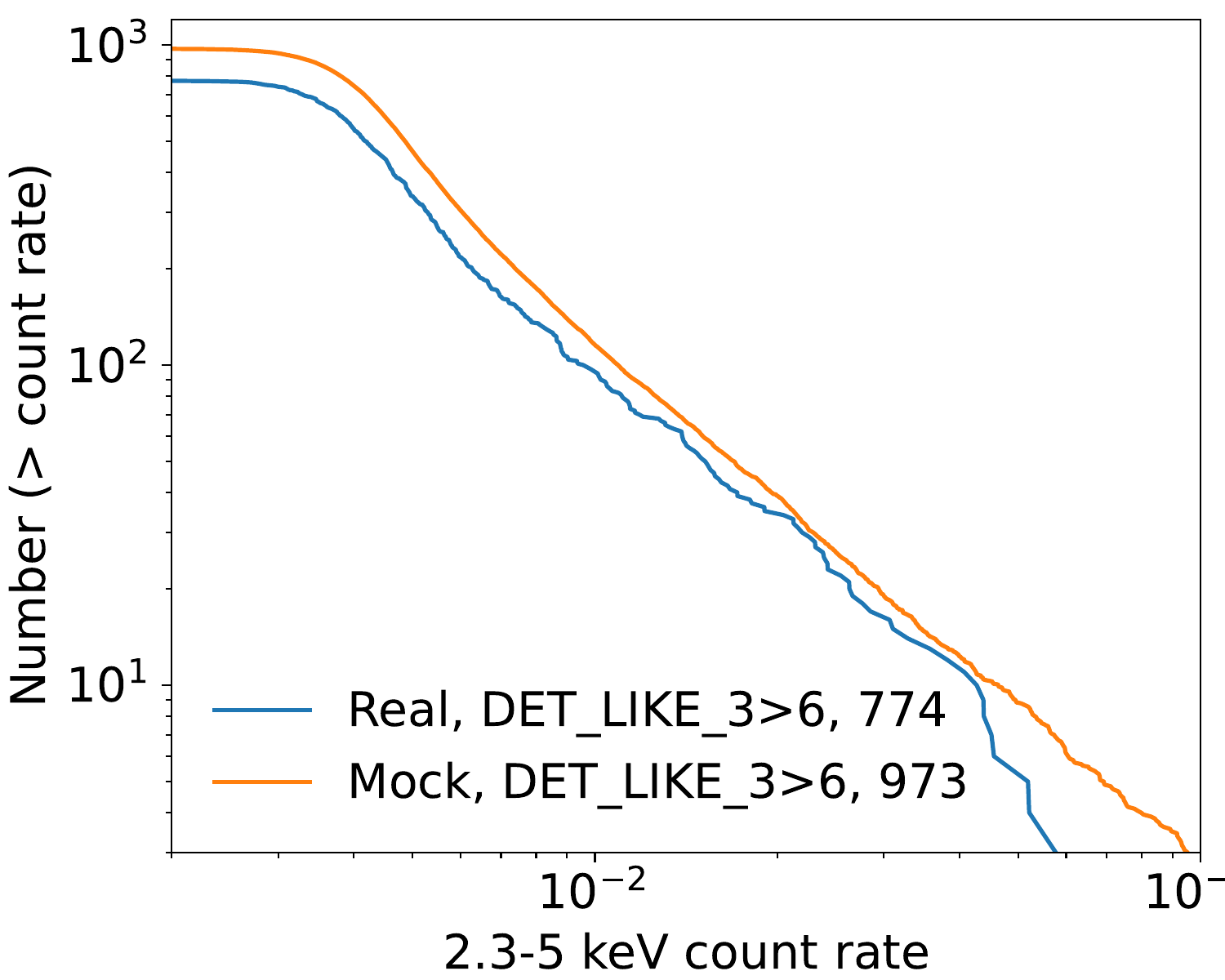}
    \put(80,50){(6)} \end{overpic}
\caption{Comparisons between the real (blue) and the mock (orange) eFEDS data.  Panel (1) compares the event PI (photon energy) distributions. Panel (2) compares the distributions of \TL{the 0.2--2.3~keV background map value (counts/pixel)} in each pixel. The dotted vertical line corresponds to the median values of the two data sets, which are identical.
  Panels (3-4) display the total number of sources in the 0.2--2.3~keV single-band-detected catalog \edit{above a given count rate (panel 3) or above a given detection likelihood (\texttt{DET\_LIKE}; panel 4)} for point sources (``PNT''; \texttt{EXT\_LIKE}=0; solid lines) and extended sources (``EXT''; \texttt{EXT\_LIKE}$>$0; dashed lines), respectively.
  Panel (5) displays the extent likelihood (\texttt{EXT\_LIKE}) distribution of the single-band-detected extended sources.
  Panel (6) displays the 2.3--5~keV (band 3) count rate of sources that are detected in this band (\texttt{DET\_LIKE\_3}$>$6) from the three-band detection. The number of sources is printed in the label of each entry.}
\label{fig:real_mock}
\end{center}
\end{figure*}

We succeeded in making the mock data as representative as possible to the real eFEDS data in many aspects, that is, X-ray background, particle background, and distributions of source fluxes and spectral shapes. The event energy distribution of the mock data is therefore similar to that of the real data, as displayed in panel 1 of Fig.~\ref{fig:real_mock}.
Compared with the real data, the background emission lines in the simulation are not sufficiently smoothed because the RMF adopted by SIXTE is slightly different from the real RMF. This minor difference in energy response has no impact on the broadband source detection.

  In panel 2 of Fig.~\ref{fig:real_mock} we compare the  \edit{(exposure-uncorrected)} background maps measured from the real and mock data using the same method \citep{Brunner2021}.
  The distribution of the background counts per pixel shows a strong peak with two tails on both sides.
  The low-value tail corresponds to the field border, where the exposure depth drops sharply. The high-value tail corresponds to a small portion of regions in which the depth is much higher than average (e.g., regions 2, 4, 6, and 8 in Fig.~\ref{fig:bkg}).
  As we treated the X-ray background and particle background separately using proper vignetting models in the simulation, the background maps are highly similar.
  They have identical median values, but the real data show a slightly larger scatter.
  This is because of both the background variation in the real data and the averaging of the mock data among the 18 realizations.

We used eSASS to detect sources in the mock eFEDS data. The source detection is described in detail in \citet{Brunner2021}. It is done in two steps, first detecting a preliminary catalog using the eSASS task \texttt{erbox} and then performing PSF fitting for each source in the preliminary catalog using the task \texttt{ermldet}.
The PSF fitting measures a detection likelihood \texttt{DET\_LIKE} for each source, which corresponds to the probability of the source being spurious (background fluctuation) in terms of probability$=$exp(-likelihood). By applying a threshold on \texttt{DET\_LIKE}, the final catalog is selected.
In addition to \texttt{DET\_LIKE}, \texttt{ermldet} also measures an extent likelihood \texttt{EXT\_LIKE}, which corresponds to the probability of the source being point-like in terms of probability$=$exp(-likelihood).
Point sources or extended sources can be selected by thresholding \texttt{EXT\_LIKE}.
As we adopted a minimum \texttt{EXT\_LIKE} parameter of $6$ for \texttt{ermldet}, all the sources with \texttt{EXT\_LIKE}$<$6 were set as \texttt{EXT\_LIKE}$=$0.
We performed the PSF-fitting source detection adopting two different sets of energy bands, a single-band PSF fitting in 0.2--2.3~keV, and a three-band (1: 0.2--0.6, 2: 0.6--2.3, and 3: 2.3--5~keV) simultaneous PSF fitting.
In the latter case, \texttt{ermldet} measured a detection likelihood \texttt{DET\_LIKE\_$n$} for each band $n$ (1, 2, or 3) and a summary detection likelihood \texttt{DET\_LIKE\_0}.
Performing post hoc aperture photometry for each source in the catalog using the eSASS task \texttt{apetool}, we also measured the probability of the source being background fluctuation based on the source and background counts in the aperture. Similarly, this probability can also be converted into an aperture-photometry-based likelihood \texttt{APE\_LIKE} ($= -\ln$ probability).

As displayed in Fig.~\ref{fig:real_mock}, the single-band detected point sources have similar distributions in the mock and real data in the soft band (panels 3 and 4).
The mock data have $\sim25\%$ more hard sources than the real data (panel 6), likely because our assumed spectral model is slightly harder than the average spectral shape of the eFEDS sources.
The detected extended sources from the real and mock data also show similar distributions, but $\sim30\%$ more extended sources are detected in the mock data.
Particularly, these mock extended sources show a strong peak at \texttt{EXT\_LIKE}$<$11 that does not exist in the real catalog.
This is because of the uncertainty in the cosmological and astrophysical model used to create the mock cluster catalog \citep{Comparat2020}.
As the clusters are drastically outnumbered by AGN and have a low number density, they do not impact the detection of point sources.

\section{Characterizing the catalog}
\label{sec:DET_dis}

\subsection{Characterizing each source}
Based on the flag of the input-source ID on each photon, we examined the input-output source association in terms of the following four characteristics:

\begin{enumerate}
\item Does a detected source have an input counterpart?\\
  The association was made within a circular aperture of a radius of 20\arcsec{} at the position of the detected source.
\TL{This radius corresponds to the 60\% enclosed-energy fraction (EEF) radius of the PSF, which is the optimized aperture photometry size that leads to the best efficiency in distinguishing between source signal and background fluctuation (see \S~\ref{sec:ape}).}
  The input source that contributes the largest number of photons in this aperture was considered the input counterpart (with the \texttt{ID\_Any}; a negative value means that the \ledit{counterpart} was not found) of the target source.
  As the fluxes of our input catalog extend far lower than the detection limit, none or only a few photons are captured by the camera for a large number of input sources. We only considered input sources for which at least 3 photons were captured at any position. Still, a spuriously detected source might coincidentally be attributed to a very faint input source. To exclude these coincident associations, we extracted all the X-ray and particle background photons into a background image and smoothed it to increase its signal-to-noise ratio (S/N), and then we defined a lower limit of the source aperture counts according to the local aperture background counts as its $2\sigma$ Poissonian upper limit (97.725\% point of the Poisson distribution). An input source was considered a meaningful counterpart only if its aperture counts exceeded this limit.

\item Does a detected source have a secondary input counterpart?\\
  If multiple input sources meeting the above requirements were found within the 20\arcsec{} aperture of a detected source, the input source that contributed with the second-highest number of photons was considered the secondary counterpart (\texttt{ID\_Any2}; a negative value means that no \ledit{secondary counterpart was found}).

\item Is an input source contaminated by another source?\\
  If a second input sources contributes at least 3 photons within 60\arcsec{} (the PSF-fitting radius used in the eFEDS source detection) of an input source, and if its number of photons is larger than the square root of the target photon counts, we considered this second source a contamination of the target source and saved it as \texttt{ID\_contam} (a negative value means that it does not exist).

\item Unique input-output association\\
  One input source might be falsely detected as multiple sources, resulting in duplicated \texttt{ID\_Any}. In these cases, we selected the input source that contributed the largest number of photons in the 20\arcsec{} aperture of the detected source and considered this input source a unique counterpart (\texttt{ID\_Uniq}). If the \texttt{ID\_Any} of a detected source was not a unique counterpart, we took the secondary input counterpart (\texttt{ID\_Any2}) if it existed and was not assigned to other detected sources. With \texttt{ID\_Uniq}, a one-to-one association between the input and output sources was built.
\end{enumerate}

\subsection{Classifying detected sources}
\label{sec:class}
\begin{figure}[hptb]
\begin{center}
\includegraphics[width=\columnwidth]{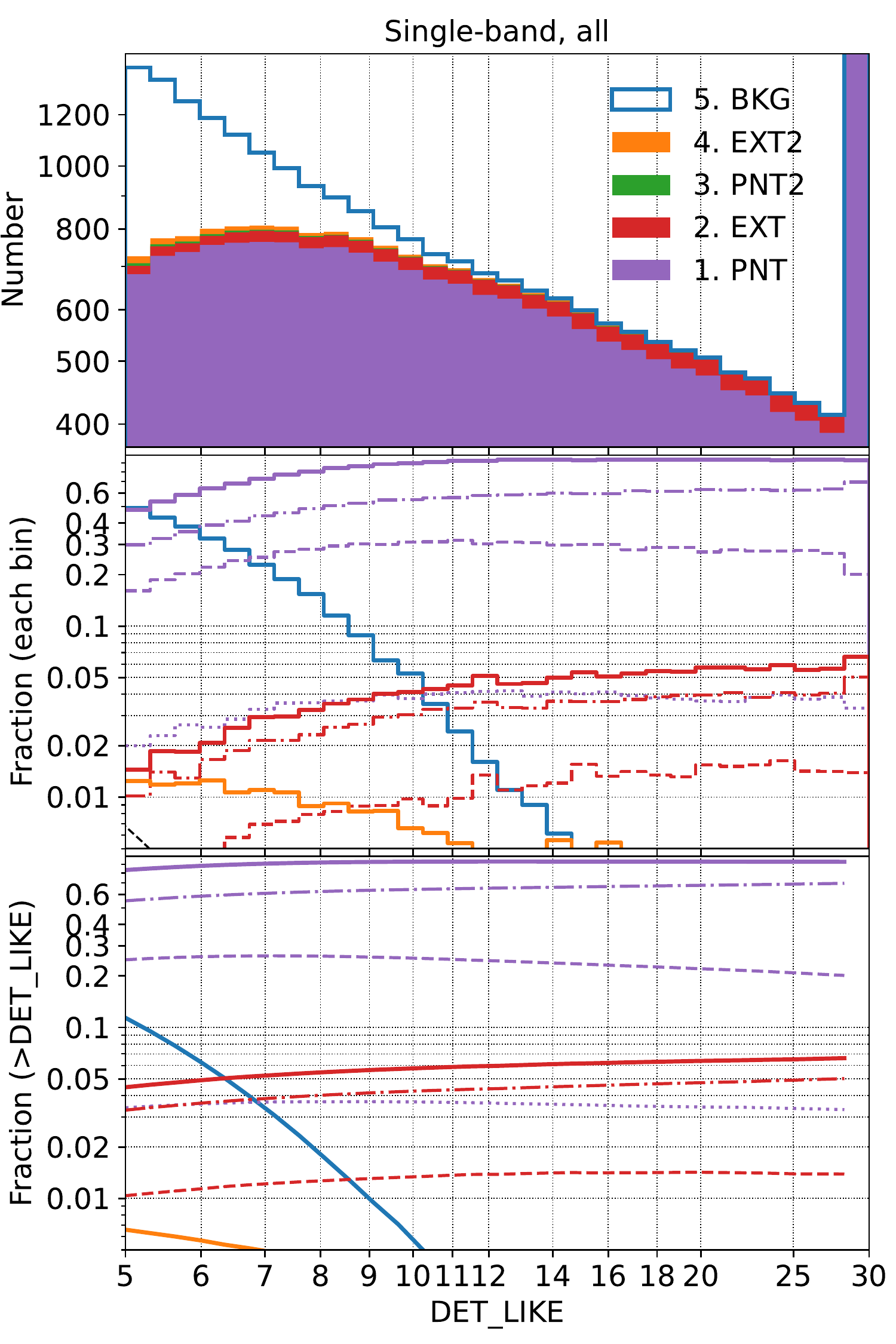}
\caption{Distributions of all the single-band-detected sources as a function of detection likelihood.
  Purple, red, green, orange, and blue correspond to the five classes ($1\sim5$) of detected sources. The top, \edit{middle, and bottom} panels display stacked histograms, differential fractions, and cumulative fractions, respectively. The dot-dashed, dashed, and dotted lines in purple and red indicate the three subclasses of no contamination, point-source contamination, and extended-source contamination, respectively. See \S~\ref{sec:class} for details.
  The dashed black line indicates probability=exp(-likelihood). It is too low and \edit{almost} drops out of the plotting scope of the \edit{middle} panel.
  }
\label{fig:DET_dis_whole}
\end{center}
\end{figure}

\begin{figure*}[hptb]
\begin{center}
\includegraphics[width=0.245\textwidth]{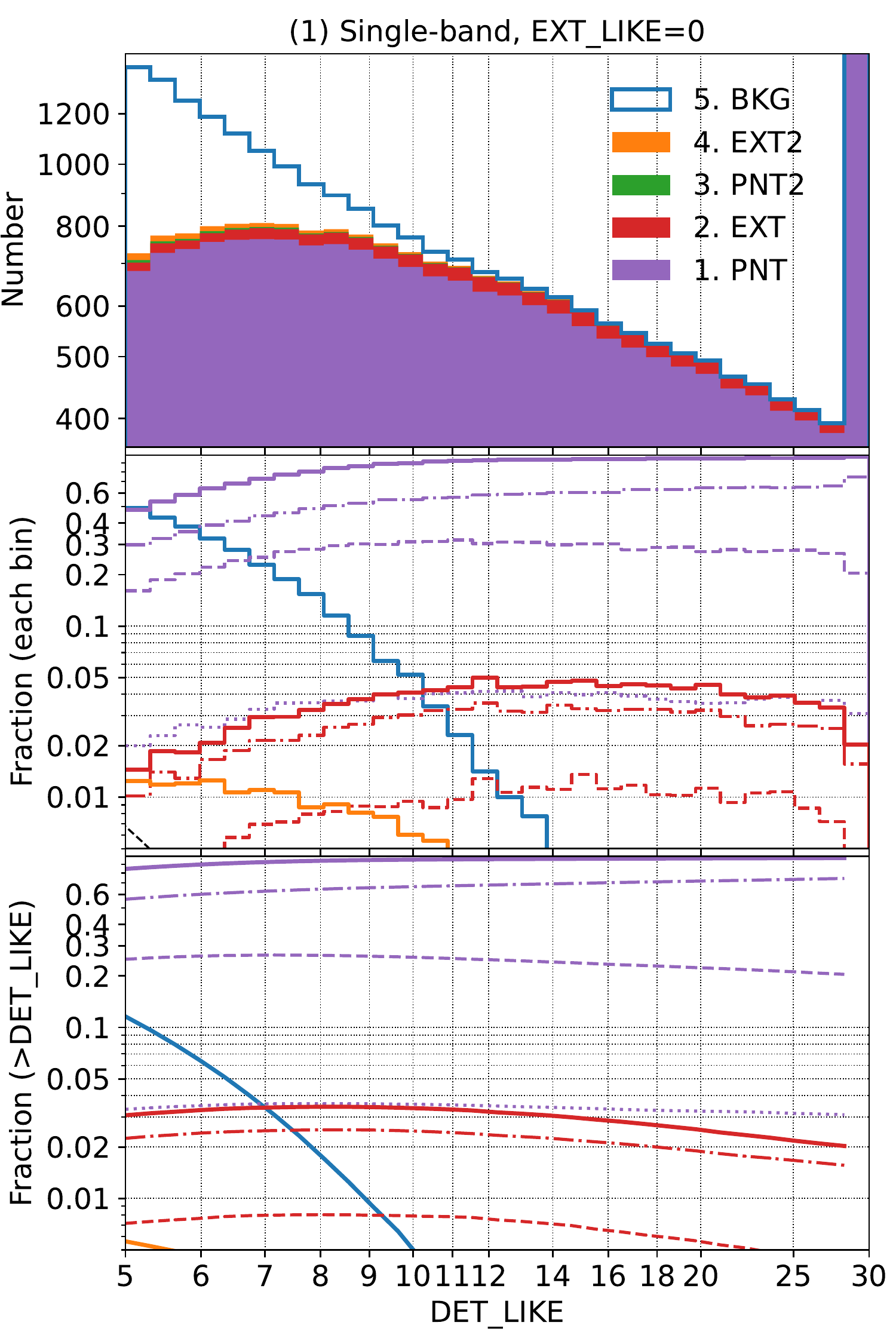}
\includegraphics[width=0.245\textwidth]{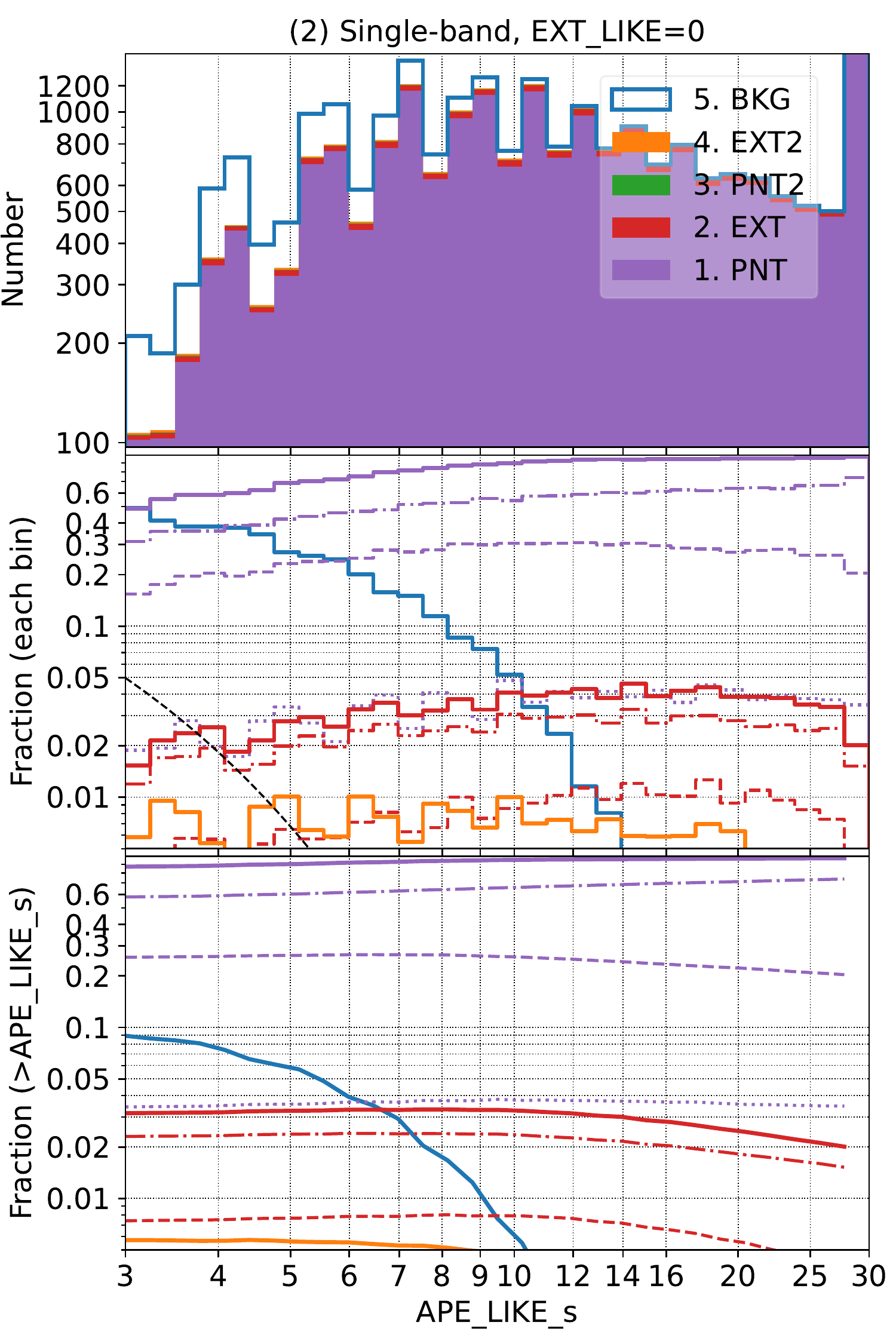}
\includegraphics[width=0.245\textwidth]{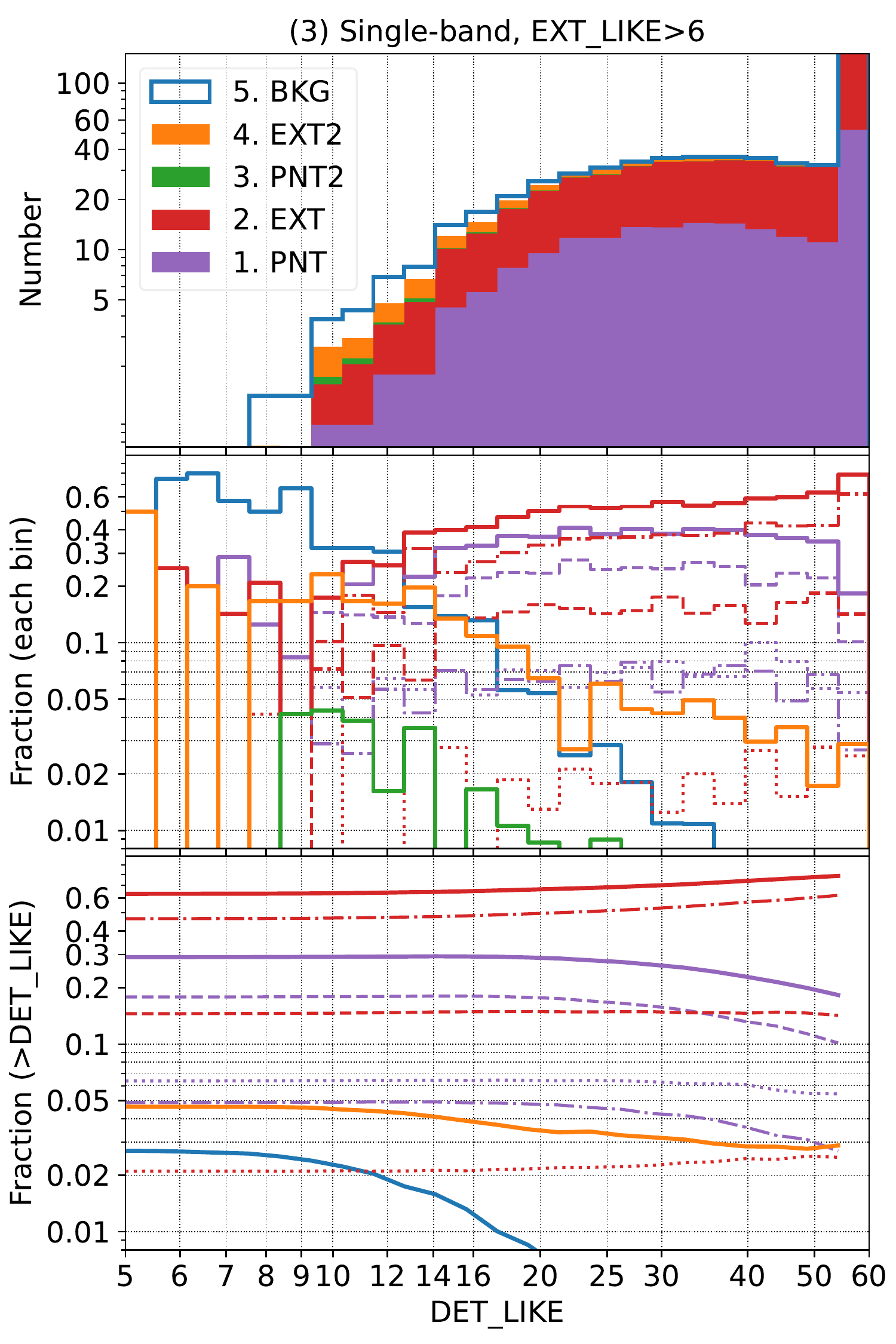}
\includegraphics[width=0.245\textwidth]{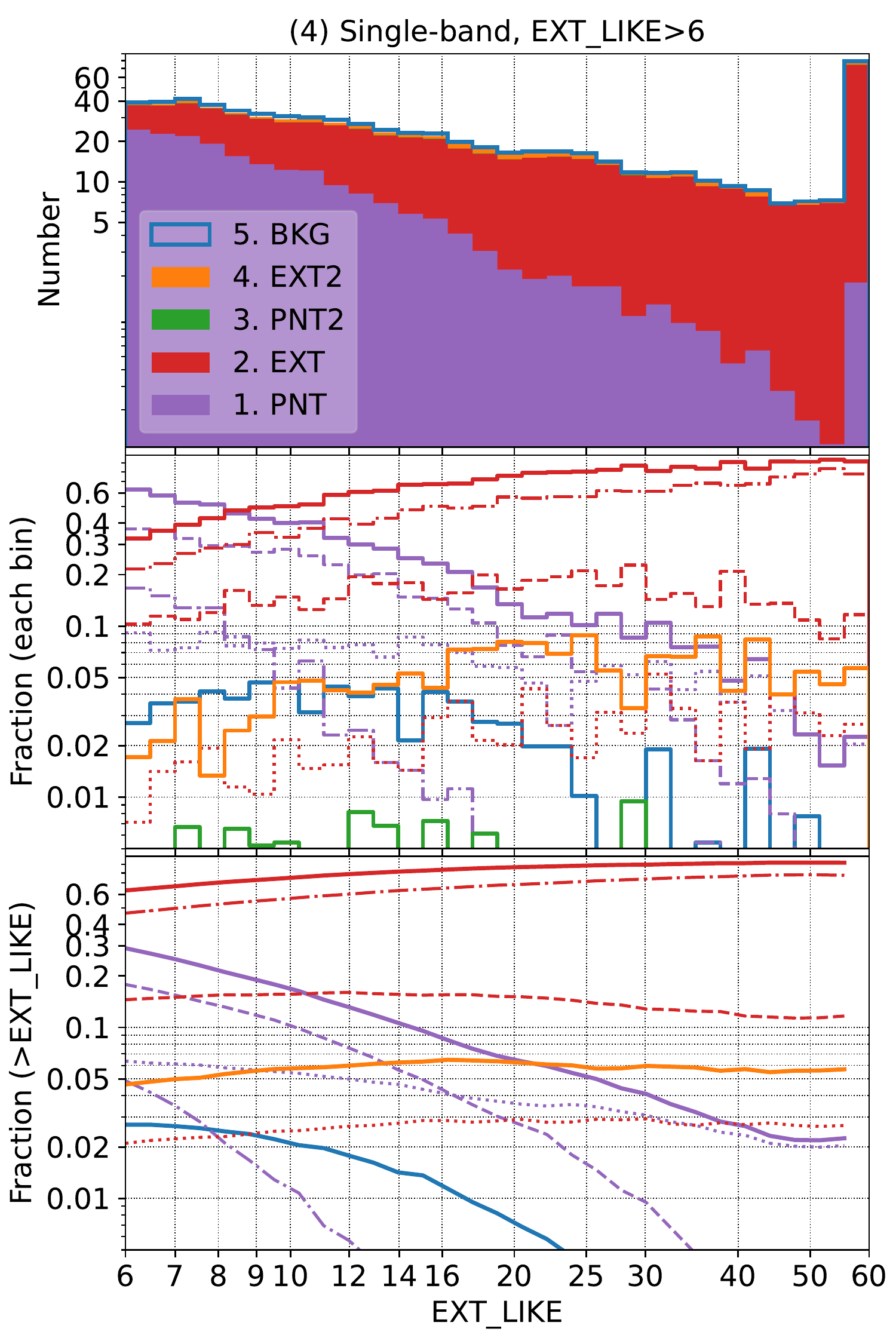}
\caption{Same as Fig.~\ref{fig:DET_dis_whole}, but for different sample selections and/or different variables.
  The panel number and selection rules are printed in the figure titles, i.e., panels (1-2) for point sources (\texttt{EXT\_LIKE}=0) and panels (3-4) for extended sources (\texttt{EXT\_LIKE}>6), both of which are selected from the single-band-detected catalog.
  The four panels display the selected subsamples as a function of (1) single-band detection likelihood (\texttt{DET\_LIKE}), (2) 0.5--2~keV aperture-photometry-based likelihood (\texttt{APE\_LIKE\_s}), (3) \texttt{DET\_LIKE}, and (4) extent likelihood (\texttt{EXT\_LIKE}).
  The dashed black lines in the \edit{middle} subpanel of panels (1-2) correspond to probability=exp(-likelihood).
  }
\label{fig:DET_dis_V18C}
\end{center}
\end{figure*}

\begin{figure*}[hptb]
\begin{center}
\includegraphics[width=0.245\textwidth]{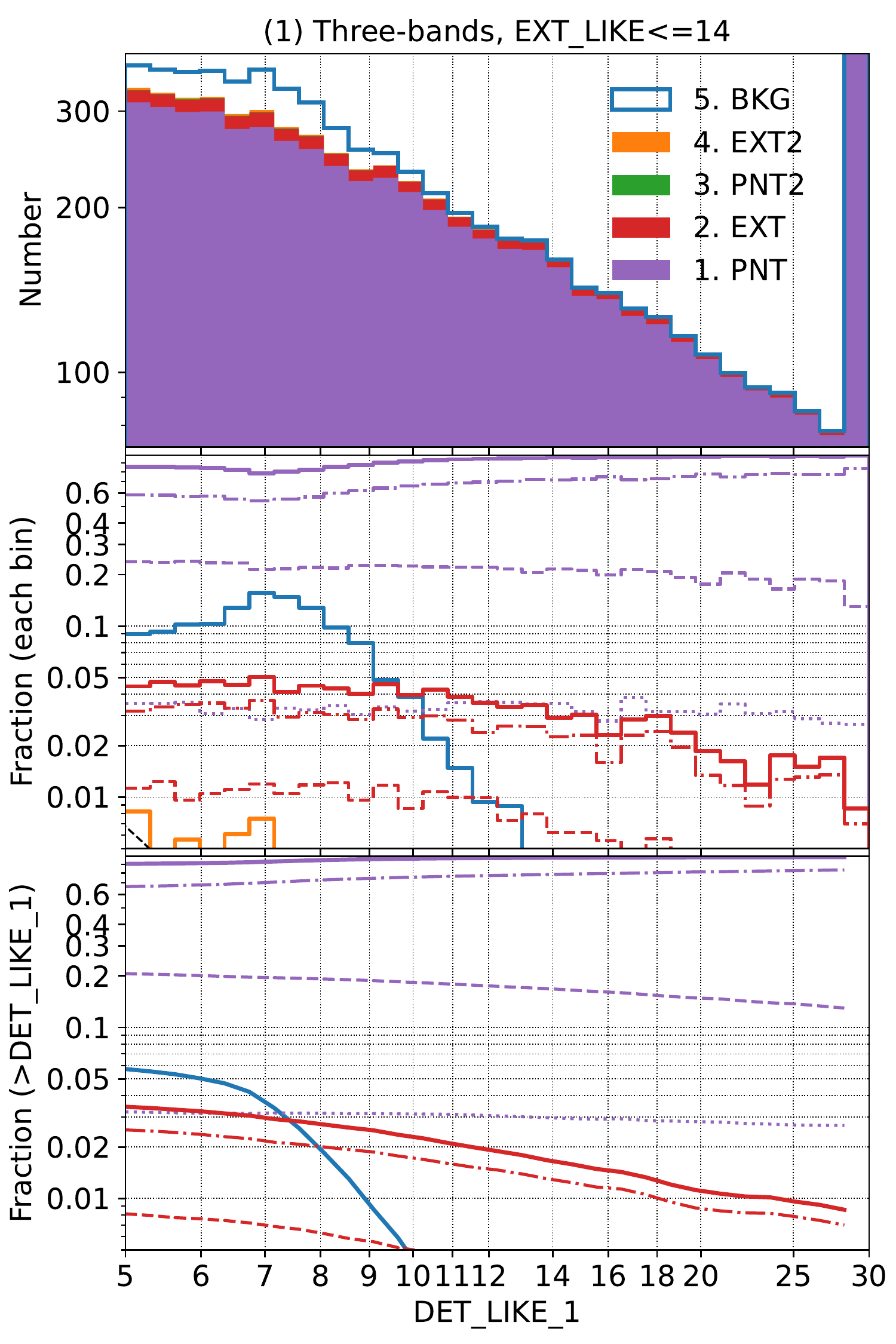}
\includegraphics[width=0.245\textwidth]{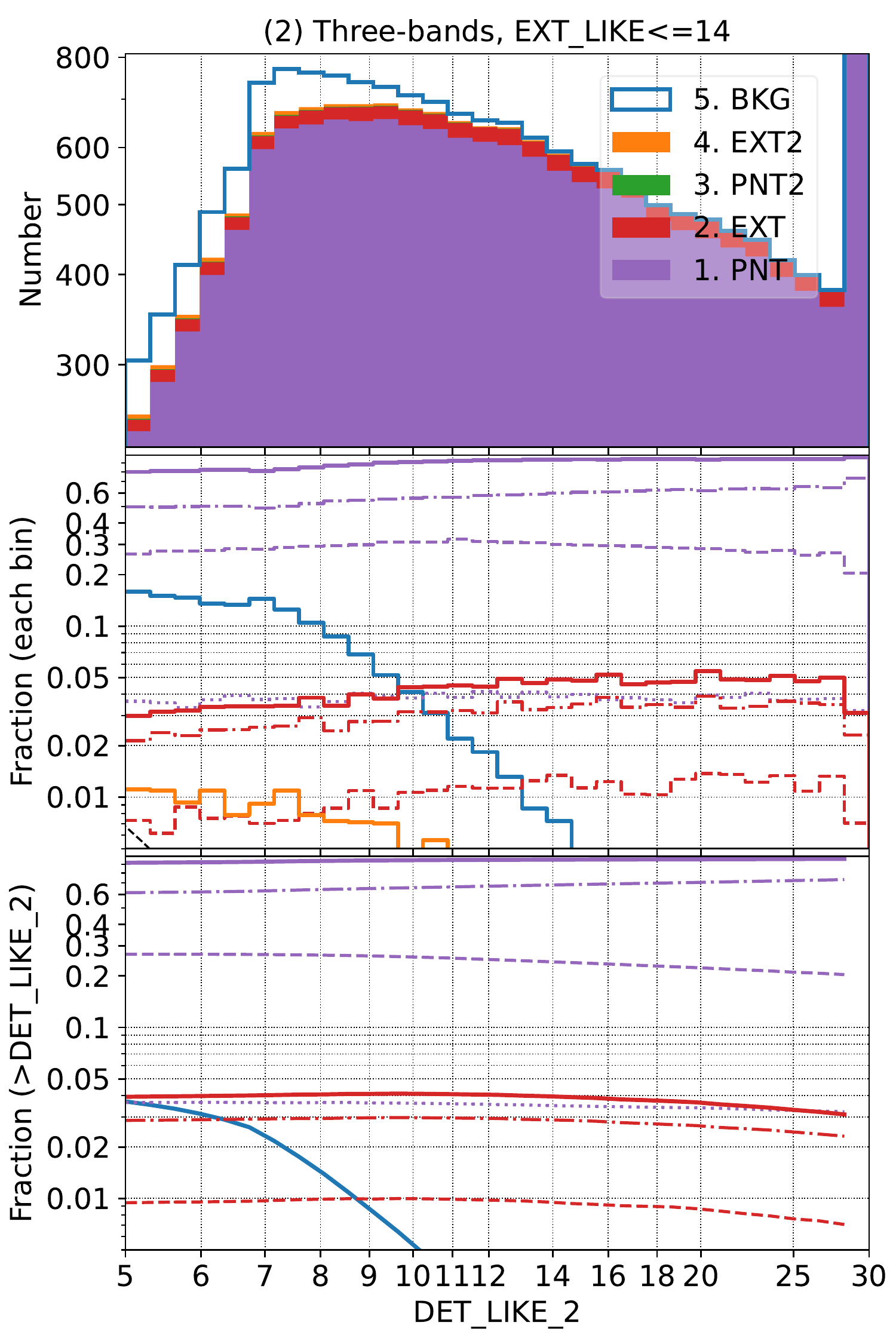}
\includegraphics[width=0.245\textwidth]{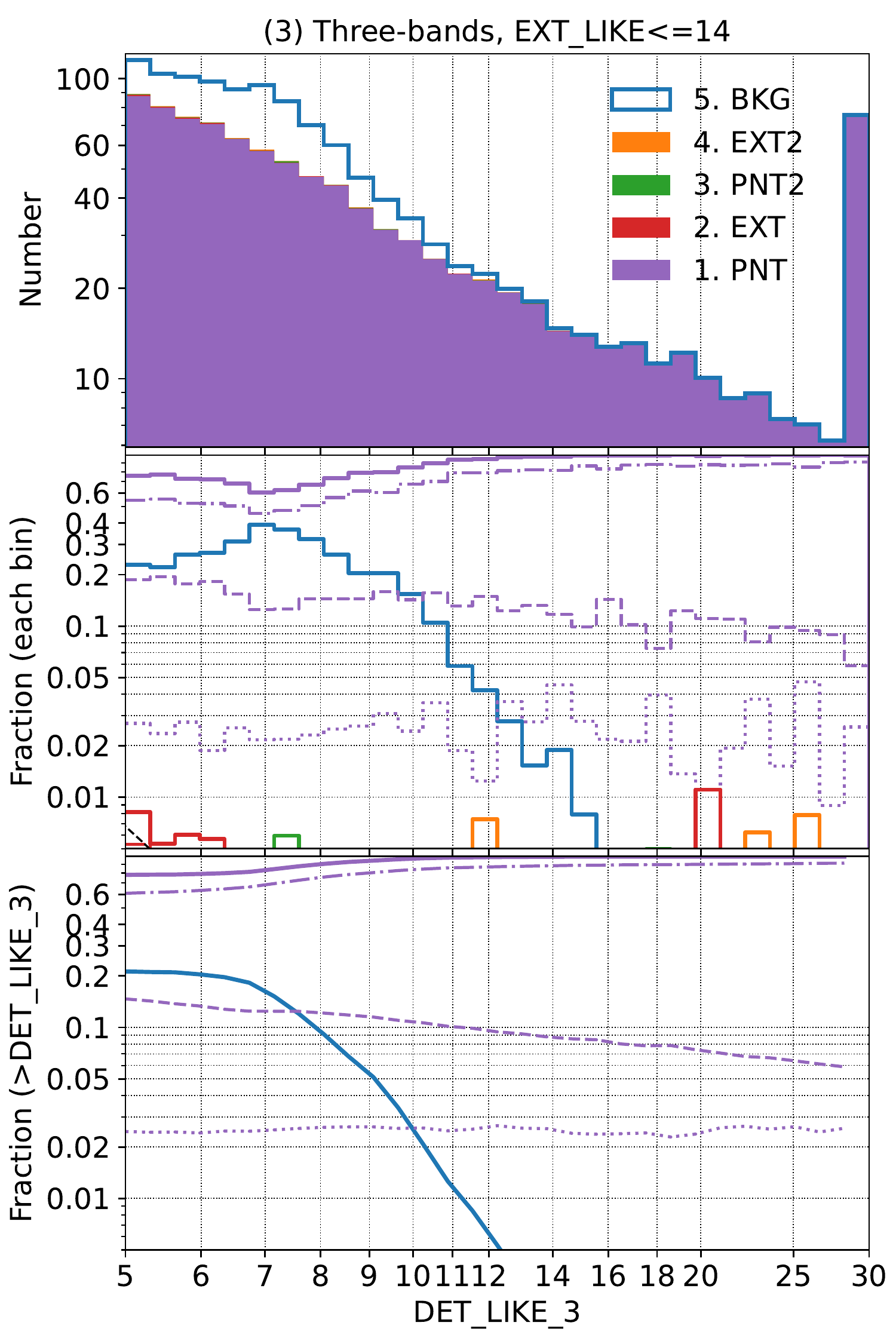}
\includegraphics[width=0.245\textwidth]{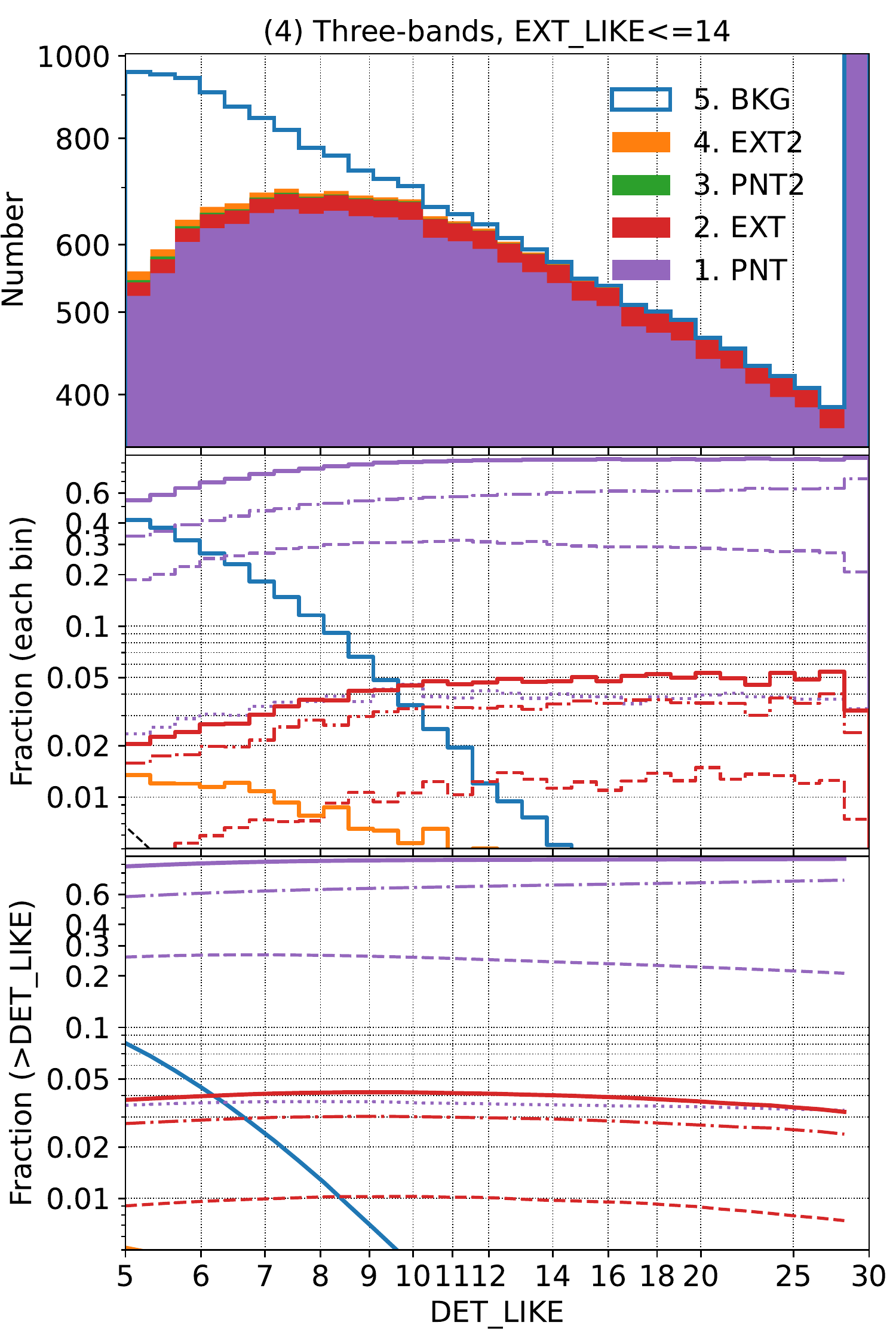}

\caption{Same as Fig.~\ref{fig:DET_dis_V18C}, but for the three-band-detected sources with \texttt{EXT\_LIKE}$\leqslant$14 and plotted as a function of the detection likelihood \texttt{DET\_LIKE\_$n$}, where $n$ indicates the energy bands 1, 2, 3, or 0.
  In panels 1--3, the \texttt{DET\_LIKE\_$n$} corresponds to the individual band detection likelihood in band 1: 0.2--0.6, band 2: 0.6--2.3, and band 3: 2.3--5 keV.
  In panel 4, the \texttt{DET\_LIKE\_0} corresponds to the three-band-summary likelihood.
  }
\label{fig:DET_dis_V18T}
\end{center}
\end{figure*}

Based on the source-matching characteristics defined above, we divided the detected sources into five classes as follows. We plot their distributions in Fig.~\ref{fig:DET_dis_whole} in five colors.

\begin{enumerate}
\item \TL{Primary counterpart of a point source (PNT,} in Fig.~\ref{fig:DET_dis_whole})\\
  A detected source is classified as the \TL{primary counterpart of} an input point source if it has a unique input counterpart (\texttt{ID\_Uniq}) of an AGN or star.

\item \TL{Primary counterpart of an extended source (EXT)} \\
  A source is classified as\TL{ the primary counterpart of} an input cluster if it has a unique input counterpart (\texttt{ID\_Uniq}) of a cluster. 

\item \TL{Secondary counterpart of a point source (PNT2)} \\
  A source with an input counterpart (\texttt{ID\_Any}) of an AGN or star but without a unique counterpart (\texttt{ID\_Uniq}$<0$) is classified as \TL{the secondary counterpart of an input point source. In these cases, a primary counterpart (class 1 or 2) for the input point source is already detected at the correct (brightest) location; the detected source contains point-source signals that correspond to a fragment of an input point source in the outer wing of the PSF.}

\item \TL{Secondary counterpart of an extended source (EXT2)} \\
  A source with an input counterpart (\texttt{ID\_Any}) of a cluster but without a unique counterpart (\texttt{ID\_Uniq}$<0$) is classified as \TL{the secondary counterpart of an input cluster. In these cases, a primary counterpart (class 1 or 2) is already detected for the input cluster at the correct (brightest) location; the detected source contains cluster photons that represent a substructure or fluctuation in the cluster.}

\item Background fluctuation (\TL{BKG}) \\
  A source without any input counterpart (\texttt{ID\_Any}$<0$) is classified as a spurious source due to background fluctuation. 
\end{enumerate}
The secondary input counterpart of a detected source \texttt{ID\_Any2} is useless here in characterizing the detected sources. It is useful, on the other hand, in characterizing the input sources. If an input source has no detected primary counterpart, it might be considered as the secondary input counterpart of any detected source. In this case, the signal of this input source is detected but not as an independent source, that is, it is blended with another brighter source.
Alternatively, another indicator of source blending is \texttt{ID\_contam}, which is defined in a larger region (60\arcsec{} instead of 20\arcsec) around each input source.
This indicator is more sensitive in selecting the cases of blended sources in the sense that all the photons involved in the PSF fitting (within 60\arcsec) were taken into account.
For input sources or primary detected counterparts (class 1 or 2) of input sources, we furthermore divided them into three subclasses: 1) without contamination (\texttt{ID\_contam}$<0$), 2) contaminated by a point source, and 3) contaminated by a cluster.

As illustrated in Fig.~\ref{fig:DET_dis_whole}, \TL{ which displays the whole single-band-detected sample as a function of \texttt{DET\_LIKE},} we can display the distributions of the five classes of detected sources in three terms: 1) a stacked histogram \TL{as a function of a particular source property}, 2) a differential fraction (in each bin of the histogram) of each class in the \TL{selected} sample, and 3) a cumulative fraction of each class in the \TL{selected} sample, calculated using subsamples above any given value of the \TL{concerned} source property.
The stacked histogram was used to choose the bins and avoid measurements of fractions with only a few sources.
The differential fraction distributions are helpful in understanding the source detection performance in various aspects and accordingly in making adjustments, for example, whether an adjustment is helpful in suppressing spurious detections at low detection likelihood (\texttt{DET\_LIKE}), distinguishing point and extended sources at low extent likelihood (\texttt{EXT\_LIKE}), or resolving blended sources at high fluxes.
The cumulative fraction distributions were used to characterize a selected sample quantitatively in terms of various types of false rates (or purity).

Given any sample selection criteria, we can display the distributions of various classes of sources as a function of any source property, for instance, detection likelihood, extent likelihood, or source flux, as illustrated in Fig.~\ref{fig:DET_dis_V18C} and Fig.~\ref{fig:DET_dis_V18T}, and as discussed Sects. \ledit{\ref{sec:spu} and \ref{sec:cla}}. Fig.~\ref{fig:DET_dis_V18C} displays the single-band-detected catalog, and Fig.~\ref{fig:DET_dis_V18T} displays the three-band-detected catalog.
We provide the single-band-detected and the three-band-detected mock catalogs here. They include the classification information (\ledit{Appendix} \S~\ref{sec:catalogs}). With this information, the fraction of any class in the eFEDS catalog under the sample selection criteria required by a particular scientific goal can be measured.

\subsection{Spurious sources}
\label{sec:spu}
By definition, the source detection likelihood, either measured through PSF fitting (\texttt{DET\_LIKE}) or through Poisson tests of the aperture photon counts (\texttt{APE\_LIKE}), reflects the probability of a source being background fluctuation, which equals exp(-likelihood). However, these definitions assume an ideal situation of a single source with pure Poissonian fluctuation, but do not account for any additional uncertainty or bias introduced during background estimation, PSF fitting, or aperture photometry, not to mention potential uncertainties of the facility hardware, calibration, and even software numerical issues. As displayed in panels 1 and 2 of Fig.~\ref{fig:DET_dis_V18C}, which display the distribution of the single-band-detected point sources, the measured spurious fractions (blue lines) are much higher than expected from the likelihood definition.

As suggested by \citet{Liu2020}, a relatively low source detection likelihood should be adopted in detecting the candidates (not necessarily the final catalog) with PSF fitting, so that many faint but potentially interesting sources can be detected, potential cases of blended faint sources can be checked by multiple PSF fitting, and faint sources can be effectively masked out when measuring the properties of nearby sources.
We adopted a threshold of \texttt{DET\_LIKE}$>$5 in eFEDS source detections.
\TL{As shown in Fig.~\ref{fig:DET_dis_whole}, this threshold corresponds to a spurious fraction of 11.5\% in the single-band-detected catalog, which is too high to be used for most scientific works.} A further selection of subsamples is needed according to particular scientific goals. To select a sample with relatively balanced completeness and purity, \texttt{DET\_LIKE}$>$6 \citep[for the main eFEDS catalog;][]{Brunner2021} might be adopted, for instance, which corresponds to a spurious fraction of 6.3\%. Finally, to select a cleaner sample, a higher threshold such as \texttt{DET\_LIKE}$>$8 would be needed (1.8\% spurious).

Fig.~\ref{fig:DET_dis_V18T} displays the likelihood distributions of the three-band-detected catalog.
The samples selected from the three-band detection based on the 0.2--0.6 and 0.6--2.3~keV individual band likelihoods seem to have relatively lower spurious fractions (panels 1 and 2 in Fig.~\ref{fig:DET_dis_V18T}) than the single-band detection. However, this is only because they result from two-pass selections, that is, first requiring the summary likelihood \texttt{DET\_LIKE\_0}$>$5 and then the individual band likelihood $>$5, and does not necessarily indicate a higher selection efficiency in the three-band detection.
Even with the two-pass selection, the sample selected with the 2.3--5~keV individual band likelihood still has a much higher spurious fraction because of the relatively small effective area and relatively high background in the hard band of eROSITA. Selecting sources with \texttt{DET\_LIKE\_3}$>$10 and \texttt{EXT\_LIKE}$<14$ (for the hard eFEDS sample; Nandra et al. in prep.) leads to a spurious fraction of \TL{2.5\%}.
When \texttt{DET\_LIKE\_0}$>$5 is adopted, the spurious fraction appears lower in the three-band detection than in the single-band detection. This is because when summing the individual-band likelihoods (\texttt{DET\_LIKE\_1},\texttt{DET\_LIKE\_2},\texttt{DET\_LIKE\_3}) to calculate \texttt{DET\_LIKE\_0}, the additional degrees of freedom used in the summing reduce the likelihood of a source. Therefore, \texttt{DET\_LIKE\_0}$>$5 is a stricter rule in the three-band detection than in the single-band detection (see also the discussion in \S~\ref{sec:effi}.)

\subsection{Blending and misclassification}
\label{sec:cla}

We have so far only discussed the spurious fraction (or false rate) introduced in the detection.
When classification is involved, ``false rate'' could also refer to the fraction of misclassified sources.
For a deep survey without high spatial resolution, it is hard to distinguish between a compact cluster and a point source, between blended point sources and an individual cluster, or between a point source in a cluster and the substructure of the cluster.
The photon-flag-based input-output association allows us to investigate these cases.
As shown in panels 1 and 2 in Fig.~\ref{fig:DET_dis_V18C}, the single-band-detected point-source catalog (\texttt{EXT\_LIKE}$=$0 and \texttt{DET\_LIKE}$>$5) contains $\sim3\%$ clusters (class 2).
It can be considered as a prediction of the fraction of clusters in the real eFEDS point-source catalog, which is only an approximation because the distributions of flux and brightness profile of clusters and groups are highly uncertain in the faint and compact regime.
The extended source catalog selected with \texttt{EXT\_LIKE}$>$6 also contains a significant fraction (29.1\%) of AGN and stars, as shown in panels 3 and 4 in Fig.~\ref{fig:DET_dis_V18C}.

Another problem beyond source detection is source blending.
As shown in panel 1 in Fig.~\ref{fig:DET_dis_V18C}, of the 84.6\% genuine point sources (class 1) in the single-band-detected point-source catalog, 30\% (25.0\% of the whole point-source catalog) are contaminated by nearby point sources and 4\% (3.3\% of the whole catalog) are contaminated by nearby clusters.
The genuine clusters also contain a significant fraction (23\%) that is contaminated by nearby point sources (Fig.~\ref{fig:DET_dis_V18C} panel 3).
As shown in panel 3 in Fig.~\ref{fig:DET_dis_V18C},  the 29.1\% genuine point sources (class 1) in the single-band-detected extended-source catalog include 61.2\% (17.8\% of the whole extended-source catalog) that are contaminated by nearby point sources and 22.0\% (6.4\% of the whole catalog) that are contaminated by nearby clusters.
The much higher fractions of blended cases in the misclassified point sources indicate that blending of nearby sources is a main reason for this misclassification.
Source blending causes not only misclassifications, it also results in sample incompleteness (fewer sources).
Unless all the blended sources are fitted simultaneously with proper models, which is not usually the case, the source flux might be overestimated because of the contamination (see also the discussion in \S~\ref{sec:logNlogS}).

Panels 3 and 4 in Fig.~\ref{fig:DET_dis_V18C} display the single-band-detected extended sources as a function of detection likelihood \texttt{DET\_LIKE} and extent likelihood \texttt{EXT\_LIKE}.
Point sources are misclassified as extended (class 1), mostly because of source blending, as discussed above. This occurs at any detection likelihood (or brightness).
The extent likelihood of these false clusters is significantly lower than that of genuine clusters (class 2), however.
These false clusters included only a few separated (nonblended) point sources (dash-dotted purple lines) that are concentrate at low \texttt{EXT\_LIKE}; there are more cases of blended point sources (dashed purple lines) and of cluster-contaminated point sources (dotted purple lines), and \ledit{they} have largely different \texttt{EXT\_LIKE} distributions. The blended point sources mostly result in false clusters at low \texttt{EXT\_LIKE}. At high \texttt{EXT\_LIKE} (e.g., $>$20), most of the misclassified point sources have underlying cluster emission.
Because we identified the input counterpart of detected sources using the photons within a small radius of 20\arcsec, in the cases of point sources inside clusters, the point sources with more concentrated signals are more easily considered to be the primary counterpart than clusters whose signals are diffusely distributed. These cases of cluster-contaminated AGN can also be considered correctly identified clusters, but with an AGN inside.

In order to evaluate the spurious fraction in the real eFEDS extended source catalog, we predicted the number of spurious clusters in the single-band-detected catalog.
We defined spurious clusters as detected sources that are not attributed to any input clusters (either \texttt{ID\_Any} or \texttt{ID\_Any2}).
In the eFEDS 90\% area region, we detect $196.7$ spurious clusters with \texttt{EXT\_LIKE}$>$6. However, this value must be higher than the real eFEDS data because the detected extended sources from the mock data show a strong peak at the lowest \texttt{EXT\_LIKE} ($<11$) that does not exist in the real data, as shown in Fig.~\ref{fig:real_mock}.
At \texttt{EXT\_LIKE}$>$12 and \texttt{DET\_LIKE}$>$12, the simulation predicts $42.9$ spurious clusters.
At \texttt{EXT\_LIKE}$>$20 and \texttt{DET\_LIKE}$>$20, the simulation predicts $9.4$ spurious clusters.
These values are more reliable.
\ledit{Another issue to point out about the simulation-predicted spurious clusters is that some genuine clusters detectable in the optical band might be considered spurious because its X-ray emission was not significant enough in the small 20\arcsec{} aperture.}

\section{Source detection efficiency}
\label{sec:efficiency}
\subsection{Source detection completeness}
\label{sec:compl}

\begin{figure*}[hptb]
  \centering
\includegraphics[width=0.34\textwidth]{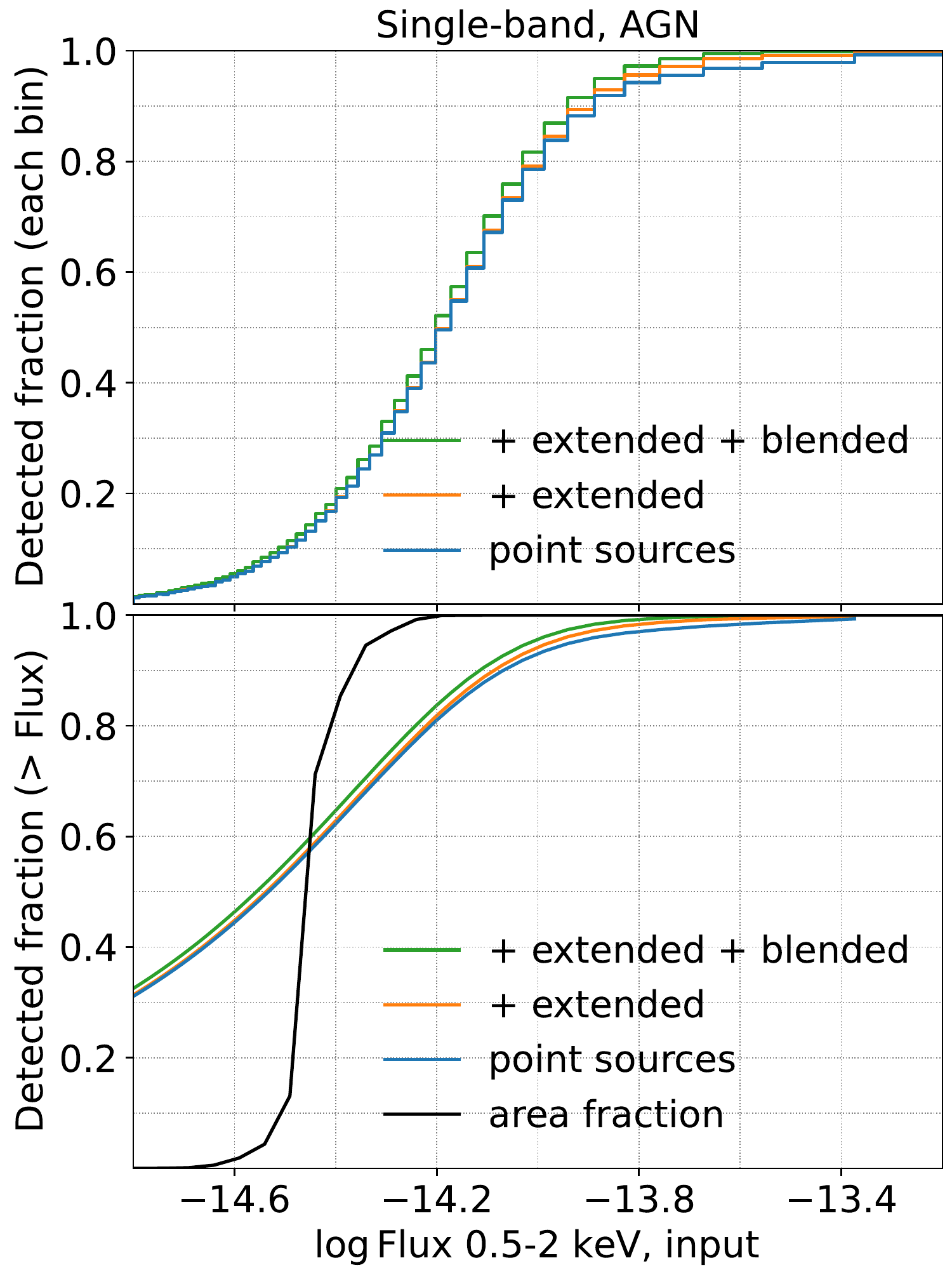}
\includegraphics[width=0.34\textwidth]{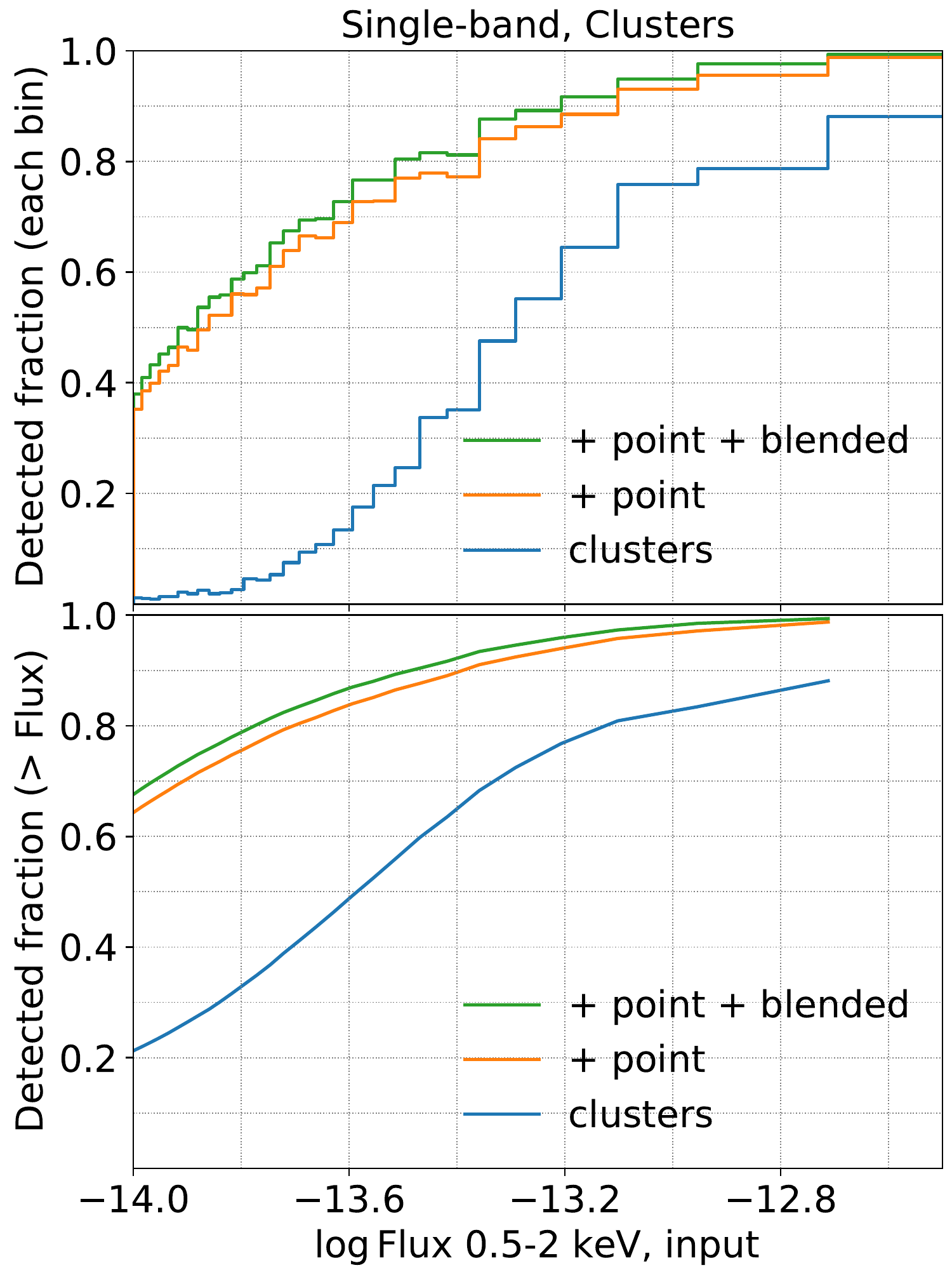}
\caption{
   Completeness of AGN (left) and completeness of clusters (right) in the single-band-detected catalog as a function of the input 0.5-2 keV flux in differential (top) and cumulative manners (bottom).
   The blue lines indicate sources that are detected (as \texttt{ID\_Uniq}) and correctly classified as point sources (in the left panel) or as extended sources (in the right panel).
   The orange lines indicate sources that are detected (as \texttt{ID\_Uniq}) regardless of classification, so that extended sources are also included in the left panel and point sources are also included in the right panel.
   The green lines indicate sources that are detected as either primary (\texttt{ID\_Uniq}) or secondary (\texttt{ID\_Any2}) counterparts, so that the cases of blended sources are included.
   The black line in the left panel shows the sensitivity curve of the single-band-detected catalog calculated using \texttt{ersensmap} in terms of fractional area at a given flux limit.
   }
\label{fig:comp_flux}   
\includegraphics[width=0.34\textwidth]{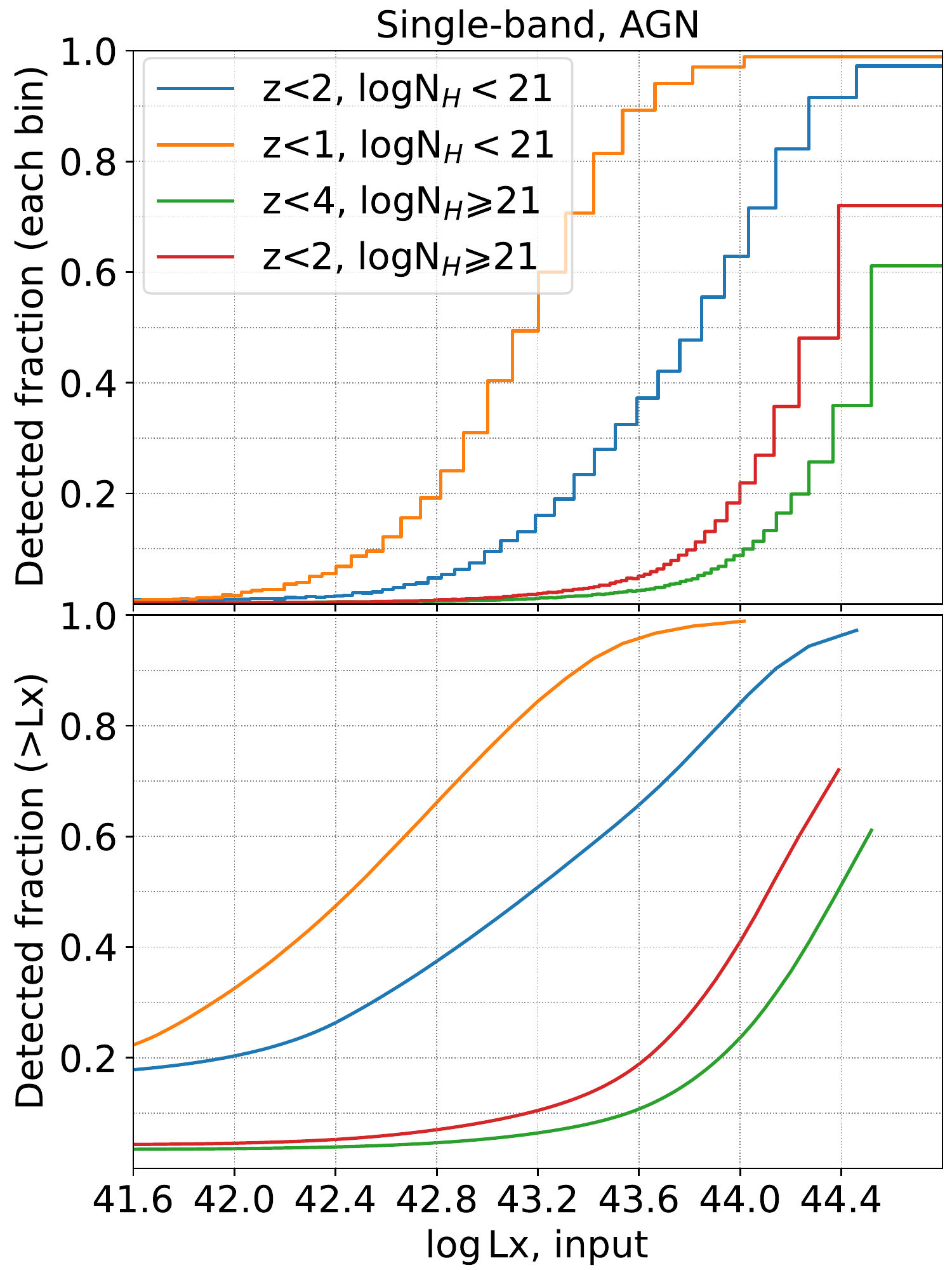}
\includegraphics[width=0.34\textwidth]{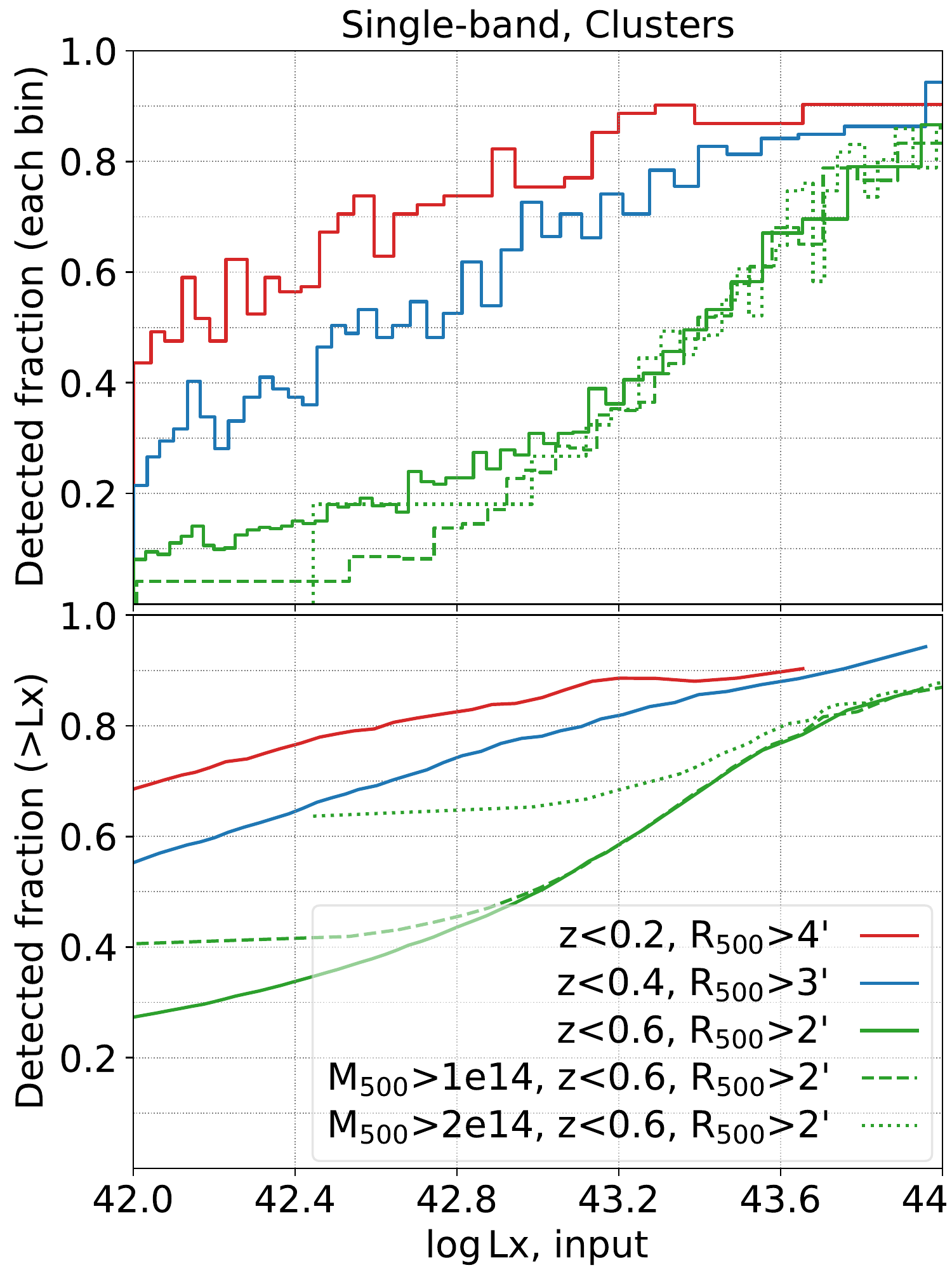}
\caption{
   Completeness of AGN (left) and clusters (right) in the single-band-detected catalog as a function of the intrinsic 0.5--2~keV luminosity in differential (top) and cumulative manners (bottom).
   Only the sources that are detected as primary counterparts (\texttt{ID\_Uniq}) and are correctly classified as point sources (for AGN; left) or extended sources (for clusters; right) are considered.
   The left panel displays four input AGN subsamples selected with different redshift and \N{H} thresholds.
   The right panel displays input cluster subsamples selected with different thresholds of redshift, apparent scale $R_{500}$ , and halo mass $M_{500}$ (in $M_\sun$).
   }
\label{fig:comp_Lx}
\end{figure*}

\begin{figure*}[hptb]
\begin{center}
\includegraphics[width=0.38\textwidth]{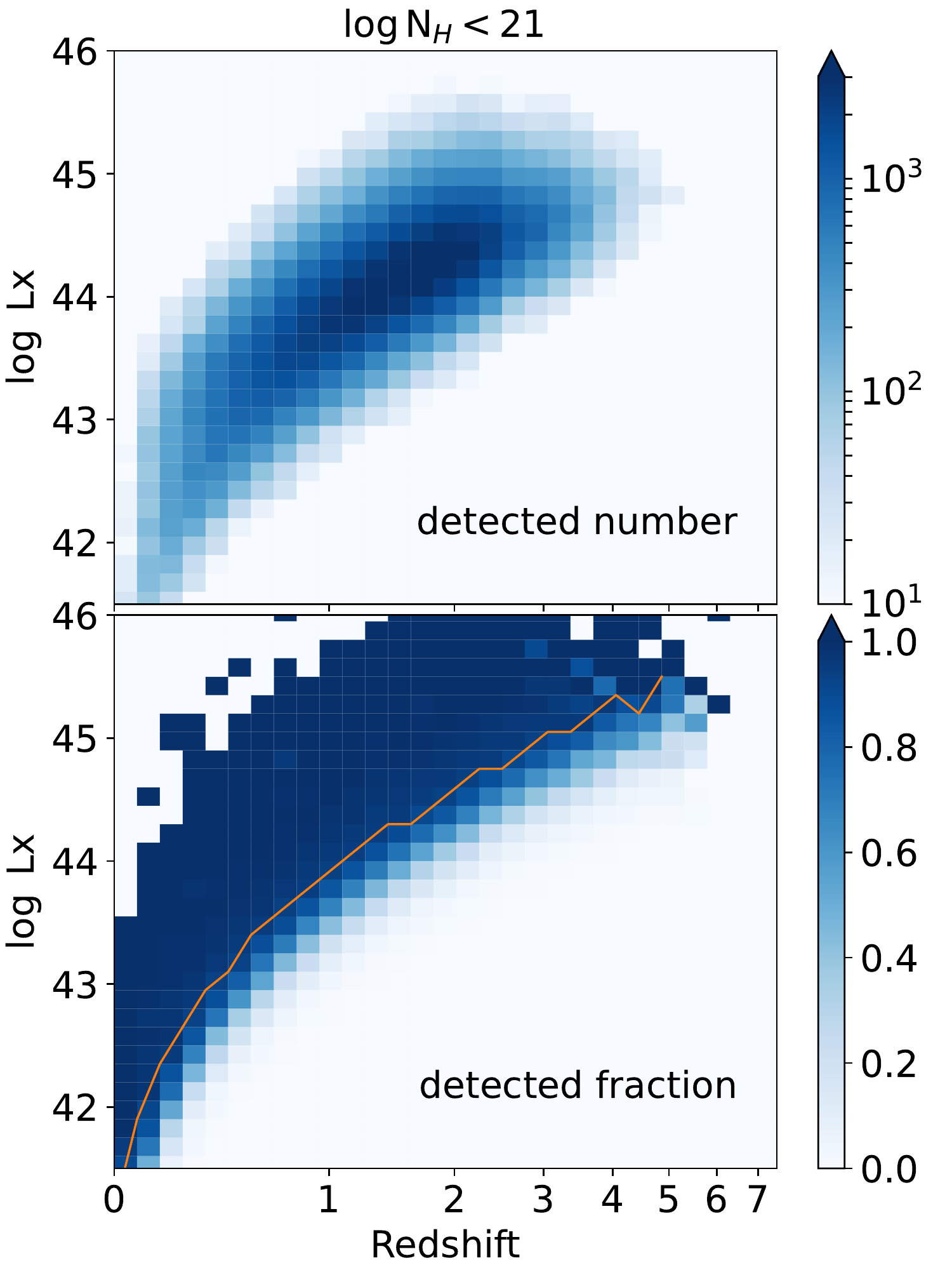}
\includegraphics[width=0.38\textwidth]{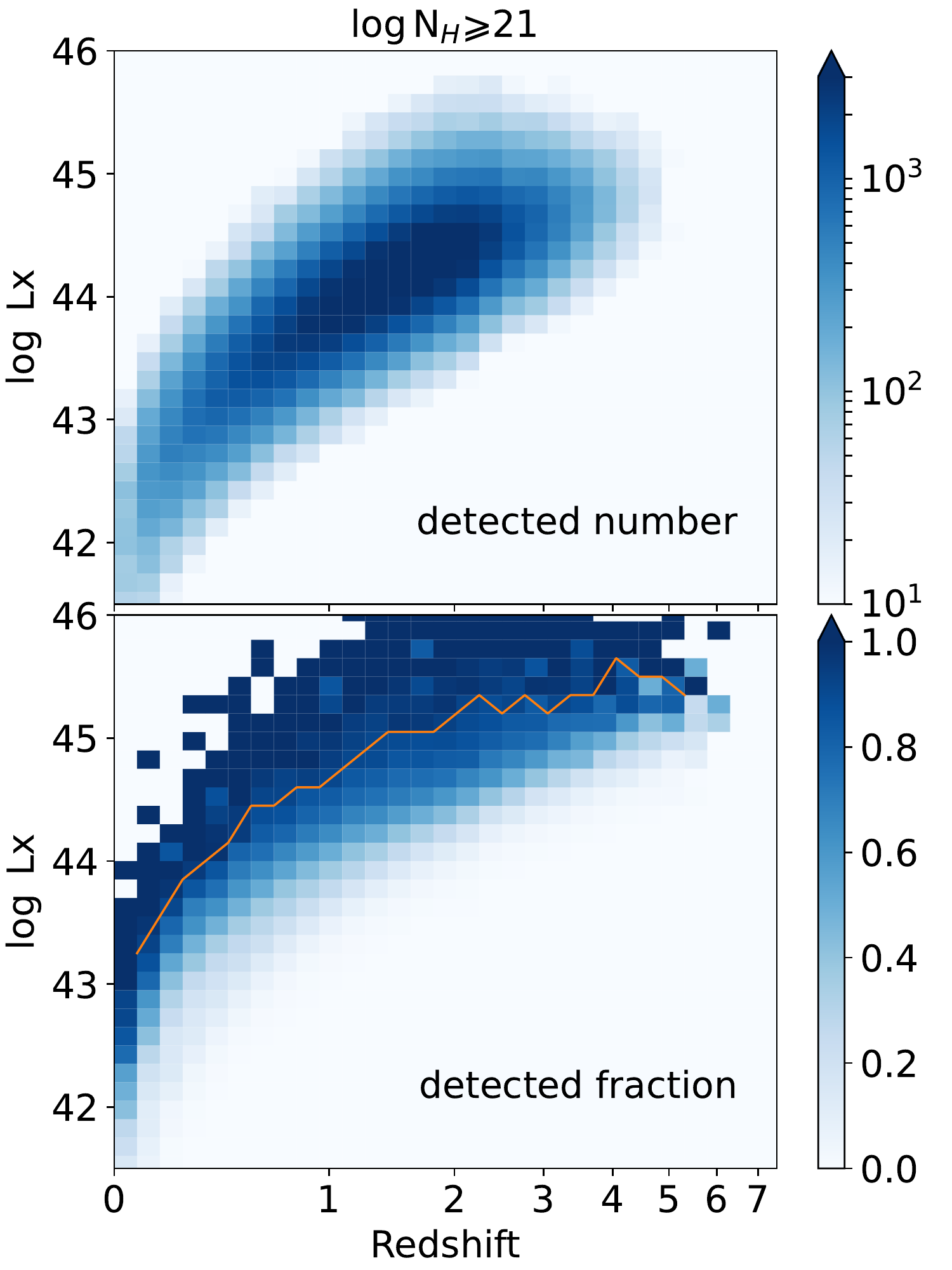}
 \caption{
   Source numbers (top) and the detection completeness (bottom) in a grid of 0.5--2~keV intrinsic luminosity $L_X$ and redshift $z$ for unobscured ($\log$N$_H<$21, left panel) and obscured ($\log$N$_H\geqslant$21, right panel) AGN.
   The overlaid orange lines indicate 90\% completeness.
   }
\label{fig:comp_Lz}
\end{center}
\end{figure*}

The source detection completeness is defined as the detected fraction of sources in the input catalog.
We can also call it selection function or sensitivity.
According to various scientific goals, the completeness may be evaluated in various manners, for example, as the completeness in a particular flux range or above a flux threshold, the completeness according a particular sample selection filter (e.g., hard-band selection, extent selection), and as the completeness for a particular input population (e.g., obscured AGN).
\TL{
A few examples are shown in this section. All of them can easily be drawn using the mock input catalogs that we present with this paper.}

In the left panel in Fig.~\ref{fig:comp_flux} we display a few levels of AGN completeness as a function of source flux.
At the first level, an AGN is detected as the primary counterpart of a source in the point-source catalog, that is, it is detected and correctly classified as a point source (\texttt{EXT\_LIKE}$=$0).
At the second level, an AGN is detected as the primary counterpart regardless of the extent classification, so that the AGN misclassified as clusters are also included. This \ledit{means} an increase of $\sim1\%$  in the completeness at fluxes above $10^{-14}$ \egs.
At the third level, an AGN is detected either as the primary or as the secondary counterpart, so that the cases of blended AGN are included. This means a further $1\%\sim2\%$ increase.
  Similarly, we plot the cluster completeness in the right panel of Fig.~\ref{fig:comp_flux}. At the first level, the fraction of clusters that are detected and correctly classified is not high (only exceeds 80\% at flux $>10^{-13.1}$ \egs). At the second level, however, the completeness is much higher when misclassified sources are included, indicating that the main problem is classification and not detection.
  At the third level, a small fraction (2\%$\sim$3\%) of blended cases are also included.
  There are $46$ clusters (in $18$ realizations) with fluxes $>10^{-13}$ \egs \ that are not detected (either \texttt{ID\_Any} or \texttt{ID\_Any2}) because they have relatively larger scales and thus lower surface brightness than the other clusters.

  In Fig.~\ref{fig:comp_Lx} we display the completeness of a few AGN and cluster subsamples as a function of the 0.5--2~keV luminosity.
  Here we define completeness at the first level, that is, detected and correctly classified.
  For AGN, we selected a few subsamples with different redshift and \N{H} ranges. The huge differences between them clearly show the selection bias against high-z and obscured sources. For clusters, we selected a few subsamples with different ranges of redshift and $R_{500}$ apparent size, which are correlated. The huge differences clearly show the selection bias against high-z and compact sources.
  We also compared the completeness above a few mass ($M_{500}$) thresholds among the sources with $z<0.6$ and $R_{500}>2\arcmin$. As shown in the upper panel, halo mass does not directly impact cluster detection. Higher-mass cluster samples have a higher completeness, as shown in the lower panel, because the cluster mass is strongly correlated with the X-ray luminosity.

  Fig.~\ref{fig:comp_Lx} only illustrates the AGN and cluster selection functions roughly. To quantify the selection functions accurately, we need to measure the detected fraction in finer parameter bins. Taking AGN, for example, we display the completeness distributions for obscured and unobscured AGN in the space of intrinsic luminosity and redshift in Fig.~\ref{fig:comp_Lz}.
  The large sample size allows us to plot these 2D distributions easily for AGN through a simple binning. To quantify the cluster selection function in narrow parameter bins, more sophisticated sampling methods are needed, for example, through Gaussian process \citep[see][]{Liu2021_cluster}.
  We provide the mock AGN and cluster catalogs with the information of input-output association (Appendix \S~\ref{sec:catalogs}). With these catalogs, the selection function may be quantified as needed by specific scientific goals.

Using the eSASS task \texttt{ersensmap}, we calculated the flux limit corresponding to the point-source detection likelihood threshold (\texttt{DET\_LIKE}$=$5) and thus the sky coverage area curve as a function of flux limit. When this curve is normalized to a total area of 1 (solid black line in Fig.~\ref{fig:comp_flux}), it also predicts a detectable fraction that only reflects the detectable flux limit across the whole field, however.
\TL{At high fluxes, the actual detected fraction (sensitivity curve) derived from simulation} is lower than that of the \texttt{ersensmap} flux limit because a source above the detectable flux limit might still be missed because of the fluctuation and measurement uncertainty in the source and background or because of blending with nearby sources.
At low fluxes, the simulation-derived sensitivity curve extends far lower than the flux limit because fluctuation might also make sources below the flux limit detectable in some cases (see also the discussion in \S~\ref{sec:logNlogS}).

\subsection{Source detection efficiency}
\label{sec:effi}

The essential procedure of source detection is source selection from candidates based on the likelihood of each candidate. A low likelihood threshold leads to high completeness (or sensitivity). However, it also leads to a lower purity. A source detection operation has a high efficiency if it results in high completeness and high purity.

For the eROSITA source detection, \TL{we performed a source selection} based on the PSF-fitting likelihood (\texttt{DET\_LIKE}). In the case of multiband PSF fitting, the combined likelihood (\texttt{DET\_LIKE\_0}) was used.
All the settings that were adopted in the whole detection procedure, for example, choice of energy bands, estimation of background, and PSF-fitting region size, could affect the final measurement of \texttt{DET\_LIKE}.
\TL{Optimizing the source detection is just optimizing the \texttt{DET\_LIKE} measurement with the aim that a threshold on it leads to the best efficiency.
  }
  In this section, we compare different settings by means of the completeness--contamination parametric curves as a function of \texttt{DET\_LIKE}, where completeness can be measured as the detected fraction in a given range of input flux (as displayed in Fig.~\ref{fig:comp_flux}) and contamination can be measured as the fraction of spurious sources in the selected sample (blue cumulative fraction distribution displayed in Fig.~\ref{fig:DET_dis_whole}).

We tested four \TL{sets} of energy bands for the source detection and describe them below.
\begin{enumerate}
\item Single-band detection\\
  We ran source detection in a single 0.2-2.3 keV band, which is the most sensitive band of eROSITA. It spans from the lowest energy to the turning point of the eROSITA effective area curve, above which the effective area is much lower.
\item Three-band detection\\
  To cope with the wide range of spectral shapes of all types of AGN and stars, we ran the PSF fitting simultaneously in three bands: in 0.2-0.6, 0.6-2.3, and 2.3-5 keV. In this way, we did not miss the sources that are significant only in the soft 0.2-0.6 keV band or in the hard 2.3-5 keV band.
\item Three soft band detection\\
  In principle, dividing a broad band into multiple narrow bands facilitates detecting sources with a wide range of spectral shapes. However, it might cause additional complexities in combining the detection likelihoods from each individual band. To test this issue, we divided the 0.2-2.3 keV band into three narrow bands: 0.2-0.6, 0.6-1.1, and 1.1-2.3 keV.
\item Four-band detection\\
  eROSITA could detect hard X-ray photons up to 8 keV but with a relatively low sensitivity above 2.3 keV. We have added the 2.3-5 keV band to the three-band detection. In addition to these three bands, we further added the ultra-hard band 5-8 keV to the four-band detection in order to detect ultra-hard sources.
\end{enumerate}

\begin{figure}[hptb]
\begin{center}
\includegraphics[width=0.9\columnwidth]{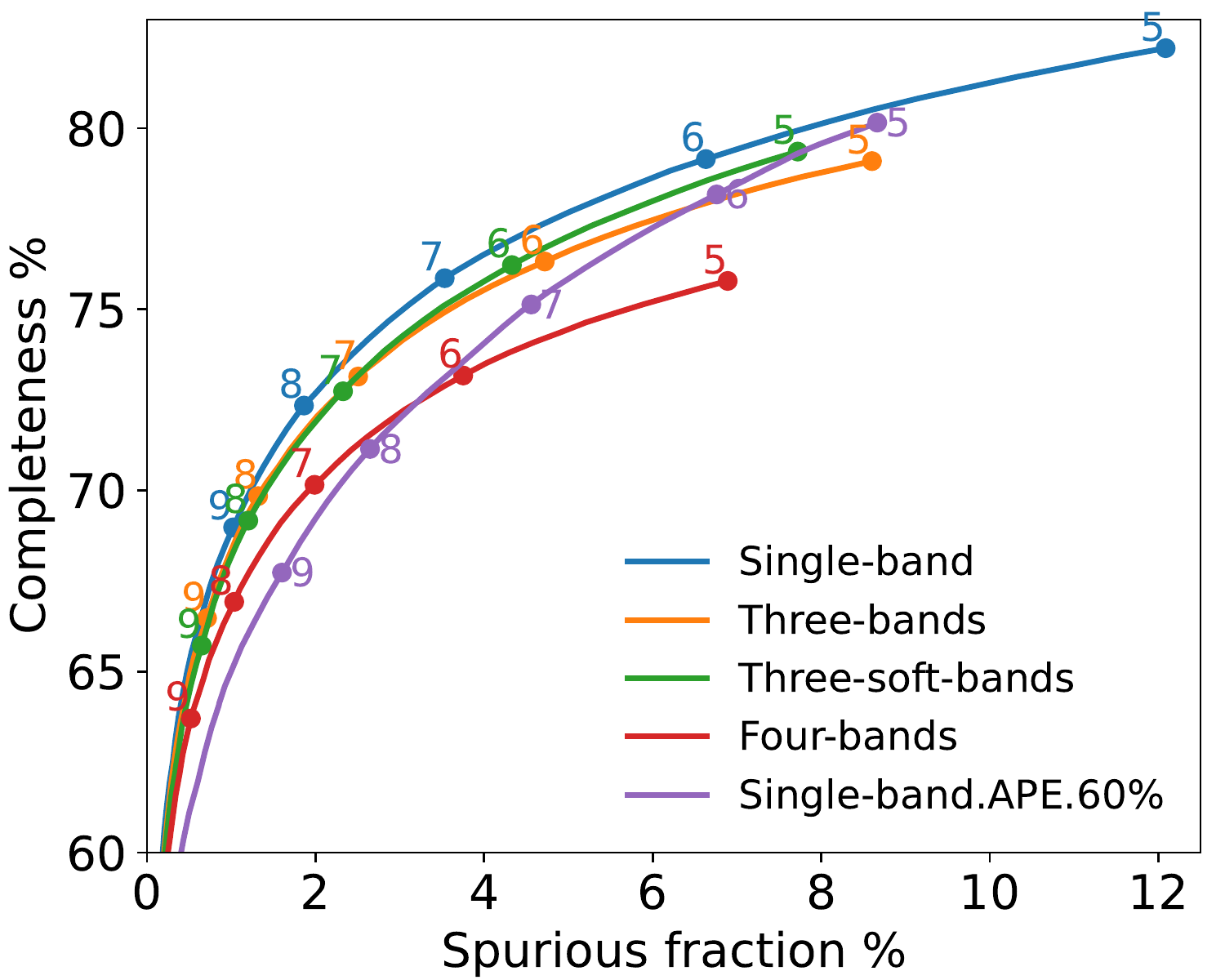}
\caption{Completeness--contamination parametric curves as a function of PSF-fitting detection likelihood \texttt{DET\_LIKE} measured from the four sets of energy bands (blue, orange, green, and red lines).
  The purple line indicates the selection based on the aperture Poissonian likelihood measured in the 0.2-2.3 keV band within 60\% EEF.
  The points corresponding to the serial of \texttt{DET\_LIKE} values are marked on the lines.}
\label{fig:com_con_bands}
\end{center}
\end{figure}

Fig.~\ref{fig:com_con_bands} displays the completeness--contamination curves \TL{corresponding to these four sets} of energy bands.
Here the completeness is defined as the detected fraction above an input-flux limit of $10^{-14.2}$ \egs , and the spurious fraction corresponds to the sources due to background fluctuation.
A curve with relatively higher completeness and lower contamination \TL{indicates} a higher source detection efficiency.
Although the hard band ($>$2.3 keV) provides valuable information, its small effective area and high background result in a negative impact on the source detection efficiency.
When the hard band between $2.3$ and $5$ keV (orange line in the figure) is included, it is not as efficient as using only the single $0.2-2.3$ keV band, where eROSITA has the largest effective area.
Including the ultra-hard band above 5 keV (red line) has a significant negative impact on the efficiency and thus should be avoided.

Dividing the single band into three narrow bands (green line) could in principle help detect sources with extreme spectral shapes, which make them only visible in the very soft or very hard band.
However, such cases are rare in the simulation and in reality, thus it does not make any improvement in the sense of the \TL{overall efficiency.
Based on the same data (in the same band), these two detections show similar efficiency curves resulting from significantly different \texttt{DET\_LIKE} measurements.
The three soft band likelihood is relatively lower and thus leads to relatively lower completeness at a given likelihood threshold.
}
This is because the likelihood value of a source is reduced when the individual band likelihoods are combined into a summary value.

  In addition to the choice of energy bands, we also tested the impact of the essential parameter of PSF fitting, the extraction radius.
  A radius of 15 pixels (60\arcsec) was adopted when the real eFEDS catalog was detected \citep{Brunner2021}.
  With a given background, a point source has an optimized radius in which the S/N is maximized. Reducing it will cause too much loss of source signal, and enlarging it will include too much background noise.
  As we discuss in \S~\ref{sec:ape}, through investigation of aperture photometry, we find an optimized aperture radius of 60\% EEF ($\sim20\arcsec$) for the 0.2--2.3~keV band.
  \citet{Liu2021_AGN} also measured an optimized source spectrum extraction radius that maximized the S/N of each eFEDS source in the 0.2--8~keV band using the eSASS task \texttt{srctool}. The median radius is 28\arcsec.
  Therefore, the adopted 60\arcsec{} PSF-fitting radius is sufficiently large.
  We performed the source detection procedure by changing the radius to a lower value of 12 pixels and a higher value of 18 pixels.
  Then we compared the three cases using the completeness-contamination curve as \ledit{done above}. Although the smaller radius leads to slightly higher likelihoods because the S/N is higher, the efficiency is almost identical in the three cases.
  In other words, the reduced PSF-fitting radius leads to both higher completeness and higher contamination.
  If completeness is taken as the main figure of merit of the catalog and when spurious sources are allowed to be eliminated through multiband follow-up, a smaller PSF-fitting radius might be adopted in order to detect more faint point sources.

    We also tested whether excluding the field of view (FOV) border outside a radius of 180 pixels in the single-band detection improves the detection efficiency. Here the spatial resolution is poor, the background is relatively high, and the calibration is relatively uncertain. Excluding the border region reduces both source signal and background. By comparing the completeness-contamination curve, we did not find any improvement, indicating that at least in the 0.2--2.3~keV band, the border region does not contribute negatively to the whole data set.

    In addition, we tested the idea of repeating the PSF fitting using the PSF-fitting selected catalog as input instead of using the dirty (with many spurious sources) preliminary catalog. By comparing the completeness-contamination curve, we found no improvement either.

\subsection{Classification efficiency}
\begin{figure}[hptb]
\begin{center}
\includegraphics[width=0.9\columnwidth]{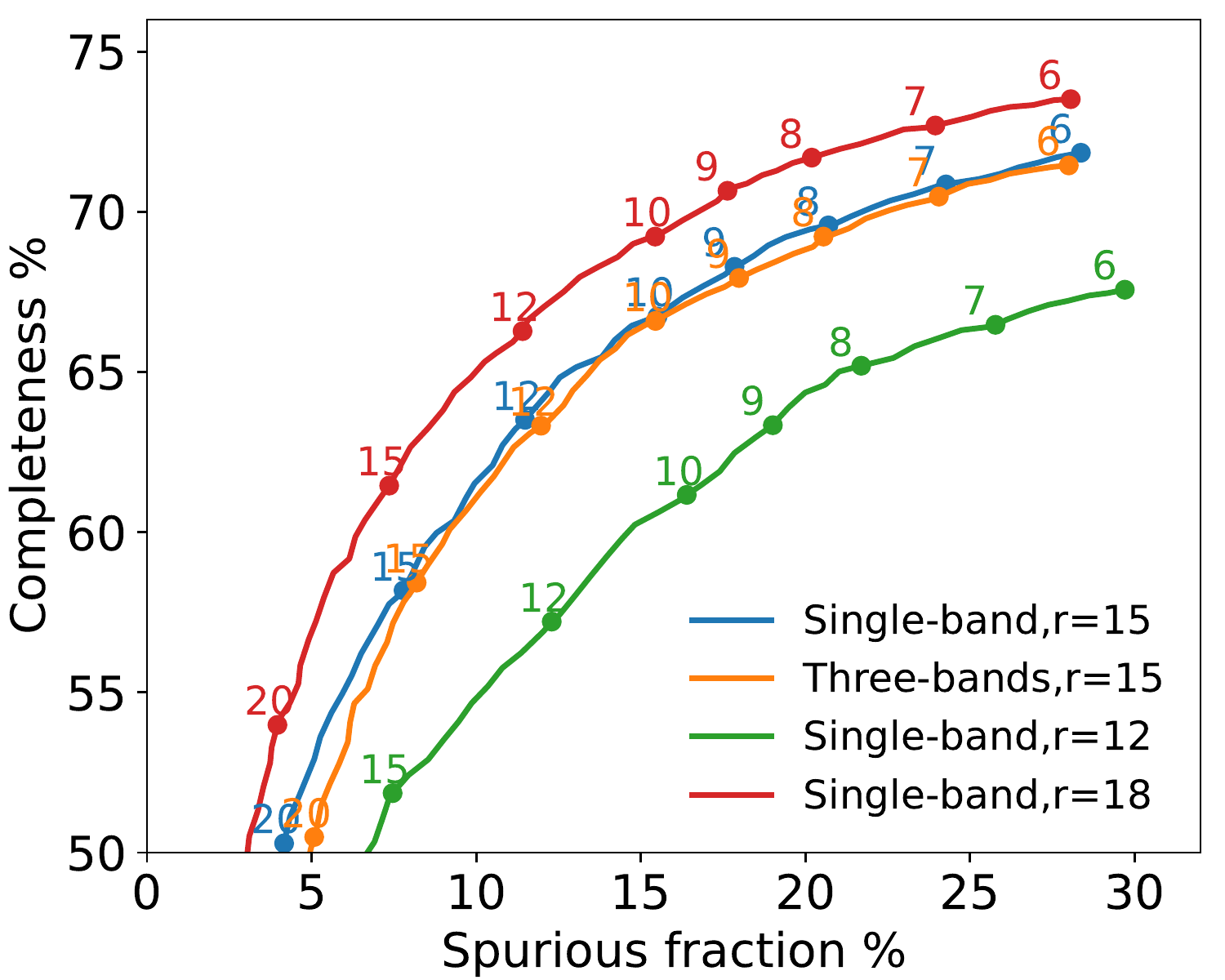}
\caption{Completeness--contamination parametric curves as a function of extent likelihood \texttt{EXT\_LIKE} measured from single-band PSF fitting adopting three different radii (orange: 12 pixels; blue: 15 pixels; red: 18 pixels) and from three-band PSF fitting with a radius of 15 pixels (orange). 
  The points corresponding to the serial of discrete \texttt{EXT\_LIKE} values are marked on the lines.}
\label{fig:com_con_EXT}
\end{center}
\end{figure}

As discussed in \S~\ref{sec:compl}, the main problem for galaxy clusters is source classification, that is, distinguishing between point and extended sources, and not detection. In this section, we investigate the classification efficiency through the completeness-contamination parametric curve as a function of extent likelihood \texttt{EXT\_LIKE} instead of detection likelihood \texttt{DET\_LIKE}.
Here, completeness is defined as the fraction of detected clusters with input 0.5--2~keV fluxes above $5\times 10^{-14}$ \egs , and the spurious fraction corresponds to the extended sources that are not associated with any input clusters (either \texttt{ID\_Any} or \texttt{ID\_Any2}).
This completeness-contamination curve reflects the combined efficiency of source detection and classification. Because most of the clusters are significantly detected (panel 3 in Fig.~\ref{fig:DET_dis_V18C}), the main factor is classification.
As displayed in Fig.~\ref{fig:com_con_EXT}, the efficiency is highly dependent on the PSF-fitting radius. The larger the region involved in the PSF fitting, the more efficiently clusters and point sources can be distinguished.
Comparison of the three cases at the same levels of \texttt{EXT\_LIKE} thresholds shows that the PSF-fitting radius impacts the completeness more than the contamination.
This is because point sources that are misclassified as extended sources (mostly blended, discussed in \S~\ref{sec:cla}) have small scales not much larger than the PSF HEW. Some compact clusters are hard to be distinguished from point sources unless the PSF-fitting radius is extended to a large scale \ledit{at} which the PSF drops to extremely low values and thus the difference between PSF and the cluster profile can be revealed.
We recommend using a larger PSF-fitting radius or performing post hoc analysis within a larger radius in future eROSITA surveys to improve the completeness of the galaxy clusters.

We also compare the single-band detection with the three-band detection in Fig.~\ref{fig:com_con_EXT}. Adding the 2.3--5~keV band is not helpful for the detection and classification of clusters because the cluster emission is more prominent in the soft band.

\section{AGN number counts}
\label{sec:ape_logNlogS}
\subsection{Aperture-photometry-based likelihood}
\label{sec:ape}
\TL{In addition to the PSF-fitting likelihood \texttt{DET\_LIKE},} the eSASS task \texttt{apetool} also measures a likelihood  for each source by comparing the aperture source photon counts with the Poisson distribution of the aperture background counts.
This likelihood can also be used for source selection.
Fig.~\ref{fig:com_con_bands} compares the selection efficiencies based on the 0.2--2.3~keV PSF-fitting likelihood and the 0.2--2.3~keV aperture Poissonian likelihood (\texttt{APE\_LIKE}).
We only ran aperture photometry for the sources selected with the PSFfitting (\texttt{DET\_LIKE}$\geqslant$5), so that the \texttt{APE\_LIKE} selection is in fact a two-step filtering.
When a low \texttt{APE\_LIKE} threshold of $5$ is adopted, this two-step selection shows a similar completeness and purity as the PSF-fitting likelihood selection because the \texttt{DET\_LIKE}$\geqslant$5 preselection plays the main role.
At a high likelihood (e.g., $>$8), where the impact of the \texttt{DET\_LIKE}$>=5$ preselection becomes minor, the aperture likelihood selection shows a relatively lower efficiency.
Compared with the aperture likelihood based on source and background aperture counts, making use of additional knowledge about the source image profile and the PSF model therefore improves the efficiency of source detection.
We used the PSF-fitting likelihood for the source detection and only used the aperture photometry results to recover the number counts of point sources.

\begin{figure}[hptb]
\begin{center}
\includegraphics[width=0.9\columnwidth]{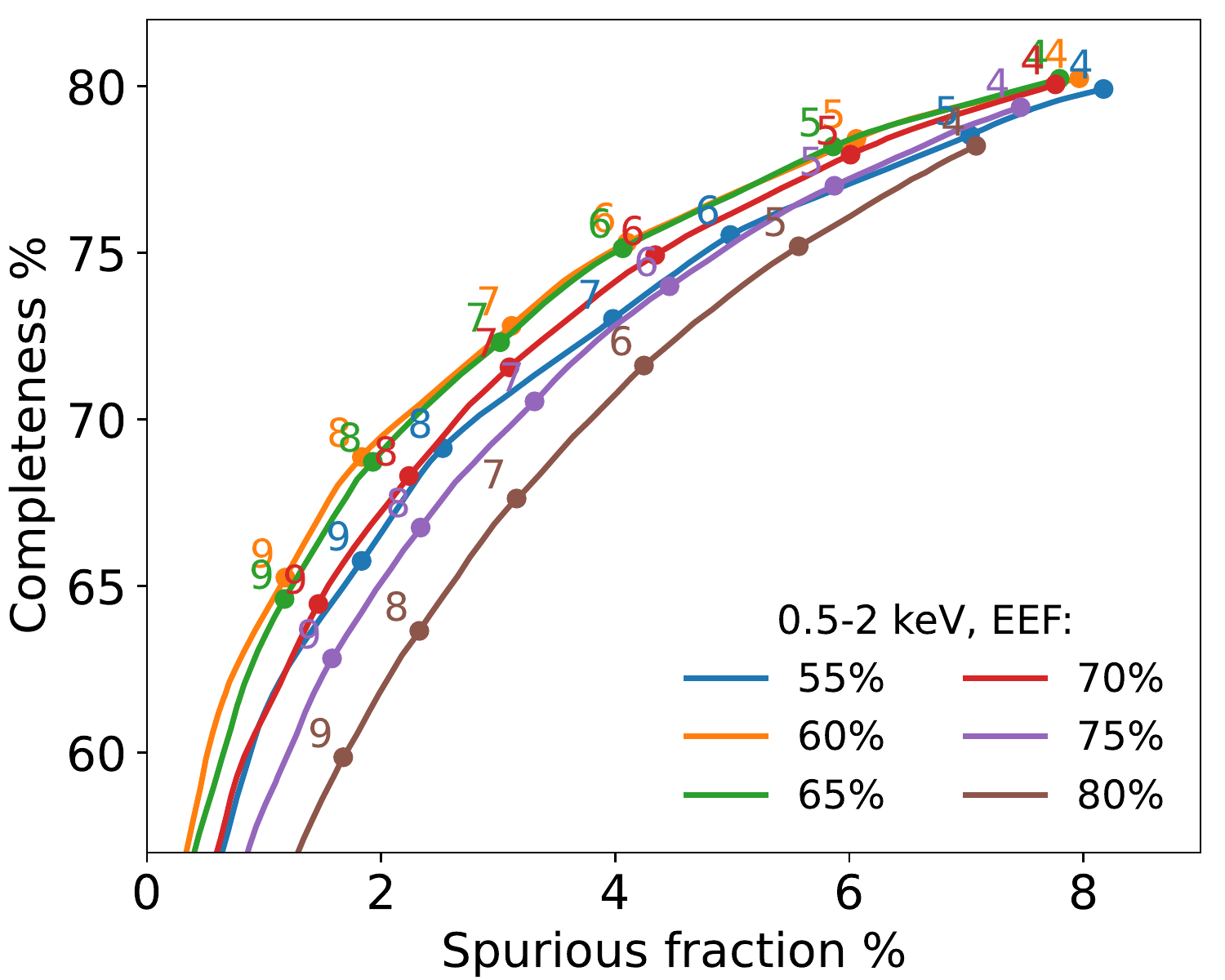}
\includegraphics[width=0.9\columnwidth]{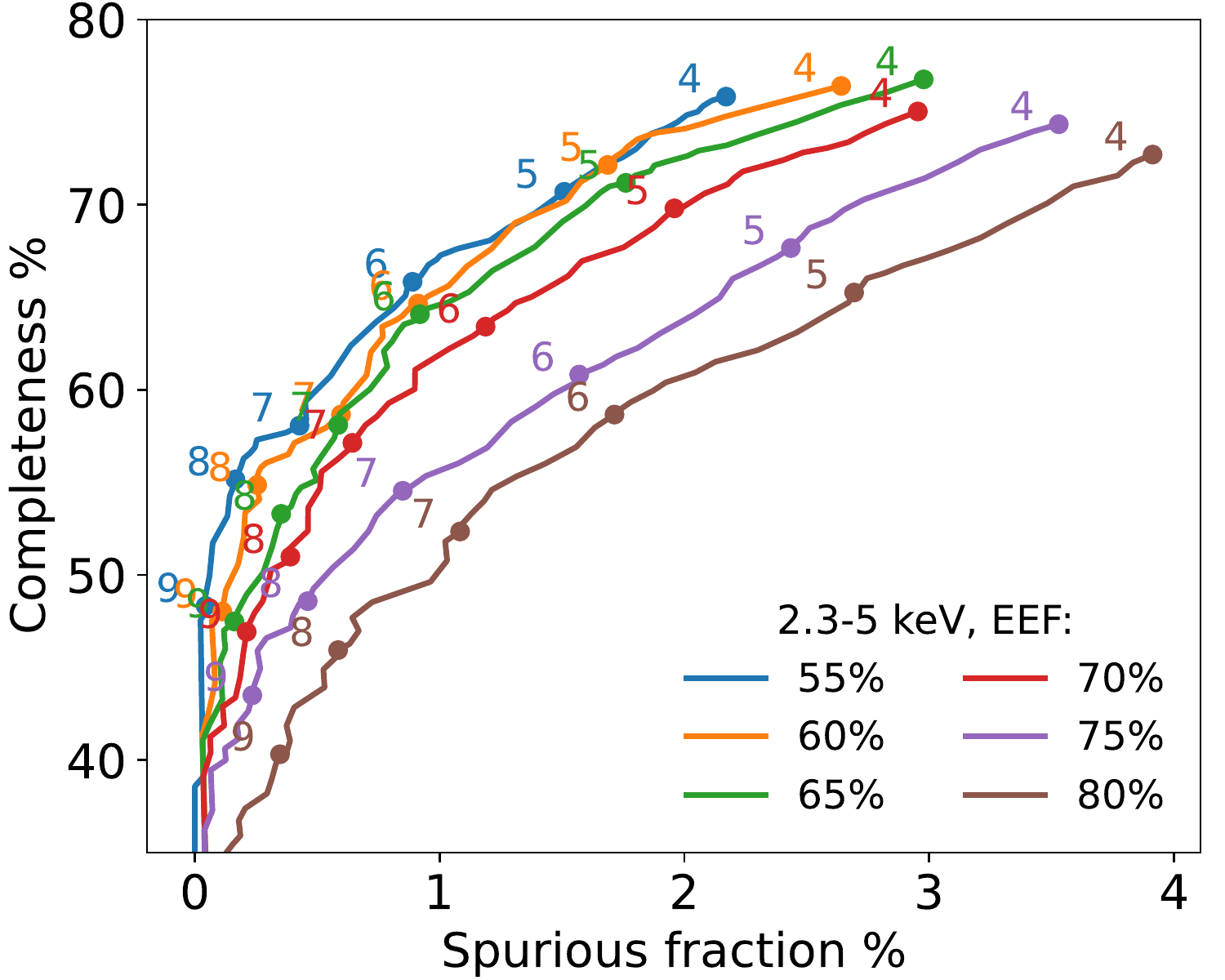}
\caption{Completeness--contamination parametric curves as a function of aperture Poissonian likelihood \texttt{APE\_LIKE} measured within a serial of aperture radius (in terms of EEF) in the 0.5-2 keV band (top) and in the 2.3-5 keV band (bottom). The points corresponding to the serial of discrete \texttt{APE\_LIKE} values are marked on the lines.}
\label{fig:com_con_APE}
\end{center}
\end{figure}

The essential parameter in aperture photometry is the aperture radius. We ran \texttt{apetool} with \ledit{radii of EEFs of 55\%, 60\%, 65\%, 70\%, 75\%, and 80\%} in the soft 0.5--2~keV band and the hard 2.3--5~keV band,
\TL{and compare the completeness--contamination parametric curves as a function of the \texttt{APE\_LIKE} in Fig.~\ref{fig:com_con_APE}.
}
In the soft band, an EEF of 60\% or 65\% leads to the best efficiency in distinguishing source signal from background. \TL{This aperture can be adopted as the optimized size for aperture photometry.}
In the hard band, although the PSF has a wider shape, we find that the most efficient aperture (55\% or 60\%) is smaller than that in the soft band. This is because of the higher noise in the hard band: a smaller region around the PSF core leads to a higher S/N inside it.
Based on these tests, we suggest adopting 60\% EEF, \TL{which corresponds to $\sim20\arcsec$}, in the aperture photometry for eFEDS.

\subsection{AGN number counts}
\label{sec:logNlogS}

\begin{figure}[htbp]
\begin{center}
\includegraphics[width=\columnwidth]{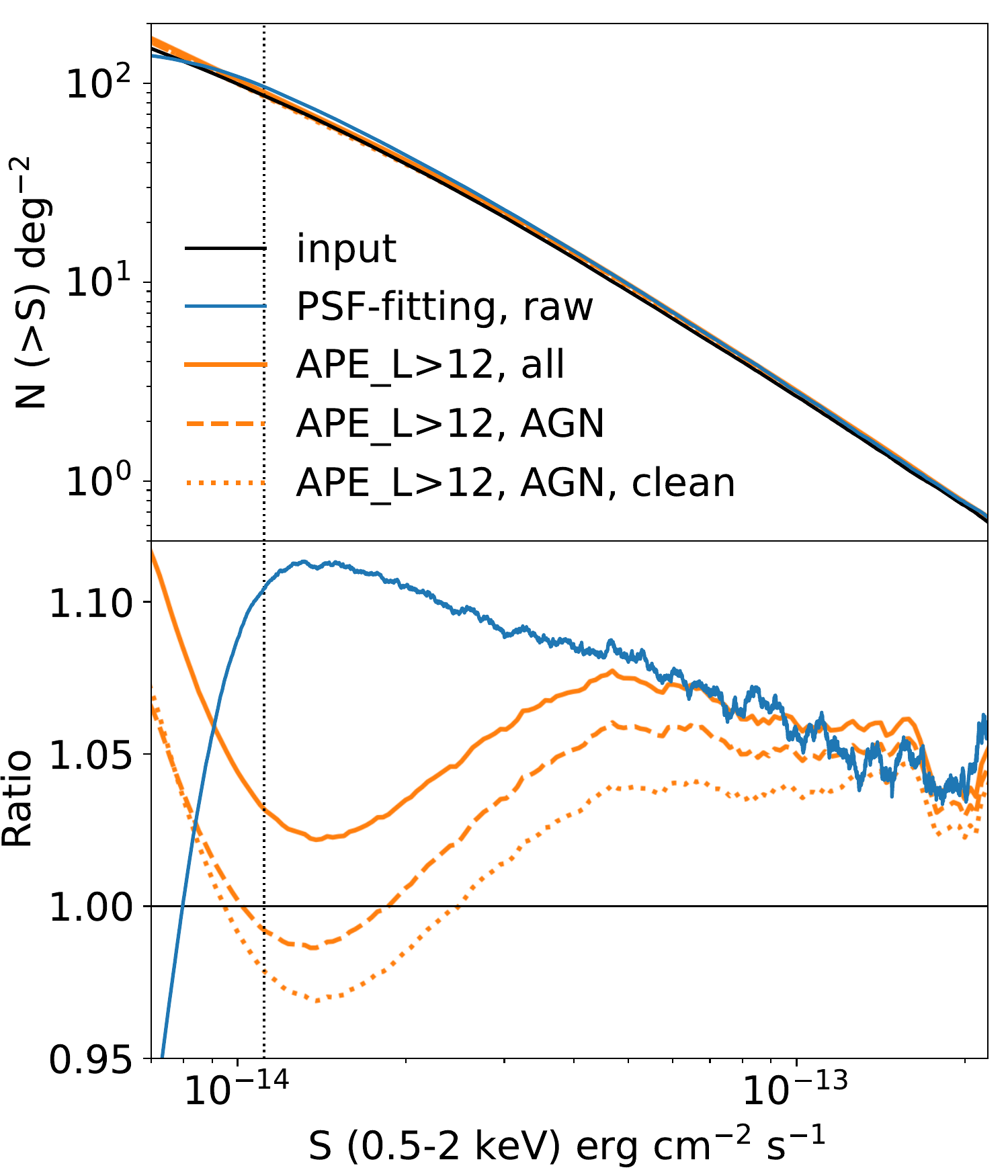}
\caption{Iput (solid black line) and output number counts of AGN (upper panel) and the output-to-input ratios (lower panel).
  The blue lines correspond to the raw cumulative distribution of 0.5-2~keV fluxes measured by forced PSF fitting divided by the total area.
  The solid orange line shows the number counts we constructed based on the aperture Poissonian likelihood \texttt{APE\_LIKE} using all the sources with \texttt{APE\_LIKE}$>$12.
  When only the genuine AGN instead of all the sources (including misclassified clusters) is used, the result is displayed as the dashed orange line.
  After a further flux correction when constructing the number counts of genuine AGN, the result is displayed as the dotted orange line.
  The vertical dotted line ($1.1\times10^{-14}$ \egs) corresponds to a sensitivity of 50\%.
  }
\label{fig:logNlogS}
\end{center}
\end{figure}

\TL{
  As an extragalactic field, eFEDS has a majority of AGN in the X-ray sources \citep{Liu2021_AGN}. 
  From the single-band-detected sources located inside the 90\%-area region, we selected \ledit{an AGN catalog as significantly detected (\texttt{DET\_LIKE}$\geqslant$8) point sources} (\texttt{EXT\_LIKE}=0) that are not attributed to any input stars (according to \texttt{ID\_Any}).
  We constructed the cumulative distribution of their 0.5-2~keV fluxes measured by forced PSF fitting and divided it by the observed area (126.6 degree$^2$). \ledit{These raw number counts per degree$^2$ are compared with the input AGN number counts in Fig.~\ref{fig:logNlogS}.
Their ratios} as displayed in the lower panel of Fig.~\ref{fig:logNlogS} can be used to convert the raw distribution of the real eFEDS catalog into its intrinsic number count \ledit{distribution}.
The output catalog is significantly overpopulated. This has three reasons: 1) catalog contamination, especially from galaxy clusters that are misclassified as AGN, 2) source flux overestimation caused by blended nearby sources, and 3) Eddington bias due to Poisson fluctuation.
These factors are so strong that they lead to significantly more output sources than the input after compensating for source detection incompleteness (Fig.~\ref{fig:comp_flux}).
}

Then we tested the method we used to construct the point source number counts for the real eFEDS catalog \citep{Brunner2021}.
We measured an \texttt{APE\_LIKE} for each source using \texttt{apetool}.
Selecting sources above an \texttt{APE\_LIKE} threshold (e.g., 12), we stacked them to construct the number counts considering each source as a random variable following a probability density distribution of flux rather than a single count \citep{Georgakakis2008}. 
In this way, the probability of a source having a flux below the detection limit was taken into account and the Eddington bias was corrected for.
As displayed in Fig.~\ref{fig:logNlogS}, this method (solid orange line) largely eliminates the bump of the raw distribution (blue line) near the detection limit and leads to a number count distribution that is more consistent with the input, although still higher than it by a few percent.

  We listed three reasons for the overpopulated output catalog above.
  As the third problem (Eddington bias) has already been addressed by the method based on aperture photometry, the main reasons for the number count overestimation are the first two, that is, contamination and source blending.
  These two issues are impossible to address for real data; they can only be quantified through simulation.
We explain these two factors through the following two experiments.
First, when constructing the number counts, we selected only the sources that are attributed to genuine AGN (according to \texttt{ID\_Uniq}). 
The spurious AGN due to compact clusters, fragments of AGN or cluster wing region, or background fluctuations were excluded. Above a likelihood threshold of \texttt{APE\_LIKE}$>$12, the main source of spurious AGN is the contamination of clusters (Fig.~\ref{fig:DET_dis_V18C}).
As displayed in Fig.~\ref{fig:logNlogS}, excluding this contamination improves the derived number counts significantly.
Second, we corrected the source flux for the blending of nearby sources.
When a detected source has a secondary counterpart (\texttt{ID\_Any2}) and this secondary counterpart is not identified as the unique counterpart (\texttt{ID\_Uniq}) of any detected source, we calculated  a flux correction factor for each of
these sources as the ratio of the counts of the primary counterpart
and the total counts of the two sources because we know the photon counts of both of them in the 20\arcsec{} circular region (approximately the same size for aperture photometry) of the detected source.
As displayed in Fig.~\ref{fig:logNlogS}, when this flux correction is applied to the relevant sources in the subsample of genuine AGN, the derived number counts are further reduced, confirming that source flux enlargement caused by blending causes an overestimation of number counts.
It is impossible to recover the input number counts precisely because of the additional uncertainty in the flux measurement. The input sources have a variety of spectral shapes, while for the output sources, a fixed spectral shape was assumed when we converted the count rate into flux.

  The simulation provides a correction curve (blue line in the lower panel of Fig.~\ref{fig:logNlogS}) that can be used to convert the raw distribution of the real catalog into the intrinsic number counts and to predict that the intrinsic number counts derived in this way are lower than the counts derived using the method based on aperture photometry. This prediction has been confirmed with real eFEDS data \citep{Brunner2021}. It does not indicate any problem with the method based on aperture photometry itself because the information of contamination and blending is invisible for the real data and can only be revealed through simulation. Therefore, we recommend taking these factors into account \ledit{only} when an accurate measurement of AGN luminosity function is needed.

\section{Discussions and conclusions}
\label{sec:conclusion}
\subsection{eFEDS source detection strategy}

X-ray source detection can be considered as a two-step procedure: first, a search for source candidates, and second, a down-selection of these candidates to exclude those with a high false-positive rate. The detected sources \ledit{are often characterized either simultaneously in the down-selection \citep[e.g.,][]{Hasinger1994,Liu2020} or separately in an additional process \citep[e.g.,][]{Liu2013}, including brightness measurement and classification of point or extended source}. The goal of the first step is to find possible candidates in as complete a way as possible. This can be accomplished using various algorithms such as the sliding-box detection algorithm \citep{Deponte1993, Calderwood2001}, the Mexican hat wavelet detection algorithm \citep{Freeman2002}, or Voronoi tessellation plus a friends-of-friends detection algorithm \citep{Ebeling1993, Liu2013}. The sample selection in the second step determines the essential figure of merit for the source detection, that is, the efficiency with which sources can be distinguished from background fluctuations. The false-positive rate of a source candidate is usually expressed as the source likelihood, which is used to threshold samples. The ideal source detection strategy should define a source likelihood such that a threshold on the likelihood leads to both an optimized completeness and an optimized purity.

With the eSASS tasks, we can measure two types of source likelihoods.
The PSF-fitting likelihood is measured through a maximum-likelihood PSF-fitting, which uses an algorithm that was first developed for ROSAT \citep{Hasinger1994} and is still in use for \XMM{} \citep[e.g.,][]{Liu2020}. The aperture-based likelihood is calculated through a comparison of the source and background photon counts extracted in a small aperture \citep{Georgakakis2008} to the expectation of Poisson statistics.
By comparing the source detection efficiency based on these two types of likelihood, we confirm the good performance of the PSF-fitting algorithm in the sense that involving the known PSF shape information leads to a higher efficiency than using only aperture photometry results (Fig.~\ref{fig:com_con_bands}).
We also remark that in a broad sense, the likelihood is defined by all the methods and parameters adopted in the procedure of source detection because they all affect it; and the likelihood definition based either on PSF-fitting or on aperture photometry underestimates the false-positive rate (Fig.~\ref{fig:DET_dis_whole}, Fig.~\ref{fig:DET_dis_V18C}) because they are based on the assumption that there are no additional uncertainties.

  Based on the experience and rich legacy of previous X-ray surveys, the eSASS scheme uses a multistage source detection procedure around a core algorithm of PSF fitting, that is, it first detects a preliminary catalog and then performs PSF fitting to make a final selection.
Since the preliminary catalog requires high completeness but not high purity, which can be controlled in subsequent steps, a simple sliding-box algorithm is used for it, and in this work, we focused on the PSF-fitting step.
A first choice must be made as to whether to use all the data or not.
The simple combination of more data does not necessarily result in a higher contrast of source to background, as it might introduce more noise rather than signal.
For instance, at the border of the FOV, Wolter-I X-ray telescopes suffer from strong vignetting and degraded spatial resolution, and at high energies, X-ray telescopes usually capture fewer X-ray photons but gain more noise due to instrumental background.
To understand how these effects contribute to the data set as a whole, we measured and compared the source detection efficiencies through simulations adopting each of the candidate strategies (Fig.~\ref{fig:com_con_bands}).
We find that including hard bands in the detection (three band and four band) introduces more noise and thus reduces the overall efficiency.
Also considering that a single-band detection leads to a more straightforward estimation of selection function, we chose the single-band detection to create the main eFEDS catalog \citep{Brunner2021}.
However, to detect hard sources (especially obscured AGN), we also make use of the three-band (0.2-0.6, 0.6-2.3, and 2.3-5 keV) detection, and select \ledit{sources based on} the 2.3--5~keV individual band detection likelihood \citep[][Nandra et al. in prep.;]{Brunner2021}.
The key parameter of the PSF fitting is the fitting aperture radius.
We find that the choice of this radius affects the measured detection likelihood, but does not impact the detection efficiency of point sources.
Enlarging the PSF-fitting radius, however, might improve the efficiency of identifying clusters (Fig.~\ref{fig:com_con_EXT}).
A moderate radius of $15$ pixels (60\arcsec) was adopted when the real eFEDS catalog was created \citep{Brunner2021}. For future eROSITA surveys, PSF fitting with a larger radius or a post hoc analysis with a larger radius is recommended.
For aperture photometry, the simulation suggests that an optimized aperture radius corresponds to 60\% EEF.
Excluding the FOV border or repeating the PSF fitting might improve the detection of faint sources, but no visible impact is found for the whole eFEDS sample in the soft band.

The eFEDS catalog is dominated by soft point sources \citep{Liu2021_AGN}, containing only $<2\%$ galaxy clusters \citep{Liu2021_cluster}.
When we chose the source detection strategy for eFEDS, we aimed at the overall quality of the entire X-ray catalog rather than focusing on any particular type of astronomical object.
We remark that the method described in this paper allows us to fine-tune the detection accordingly to improve the detection of any particular type of objects such as obscured AGN or galaxy clusters, and different types of objects will require different strategies or parameters.
We also remark that the simulation results reported in this paper are subject to the observing strategy of eFEDS.
There will be differences in pointing-mode observations and in the eRASS observations of regions with a largely different exposure depth, for instance, the south ecliptic pole \citep[see also the discussion in][]{Clerc2018}.
The results of pointing-mode simulation are reported in a following paper about the extragalactic serendipitous X-ray catalog detected from several pointing-mode observations (Liu T. et al. in prep.).

\subsection{Characterizing the eFEDS catalog}

In order to simulate eFEDS as representative of the real data as possible, we extracted the background spectrum from the real eFEDS data and decomposed it into an X-ray component and a particle component. The two components were simulated separately with and without vignetting.
We compared the mock data with the real data in a few aspects in Fig.~\ref{fig:real_mock}. The sources and the background are largely similar \ledit{between them}, except that the cluster surface profile distribution has a relatively large uncertainty and possibly differs from the real eFEDS clusters.
With this detailed construction of mock data and performing on it the identical source detection pipeline used for the real eFEDS catalog \citep{Brunner2021}, we can quantify the completeness and purity of the real eFEDS catalog with the mock catalog.

We introduced a detailed strategy of analyzing detected sources in a simulation based on the origin of each photon. By verifying the source types (point source or cluster) of the primary and secondary (if exists) input counterparts of each detected source, we identified the nature of the detected sources as either correctly detected point or extended sources or due to blended sources, fragmentation of a large source, or misclassified sources.
By plotting the distributions of various classes of detected sources as a function of source properties (e.g., detection likelihood and extent likelihood), the fractions of spurious sources or any particular cases as mentioned above can be measured quantitatively.
For the single-band-detected catalog, adopting a \texttt{DET\_LIKE} threshold of 5, 6, and 8 resulted in a spurious fraction of 11.5\%, 6.3\%, and 1.8\%, respectively.
\ledit{Selecting hard sources from the three-band-detected catalog} with the 2.3--5~keV individual band likelihood \texttt{DET\_LIKE\_3}$>$10 and the extent likelihood \texttt{EXT\_LIKE}$<14$ resulted in a spurious fraction of 2.5\%.
For the single-band detection, the point source catalog (\texttt{EXT\_LIKE}$=$0) includes 3\% misclassified clusters, and the cluster catalog (\texttt{EXT\_LIKE}$>$0) includes 29\% misclassified point sources.
The fraction of spurious clusters is likely overestimated because the mock cluster catalog shows a strong peak near the lower limit of \texttt{EXT\_LIKE}, which does not exist in the real eFEDS catalog (Fig.~\ref{fig:real_mock}).
In the single-band-detected catalog, a fraction of 34\% of the correctly classified point sources is contaminated by nearby point sources or clusters; but for the misclassified point sources, this fraction increases to 83\%, indicating that source blending is the main reason for misclassifying point sources as extended.
In the eFEDS 90\% area region, the simulation predicts 42.9 spurious clusters (\texttt{EXT\_LIKE}$>$0) when we adopt \texttt{DET\_LIKE}$>$12 and \texttt{EXT\_LIKE}$>$12 and predicts 9.4 spurious clusters when we adopt \texttt{DET\_LIKE}$>$20 and \texttt{EXT\_LIKE}$>$20.
Using the mock output catalogs (Appendix \S~\ref{sec:catalogs}) presented with this paper, the fraction of any particular case in the eFEDS catalog under any particular sample selection criteria can be measured.

On the output side, we characterized the catalog purity. On the input side, we measured the detection completeness (selection function) in various manners, including the detected fraction of AGN or clusters as a function of input flux (Fig.~\ref{fig:comp_flux}), the detected fraction as a function of luminosity for AGN in particular redshift and column density ranges and for clusters in particular redshift, scale, and mass ranges (Fig.~\ref{fig:comp_Lx}), and the distribution of AGN completeness in 2D luminosity--redshift space.
The AGN completeness exceeds 80\% at 0.5--2~keV fluxes above $10^{-14.2}$ \egs, and the cluster completeness exceeds 80\% at 0.5--2~keV fluxes above $10^{-13.1}$ \egs. The main cause of the cluster incompleteness is misclassification. Including the misclassified sources, the cluster completeness exceeds 80\% at fluxes above $10^{-13.7}$ \egs.
In this paper, we only display the AGN and cluster selection functions approximately.
More sophisticated sampling methods are needed in order to construct the selection function of AGN and clusters more accurately as a function of multidimensional intrinsic source properties from the mock catalogs, which are available with this paper (Appendix \S~\ref{sec:catalogs}).
In a few accompanying papers, \ledit{detailed analyses} of the selection functions \ledit{are} performed in the demography studies of AGN (e.g., Buchner et al. in prep.) and galaxy clusters \citep[e.g.,][]{Liu2021_cluster}.

We compared the point-source number counts constructed with the method based on aperture photometry, which is used for the real eFEDS catalog, with the input number counts, and confirmed that this method can broadly recover the input number counts.
However, the recovered number counts are slightly (a few percent) higher than the input.
Our sophisticated analyzing method based on the origin of each photon enabled us to determine that this is caused by contamination of misclassified clusters and by source blending.
These factors can only be revealed by simulation and not in any analytical way.

Through the simulation, we also realized a few minor issues with the current (version c001) eROSITA PSF models.
As described in Appendix \S~\ref{sec:accuracy}, the source positions measured through PSF fitting have a tiny offset and the positional uncertainties are slightly underestimated. Therefore, we recommend post hoc astrometric corrections to the real eFEDS catalog \citep[done in][]{Brunner2021}.
We also noted a $\sim5\%$ uncertainty in the PSF fitting measured source counts (or flux) caused by the normalization of the PSF models (Appendix \S~\ref{sec:accuracy}).

\begin{acknowledgement}
This work is based on data from eROSITA, the soft X-ray instrument aboard SRG, a joint Russian-German science mission supported by the Russian Space Agency (Roskosmos), in the interests of the Russian Academy of Sciences represented by its Space Research Institute (IKI), and the Deutsches Zentrum f\"ur Luft- und Raumfahrt (DLR). The SRG spacecraft was built by Lavochkin Association (NPOL) and its subcontractors, and is operated by NPOL with support from the Max Planck Institute for Extraterrestrial Physics (MPE). The development and construction of the eROSITA X-ray instrument was led by MPE, with contributions from the Dr. Karl Remeis Observatory Bamberg \& ECAP (FAU Erlangen-Nuernberg), the University of Hamburg Observatory, the Leibniz Institute for Astrophysics Potsdam (AIP), and the Institute for Astronomy and Astrophysics of the University of T\"ubingen, with the support of DLR and the Max Planck Society. The Argelander Institute for Astronomy of the University of Bonn and the Ludwig Maximilians Universit\"at Munich also participated in the science preparation for eROSITA.

The eROSITA data shown here were processed using the eSASS/NRTA software system developed by the German eROSITA consortium.

Some of the results in this paper have been derived using the HEALPix (K.M. G\'orski et al., 2005, ApJ, 622, p759) package.
\end{acknowledgement}

\bibliographystyle{aa}
\bibliography{sim}

\begin{thebibliography}{26}
\expandafter\ifx\csname natexlab\endcsname\relax\def\natexlab#1{#1}\fi

\bibitem[{{Anderson} {et~al.}(2015){Anderson}, {Gaspari}, {White}, {Wang}, \&
  {Dai}}]{Anderson2015MNRAS.449.3806A}
{Anderson}, M.~E., {Gaspari}, M., {White}, S. D.~M., {Wang}, W., \& {Dai}, X.
  2015, \mnras, 449, 3806

\bibitem[{{Brandt} \& {Alexander}(2015)}]{Brandt2015}
{Brandt}, W.~N. \& {Alexander}, D.~M. 2015, \aapr, 23, 1

\bibitem[{{Brandt} \& {Hasinger}(2005)}]{Brandt2005}
{Brandt}, W.~N. \& {Hasinger}, G. 2005, \araa, 43, 827

\bibitem[{{Brunner} {et~al.}(2021){Brunner}, {Liu}, {Lamer}, {Georgakakis},
  {Merloni}, {Brusa}, {Bulbul}, {Dennerl}, {Friedrich}, {Liu}, {Maitra},
  {Nandra}, {Ramos-Ceja}, {Sanders}, {Stewart}, {Boller}, {Buchner}, {Clerc},
  {Comparat}, {Dwelly}, {Eckert}, {Finoguenov}, {Freyberg}, {Ghirardini},
  {Gueguen}, {Haberl}, {Kreykenbohm}, {Krumpe}, {Osterhage}, {Pacaud},
  {Predehl}, {Reiprich}, {Robrade}, {Salvato}, {Santangelo}, {Schrabback},
  {Schwope}, \& {Wilms}}]{Brunner2021}
{Brunner}, H., {Liu}, T., {Lamer}, G., {et~al.} 2021, arXiv e-prints,
  arXiv:2106.14517

\bibitem[{{Calderwood} {et~al.}(2001){Calderwood}, {Dobrzycki}, {Jessop}, \&
  {Harris}}]{Calderwood2001}
{Calderwood}, T., {Dobrzycki}, A., {Jessop}, H., \& {Harris}, D.~E. 2001, in
  Astronomical Society of the Pacific Conference Series, Vol. 238, Astronomical
  Data Analysis Software and Systems X, ed. J.~{Harnden}, F.~R., F.~A.
  {Primini}, \& H.~E. {Payne}, 443

\bibitem[{{Chuang} {et~al.}(2019){Chuang}, {Yepes}, {Kitaura},
  {Pellejero-Ibanez}, {Rodr{\'\i}guez-Torres}, {Feng}, {Metcalf}, {Wechsler},
  {Zhao}, {To}, {Alam}, {Banerjee}, {DeRose}, {Giocoli}, {Knebe}, \&
  {Reyes}}]{Chuang2019}
{Chuang}, C.-H., {Yepes}, G., {Kitaura}, F.-S., {et~al.} 2019, \mnras, 487, 48

\bibitem[{{Clerc} {et~al.}(2018){Clerc}, {Ramos-Ceja}, {Ridl}, {Lamer},
  {Brunner}, {Hofmann}, {Comparat}, {Pacaud}, {K{\"a}fer}, {Reiprich},
  {Merloni}, {Schmid}, {Brand}, {Wilms}, {Friedrich}, {Finoguenov}, {Dauser},
  \& {Kreykenbohm}}]{Clerc2018}
{Clerc}, N., {Ramos-Ceja}, M.~E., {Ridl}, J., {et~al.} 2018, \aap, 617, A92

\bibitem[{{Comparat} {et~al.}(2020){Comparat}, {Eckert}, {Finoguenov},
  {Schmidt}, {Sanders}, {Nagai}, {Lau}, {K�fer}, {Pacaud}, {Clerc},
  {Reiprich}, {Bulbul}, {Chitham}, {Chiang}, {Ghirardini}, {Gonzalez-Perez},
  {Gozaliasl}, {Fitzpatrick}, {Klypin}, {Merloni}, {Nandra}, {Liu}, {Prada},
  {Ramos-Ceja}, {Salvato}, {Seppi}, {Tempel}, \& {Yepes}}]{Comparat2020}
{Comparat}, J., {Eckert}, D., {Finoguenov}, A., {et~al.} 2020, The Open Journal
  of Astrophysics, 3, 13

\bibitem[{{Comparat} {et~al.}(2019){Comparat}, {Merloni}, {Salvato}, {Nandra},
  {Boller}, {Georgakakis}, {Finoguenov}, {Dwelly}, {Buchner}, {Del Moro},
  {Clerc}, {Wang}, {Zhao}, {Prada}, {Yepes}, {Brusa}, {Krumpe}, \&
  {Liu}}]{Comparat2019}
{Comparat}, J., {Merloni}, A., {Salvato}, M., {et~al.} 2019, \mnras, 487, 2005

\bibitem[{{Dauser} {et~al.}(2019){Dauser}, {Falkner}, {Lorenz}, {Kirsch},
  {Peille}, {Cucchetti}, {Schmid}, {Brand}, {Oertel}, {Smith}, \&
  {Wilms}}]{Dauser2019}
{Dauser}, T., {Falkner}, S., {Lorenz}, M., {et~al.} 2019, \aap, 630, A66

\bibitem[{{Dennerl} {et~al.}(2020){Dennerl}, {Andritschke}, {Br{\"a}uninger},
  {Burkert}, {Burwitz}, {Emberger}, {Freyberg}, {Friedrich}, {Gaida},
  {Granato}, {Hartner}, {von Kienlin}, {Meidinger}, {Menz}, \&
  {Predehl}}]{Dennerl2020}
{Dennerl}, K., {Andritschke}, R., {Br{\"a}uninger}, H., {et~al.} 2020, in
  Society of Photo-Optical Instrumentation Engineers (SPIE) Conference Series,
  Vol. 11444, Society of Photo-Optical Instrumentation Engineers (SPIE)
  Conference Series, 114444Q

\bibitem[{{Deponte} \& {Primini}(1993)}]{Deponte1993}
{Deponte}, J. \& {Primini}, F.~A. 1993, in Astronomical Society of the Pacific
  Conference Series, Vol.~52, Astronomical Data Analysis Software and Systems
  II, ed. R.~J. {Hanisch}, R.~J.~V. {Brissenden}, \& J.~{Barnes}, 425

\bibitem[{{Ebeling} \& {Wiedenmann}(1993)}]{Ebeling1993}
{Ebeling}, H. \& {Wiedenmann}, G. 1993, \pre, 47, 704

\bibitem[{{Freeman} {et~al.}(2002){Freeman}, {Kashyap}, {Rosner}, \&
  {Lamb}}]{Freeman2002}
{Freeman}, P.~E., {Kashyap}, V., {Rosner}, R., \& {Lamb}, D.~Q. 2002, \apjs,
  138, 185

\bibitem[{{Georgakakis} {et~al.}(2008){Georgakakis}, {Nandra}, {Laird}, {Aird},
  \& {Trichas}}]{Georgakakis2008}
{Georgakakis}, A., {Nandra}, K., {Laird}, E.~S., {Aird}, J., \& {Trichas}, M.
  2008, \mnras, 388, 1205

\bibitem[{{Ghirardini} {et~al.}(2019){Ghirardini}, {Eckert}, {Ettori},
  {Pointecouteau}, {Molendi}, {Gaspari}, {Rossetti}, {De Grandi}, {Roncarelli},
  {Bourdin}, {Mazzotta}, {Rasia}, \& {Vazza}}]{Ghirardini2019}
{Ghirardini}, V., {Eckert}, D., {Ettori}, S., {et~al.} 2019, \aap, 621, A41

\bibitem[{{Hasinger} {et~al.}(1994){Hasinger}, {Johnston}, \&
  {Verbunt}}]{Hasinger1994}
{Hasinger}, G., {Johnston}, H.~M., \& {Verbunt}, F. 1994, \aap, 288, 466

\bibitem[{{LaMassa} {et~al.}(2013){LaMassa}, {Urry}, {Cappelluti}, {Civano},
  {Ranalli}, {Glikman}, {Treister}, {Richards}, {Ballantyne}, {Stern},
  {Comastri}, {Cardamone}, {Schawinski}, {B{\"o}hringer}, {Chon}, {Murray},
  {Green}, \& {Nandra}}]{LaMassa2013}
{LaMassa}, S.~M., {Urry}, C.~M., {Cappelluti}, N., {et~al.} 2013, \mnras, 436,
  3581

\bibitem[{{Liu} {et~al.}(2021{\natexlab{a}}){Liu}, {Bulbul}, {Ghirardini},
  {Liu}, {Klein}, {Clerc}, {Oezsoy}, {Ramos-Ceja}, {Pacaud}, {Comparat},
  {Okabe}, {Bahar}, {Biffi}, {Brunner}, {Brueggen}, {Buchner}, {Ider Chitham},
  {Chiu}, {Dolag}, {Gatuzz}, {Gonzalez}, {Hoang}, {Lamer}, {Merloni}, {Nandra},
  {Oguri}, {Ota}, {Predehl}, {Reiprich}, {Salvato}, {Schrabback}, {Sanders},
  {Seppi}, \& {Thibaud}}]{Liu2021_cluster}
{Liu}, A., {Bulbul}, E., {Ghirardini}, V., {et~al.} 2021{\natexlab{a}}, arXiv
  e-prints, arXiv:2106.14518

\bibitem[{{Liu} {et~al.}(2021{\natexlab{b}}){Liu}, {Buchner}, {Nandra},
  {Merloni}, {Dwelly}, {Sanders}, {Salvato}, {Arcodia}, {Brusa}, {Wolf},
  {Georgakakis}, {Boller}, {Krumpe}, {Lamer}, {Waddell}, {Urrutia}, {Schwope},
  {Robrade}, {Wilms}, {Dauser}, {Comparat}, {Toba}, {Ichikawa}, {Iwasawa},
  {Shen}, \& {Ibarra Medel}}]{Liu2021_AGN}
{Liu}, T., {Buchner}, J., {Nandra}, K., {et~al.} 2021{\natexlab{b}}, arXiv
  e-prints, arXiv:2106.14522

\bibitem[{{Liu} {et~al.}(2020){Liu}, {Merloni}, {Simm}, {Green}, {Brandt},
  {Schneider}, {Dwelly}, {Salvato}, {Buchner}, {Shen}, {Nandra}, {Georgakakis},
  \& {Ho}}]{Liu2020}
{Liu}, T., {Merloni}, A., {Simm}, T., {et~al.} 2020, \apjs, 250, 32

\bibitem[{{Liu} {et~al.}(2013){Liu}, {Tozzi}, {Tundo}, {Moretti}, {Wang},
  {Rosati}, \& {Guglielmetti}}]{Liu2013}
{Liu}, T., {Tozzi}, P., {Tundo}, E., {et~al.} 2013, \aap, 549, A143

\bibitem[{{Merloni} {et~al.}(2012){Merloni}, {Predehl}, {Becker},
  {B{\"o}hringer}, {Boller}, {Brunner}, {Brusa}, {Dennerl}, {Freyberg},
  {Friedrich}, {Georgakakis}, {Haberl}, {Hasinger}, {Meidinger}, {Mohr},
  {Nandra}, {Rau}, {Reiprich}, {Robrade}, {Salvato}, {Santangelo}, {Sasaki},
  {Schwope}, {Wilms}, \& {German eROSITA Consortium}}]{Merloni2012}
{Merloni}, A., {Predehl}, P., {Becker}, W., {et~al.} 2012, arXiv e-prints,
  arXiv:1209.3114

\bibitem[{{Predehl} {et~al.}(2021){Predehl}, {Andritschke}, {Arefiev},
  {Babyshkin}, {Batanov}, {Becker}, {B{\"o}hringer}, {Bogomolov}, {Boller},
  {Borm}, {Bornemann}, {Br{\"a}uninger}, {Br{\"u}ggen}, {Brunner}, {Brusa},
  {Bulbul}, {Buntov}, {Burwitz}, {Burkert}, {Clerc}, {Churazov}, {Coutinho},
  {Dauser}, {Dennerl}, {Doroshenko}, {Eder}, {Emberger}, {Eraerds},
  {Finoguenov}, {Freyberg}, {Friedrich}, {Friedrich}, {F{\"u}rmetz},
  {Georgakakis}, {Gilfanov}, {Granato}, {Grossberger}, {Gueguen}, {Gureev},
  {Haberl}, {H{\"a}lker}, {Hartner}, {Hasinger}, {Huber}, {Ji}, {Kienlin},
  {Kink}, {Korotkov}, {Kreykenbohm}, {Lamer}, {Lomakin}, {Lapshov}, {Liu},
  {Maitra}, {Meidinger}, {Menz}, {Merloni}, {Mernik}, {Mican}, {Mohr},
  {M{\"u}ller}, {Nandra}, {Nazarov}, {Pacaud}, {Pavlinsky}, {Perinati},
  {Pfeffermann}, {Pietschner}, {Ramos-Ceja}, {Rau}, {Reiffers}, {Reiprich},
  {Robrade}, {Salvato}, {Sanders}, {Santangelo}, {Sasaki}, {Scheuerle},
  {Schmid}, {Schmitt}, {Schwope}, {Shirshakov}, {Steinmetz}, {Stewart},
  {Str{\"u}der}, {Sunyaev}, {Tenzer}, {Tiedemann}, {Tr{\"u}mper}, {Voron},
  {Weber}, {Wilms}, \& {Yaroshenko}}]{Predehl2021}
{Predehl}, P., {Andritschke}, R., {Arefiev}, V., {et~al.} 2021, \aap, 647, A1

\bibitem[{{Seppi} {et~al.}(2021){Seppi}, {Comparat}, {Nandra}, {Bulbul},
  {Prada}, {Klypin}, {Merloni}, {Predehl}, \& {Ider Chitham}}]{Seppi2021}
{Seppi}, R., {Comparat}, J., {Nandra}, K., {et~al.} 2021, \aap, 652, A155

\bibitem[{{Smith} {et~al.}(2001){Smith}, {Brickhouse}, {Liedahl}, \&
  {Raymond}}]{Smith2001}
{Smith}, R.~K., {Brickhouse}, N.~S., {Liedahl}, D.~A., \& {Raymond}, J.~C.
  2001, \apjl, 556, L91

\end{thebibliography}

\begin{appendix}
\section{Accuracy of the source property measurement}
\label{sec:accuracy}
\begin{figure}[hptb]
\begin{center}
\includegraphics[width=0.8\columnwidth]{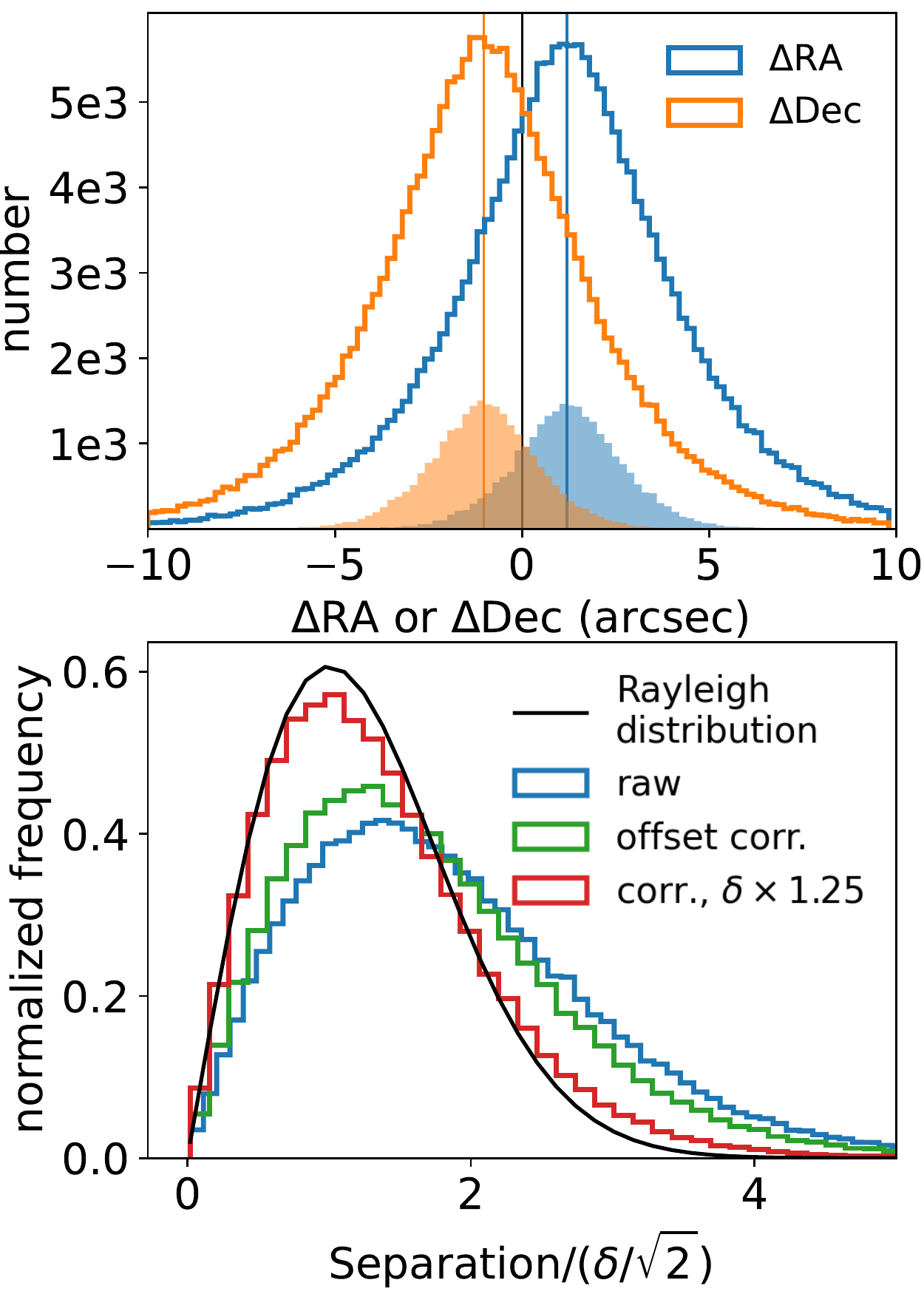}
\caption{Distributions of RA$_{out}$-RA$_{in}$ (blue, top) and Dec$_{out}$-Dec$_{in}$ (orange, top). The filled histograms indicate subsamples of very bright sources with detection likelihoods $>$100. The full samples and the subsamples have a median $\Delta$RA$=1.2\arcsec$ (vertical blue line) and a median $\Delta$Dec$=-1.0\arcsec$ (vertical orange line).
  The lower panel displays the distribution of input-output separation in terms of \TL{the ratio to} the 1D positional uncertainty in comparison with the Rayleigh distribution. The blue histogram shows the raw measurements. The distribution after correcting the detected positions for the median offsets is plotted in green. Then we further multiply the measured positional uncertainty by a factor of 1.25 and plot the distribution in red.
}
\label{fig:sep}
\end{center}
\end{figure}

\TL{
In addition to the source detection efficiency, we also tested the measurement accuracy of point-source properties.
From the single-band-detected mock catalog, we selected the point sources (\texttt{EXT\_LIKE}$=$0) that were uniquely matched to input noncontaminated (\texttt{ID\_contam}<0) point sources and compared the input and output positions and source counts.}

  The upper panel of Fig.~\ref{fig:sep} compares the input and output positions of the bright sources with \texttt{DET\_LIKE}$>$10.
Offsets of RA$_{out}-$RA$_{in}=1.2\arcsec$, Dec$_{out}-$Dec$_{in}=-1.0\arcsec$ are found, which are independent of the source brightness.\edit{
  Compared to the native eROSITA CCD pixel sizes of 9.6\arcsec, these offsets are very small.
  One possible reason is the difference between the mock data and the real data, particularly in the process of converting the accurate photon coordinates into pixel coordinates. Subpixel calibration of a photon position is achieved for eROSITA based on event patterns \citep{Dennerl2020}, but these details are not reproduced by SIXTE. Alternatively, it might also be caused by the numerical accuracy of the current PSF model, which has no subpixel resolution. To guarantee maximum positional accuracy, we suggest a post hoc astrometric correction of the X-ray catalogs by comparing the X-ray positions and optical-IR positions of distant AGN \citep[e.g., as done in][]{Brunner2021}.
  }

A good measurement of positional uncertainty is essential to identify the optical counterpart of an X-ray source from deep optical surveys, as there are often multiple candidates consistent in position within the uncertainty.
\TL{The lower panel of Fig.~\ref{fig:sep} displays the distribution of separation of all the matched input-output pairs in terms of $\sigma$ level (ratio to the positional uncertainty).
After correcting the offsets found above, this distribution is still broader than expected (the Rayleigh distribution), indicating a slight underestimation of the positional uncertainty.} We found that by enlarging the measured positional uncertainty by a factor of 1.25, the distribution becomes comparable with the Rayleigh distribution.

\TL{
For the simulation, SIXTE adopts the 2D PSF image library.
To detect the real eFEDS catalog, we performed photon-mode PSF fitting using \texttt{ermldet}, which accounts for the PSF of each photon based on a shapelet PSF library \citep{Brunner2021}.
The same photon-mode PSF fitting was performed on the mock data in this work, providing a measurement of source counts, which was corrected for the PSF loss outside the extracting circle. Comparing these measured counts with the input photon counts of each source in the soft (0.2--2.3~keV, in the single-band detection) and hard (2.3--5~keV, in the three-band detection) band, we find that the PSF-corrected source counts are $\sim 5\%$ lower than the input.
We repeated the \texttt{ermldet} PSF fitting of the mock data in image-mode, that is, adopting the 2D PSF image library, and find that the input and output source counts are perfectly consistent.
Therefore, the current (version c001) shapelet PSF library and the 2D PSF library deviate slightly in the normalization. This issue will be addressed in a future version of the eROSITA calibration database.
}

\section{Catalog description}
\label{sec:catalogs}
We provide the input AGN and cluster catalogs and the output single-band-detected and three-band-detected catalogs, whose columns are described in Table.~\ref{table:cat}.
\edit{These mock catalogs are available on the eROSITA Early Data Release website}\footnote{\url{https://erosita.mpe.mpg.de/edr/eROSITAObservations/Catalogues/}}.
The input catalogs contain a number of columns of AGN or cluster properties drawn from \citet{Comparat2019, Comparat2020} and a few columns of the output-counterpart ID (-99 means that it does not exist).
The very faint input AGN with fluxes below $5\times 10^{-16}$ \egs \ are omitted in the input AGN catalog.
The output catalogs contain a few essential columns that are the same as in the real eFEDS catalog \citep{Brunner2021} and a few additional columns of the input-counterpart ID (-99 means that it does not exist).

\begin{table*}
  \centering
  \caption{Columns of the mock input and output eFEDS catalogs}
  \begin{tabular}{p{0.2\textwidth} p{0.7\textwidth}}
   \hline

    \hline
    Name & Description \\
    \hline
    \multicolumn{2}{c}{For both input AGN and clusters}\\
    \hline
SRC\_ID             & Source ID; always $<5\times 10^6$ for AGN; always $>10^7$ for clusters \\
RA                 & Right ascension (deg) \\
DEC                & Declination (deg) \\
z                  & redshift (in redshift space) \\
dL                 & luminosity distance (cm) in the cosmology adopted by \citet{Comparat2019,Comparat2020} to create the mock catalogs \\
ID\_Uniq            & source ID in single-band catalog whose ID\_Uniq is this source \\
ID\_Any             & source ID in single-band catalog whose ID\_Any is this source \\
ID\_Any2            & source ID in single-band catalog whose ID\_Any2 is this source \\
ID\_contam          & ID of contaminating input source \\
ID\_Uniq\_T          & source ID in three-band catalog whose ID\_Uniq is this source \\
ID\_Any\_T           & source ID in three-band catalog whose ID\_Any is this source \\
ID\_Any2\_T          & source ID in three-band catalog whose ID\_Any2 is this source \\
Nreal              & number of realization (1--18) \\
    \hline
    \multicolumn{2}{c}{For input AGN only, see \citet{Comparat2019} for more details}\\
    \hline
logNH              & AGN column density (cm$^{-2}$) \\
FLUX               & 0.5-2 keV observed flux (\egs); the input flux in simulation \\
FLUX\_GalCorr       & 0.5-2 keV flux corrected for Galactic absorption (\egs) \\
Lx\_hard            & 2-10 keV intrinsic luminosity (erg/s) \\
agn\_type           & AGN type \\
galaxy\_SMHMR\_mass  & Host galaxy stellar mass (M$_{\sun}$) \\
    \hline
    \multicolumn{2}{c}{For input cluster only, see \citet{Comparat2020} for more details}\\
    \hline
kT                 & cluster tempeature (keV) \\
Lx\_soft            & rest-frame 0.5-2 keV luminosity within R500c (erg/s) \\
Lx\_soft\_2R500      & rest-frame 0.5-2 keV luminosity within twice R500c (erg/s) \\
Fx\_soft            & observed 0.5-2 keV flux within R500c \\
Fx\_soft\_2R500      & observed 0.5-2 keV flux within twice R500c; the input flux in simulation \\
M500c              & halo mass within R500c (M$_{\sun}$) \\
R500c\_arcmin       & apparent radius where the mass density is 500 times the universe critical density (arcmin) \\
R500c\_kpc          & physical radius where the mass density is 500 times the universe critical density (kpc, proper distance) \\
pid           & flag; a value of -1 means the halo is a distinct halo \\
Mvir          & halo mass within the virial radius (M$_{\sun}$) \\
Rvir          & halo virial radius (kpc) \\
Xoff          & halo offset parameter (kpc) \\
b\_to\_a\_500c   & halo ellipticity \\
    \hline
    \multicolumn{2}{c}{Single-band and three-band detected output catalogs}\\
    \hline
ID\_SRC             & source ID \\
RA                 & Right ascension (deg) \\
DEC                & Declination (deg) \\
RADEC\_ERR          & combined positional error (arcsec) \\
EXT                & source extent (arcsec) \\
EXT\_ERR            & extent error (arcsec) \\
EXT\_LIKE           & extent likelihood measured by PSF-fitting \\
ML\_RATE{\sl\_n}            & source count rate (cts/s) \\
ML\_RATE\_ERR{\sl\_n}        & count rate error (cts/s) \\
DET\_LIKE{\sl\_n}           & detection likelihood measured by PSF-fitting \\
ML\_BKG{\sl\_n}             & background at source position (cts/arcmin$^2$) \\
ML\_EXP{\sl\_n}             & vignetted exposure (s) at source position \\
ID\_Uniq            & ID of the unique input counterpart \\
ID\_Any             & ID of the brightest input counterpart, allowing duplicate \\
ID\_Any2            & ID of the secondary input counterpart \\
ID\_contam          & ID of the input source contaminating the unique input counterpart \\
Nreal              & number of realization (1--18) \\
    \hline
\end{tabular}
\tablefoot{The input AGN and clusters catalogs have some columns in common. They are listed in the first section. Some exclusive columns are listed in the second and third sections.
  The output single-band-detected and three-band-detected catalogs share basically the same columns, except for a few columns from PSF fitting with an optional suffix \_$n$ in the name. The columns with $n=$ 1, 2, or 3 correspond to the three individual energy bands 0.2--0.6, 0.6--2.3, and 2.3--5~keV used in the three-band detection; $n=$0 indicates the summary value in the three-band detection; if without the \_$n$ suffix, the column corresponds to the single-band detection.
  \label{table:cat}
}
\end{table*}

\end{appendix}
\end{document}